\documentclass[12pt,preprint]{aastex}

\newcommand{\vecrho}{\vec{\rho}}
\newcommand{\amt}{\alpha_m(t)}

\newcommand{\ip}{S(\vecrho)}
\newcommand{\hip}{\hat{S}(\vecrho)}
\newcommand{\dm}{D_m}
\newcommand{\hdm}{\hat{D}_m}

\newcommand{\mv}{\mathcal{V}}
\newcommand{\fip}{\mathcal{F}_{\hip}}

\newcommand{\pvd}{p(\mathcal{V};\ip,\dm,\sigma_{th})}

\newcommand{\tpvd}{\tilde{p}(\mathcal{V};\ip,\dm, \sigma_{th})}
\newcommand{\dtb}{\Delta t_b}

\newcommand{\ta}{\Delta\alpha}
\newcommand{\bta}{\overline{\ta}}
\newcommand{\etal}{et al.}

\slugcomment{Accepted by AJ}

\shorttitle{Further evaluation of bootstrap resampling}
\shortauthors{Kemball \etal}

\begin{document}

\title{Further evaluation of bootstrap resampling as a tool \\
       for radio-interferometric imaging fidelity assessment}

\author{Athol Kemball}
\affil{Department of Astronomy and 
  Institute for Advanced Computing Applications and Technologies/NCSA, University of
    Illinois at Urbana-Champaign, 1002 W. Green Street, Urbana, IL 61801}

\author{Adam Martinsek}
\affil{Department of Statistics, University of Illinois at
Urbana-Champaign, 725 S. Wright Street, Champaign, IL 61820}

\author{Modhurita Mitra and Hsin-Fang Chiang}
\affil{Department of Astronomy, University of Illinois at
Urbana-Champaign, 1002 W. Green Street, Urbana, IL 61801}

\begin{abstract}

We report on a broader evaluation of statistical bootstrap resampling
methods as a tool for pixel-level calibration and imaging fidelity
assessment in radio interferometry. Pixel-level imaging fidelity
assessment is a challenging problem, important for the value it holds
in robust scientific interpretation of interferometric images,
enhancement of automated pipeline reduction systems needed to broaden
the user community for these instruments, and understanding
leading-edge direction-dependent calibration and imaging challenges
for future telescopes such as the Square Kilometer Array. This new
computational approach is now possible because of advances in
statistical resampling for data with long-range dependence and the
available performance of contemporary high-performance computing
resources. We expand our earlier numerical evaluation to span a
broader domain subset in simulated image fidelity and source
brightness distribution morphologies. As before, we evaluate the statistical
performance of the bootstrap resampling methods against direct Monte
Carlo simulation. We find both model-based and subsample bootstrap
methods to continue to show significant promise for the challenging
problem of interferometric imaging fidelity assessment, when evaluated
over the broader domain subset. We report on their measured
statistical performance and guidelines for their use and application
in practice. We also examine the performance of the underlying polarization
self-calibration algorithm used in this study over a range of parallactic angle coverage.

\end{abstract}


\section{Introduction}

Radio-interferometric image formation requires a solution for both the
source brightness distribution over the image field and the
interferometric array instrumental and signal propagation effects,
estimated jointly from measurements of the electric vector spatial
coherence function of the incident radiation field measured on each
interferometer baseline \citep[and references therein]{tho01}. The
coherence data are sparsely sampled, leading to an ill-posed inverse
imaging problem that requires regularization for convergent solution
\citep{cor99}. This regularization is typically imposed as a
constraint on the properties of the source brightness distributions
during deconvolution, including positivity and compact support
\citep{hog74}, or via information or entropy measures
\citep{nar84,cor85}.

The fidelity of the resulting source brightness distribution cannot be
readily estimated given the analytic intractability of the coupled,
non-linear calibration and imaging equation, combined with the fact
that the parent probability distribution of the measured spatial
coherence function is parametrized by the source brightness
distribution and the instrumental array calibration, both unknown a
prior at the time of observation. Instead, achieved image quality is
typically estimated heuristically \citep{eke86} or by approximate
global measures. Common metrics include the ratio of the image
brightness root-mean-square (rms) measured in regions of low
brightness, $\sigma_{off}$, to the thermal noise limit, $\sigma_{th}$,
as calculated for the array from the known antenna sensitivities and
receiver and system thermal noise levels, assuming idealized
observations of an unresolved point source with perfect array
calibration \citep{wro99}. Other measures include the achieved dynamic
range $d_r$ (which is not a measure of image fidelity per se),
expressed as the ratio of the peak brightness, $I_{peak}$ to the
off-source rms, ${I_{peak}\over{\sigma_{off}}}$ \citep{per86}, or the
use of the deepest negative in a total intensity image (where Stokes
$I > 0$) to derive a scaling relation to reduce the typical
under-estimation of $\sigma_{off}$ \citep{kem93}.

These gross measures of image quality are however compromised by their
idealized underlying assumptions, including that of implied
direction-independence of image fidelity across the field. In practice
image fidelity is not constant across the field. It is
direction-dependent (equivalently, pixel-dependent) due to residual,
unmodeled instrumental calibration errors jointly or separately in the
visibility- or image-plane, and the interaction of these residual
errors with non-linear deconvolution effects, amongst other
factors. As a result, the assessment of pixel-dependent image fidelity
is a general problem in radio interferometry. It is not confined to
the important problem of calibrating direction-dependent instrumental
errors specifically, as described for example by \citet{bha08}.

Although a challenging and largely unsolved problem, understanding
pixel-level radio-interferometric imaging fidelity is essential for
current and future telescope arrays. Current instruments need to be
made more accessible to the larger astronomical community; in
particular their user base needs to expand beyond those who have
invested the substantial time and effort required to acquire expertise
in effective radio-interferometric calibration and imaging data
reduction processes (heuristics as summarized for example by \citet{per89}). This
is best achieved through the provision of automated pipeline reduction
systems; these require associated estimates of image fidelity to allow
effective, broad-based community scientific interpretation and
analysis.

For leading-edge future arrays, such as the Square Kilometer Array
(SKA\footnote{http://www.skatelescope.org}), solving the problem of
quantitative pixel-level image fidelity assessment is key to their
effective design. Any interferometer that cannot reach its target
thermal noise limit in a representative integration needed to meet its
key science goals is dynamic-range limited. This is important for many
interferometers but is particularly acute for the SKA given its high
sensitivity. As a result, the SKA has stringent imaging dynamic range
requirements \citep{sch07}, particularly in continuum observing modes
where $d_r \sim 10^6$ will be required routinely across wide image fields
in rapid survey modes and $d_r \sim 10^7$ for individual, targeted
fields. This dynamic range is not achieved routinely by contemporary
radio-interferometric arrays except for the innermost pixels in a
handful of images, and then only after extensive custom reduction by
the most skilled radio interferometry practitioners. With increasing
projected radial distance from the field center the dynamic range may
typically decline by several orders of magnitude, for reasons noted
above. Contemporary examples of high dynamic-range and
high-sensitivity observations in radio interferometry, and their
associated challenges, are provided by \citet{gel00}, \citet{deb05},
and \citet{nor05}.

Exponential advances in currently available and anticipated
high-performance computing (HPC) capabilities and resources
\citep{bad08} allow fundamentally new approaches to the problem of
pixel-level radio-interferometric imaging fidelity assessment. Two
complementary approaches have recently been reported: a frequentist
statistical resampling method \citep[Paper I]{kem05} and a Bayesian
imaging technique \citep{sut06}. In Paper I, we described the first
application of statistical bootstrap resampling techniques to the
problem of interferometric imaging fidelity assessment. Statistical
resampling is an active area in contemporary statistics research
\citep{efr03,dav03}, and general reviews are provided in recent
monographs by \citet{dav97}, \citet{che99}, \citet{pol99},
\citet{lah03}, and \citet{zou04}.

In the context of future interferometer arrays, such as the SKA
discussed above, the statistical fidelity assessment methods discussed
here are particularly important in understanding the optimal design of
these telescopes and the contributing factors to their dynamic-range
imaging performance. Specifically, although resampling methods may prove
too computationally expensive for real-time calibration and imaging at
a telescope as computationally demanding as the SKA, they are very
valuable tools during SKA design and development. These methods can
also be used in off-line analysis of data from interferometer arrays
with smaller numbers of elements and lower associated computational
costs for calibration and imaging.

We note that resampling techniques as described here are not strongly sensitive to the specific calibration and imaging algorithm in use; their goal is instead to provide a measure of the statistical properties of the underlying calibration and imaging estimator. Our initial evaluation in Paper I considered fidelity assessment of
polarization calibration for Very Long Baseline Interferometry (VLBI)
arrays, a representative radio-interferometric calibration and imaging
problem of interest in its own right \citep{kem99}. The statistical
performance of both model-based and subsample bootstrap resampling was
evaluated by inter-comparison of the bootstrap results with those
obtained by direct Monte Carlo simulation for a single fixed
two-component polarized source model and array configuration. Our
initial study found both bootstrap resampling techniques to be
computationally tractable with modern HPC resources and to have good
statistical performance for image variance estimation for the single
source model and array configuration considered in that initial study.

In the current paper we describe results from broader evaluation of
the applicability of statistical resampling to radio-interferometric
imaging fidelity assessment. We extend the scope of our original
evaluation in Paper I by expanding the problem domain along two axes,
namely source model structure and array parallactic angle
coverage. Both parameters vary the expected degree of systematic error
in the polarimetric calibration and imaging estimator used in the
study and therefore the pixel-level morphology and magnitude of the
image fidelity distribution over which the bootstrap resampling
methods can be evaluated.

In the expanded evaluation reported here, we find here that the model-based
and sub-sample bootstrap resampling methods for data with long-range
statistical dependence, used in Paper I, continue to show good
statistical performance when applied over a wider range of observing
configurations and source models. We refine guidelines and rules for
their general use in radio interferometry and report initial results
on their applicability to image bias estimation.

The paper is structured as follows. In Section 2 the theory of
bootstrap resampling is summarized, as applied to the interferometric
calibration and imaging problem under examination. Section 3 describes
the numerical simulation methods used to measure the statistical
performance of the bootstrap resampling techniques under
evaluation. Simulation results are presented in Section 4 and
discussed in Section 5. In Section 6, we summarize the conclusions
from the current work.

\section{Theory}

In this section, we recap the key elements of the statistical
resampling method used in the current study for fidelity assessment of
interferometric polarization calibration and imaging. A full
description of the theory can be found in Paper I.

\subsection{The imaging equation for radio interferometry}

We adopt the radio-interferometric imaging equation developed by
\citet{ham96a}, \citet{sau96}, \citet{ham96b} and \citet{ham00}, and
generalized for image-plane effects by \citet{cor95} and
\citet{nor95}:

\begin{equation}
V_{mn} = \prod_{\kappa} \left[G_m^{\kappa} \otimes G_n^{\kappa*}\right] \int_{\Omega} \prod_{\kappa} \left[T_m^{\kappa}(\vecrho) \otimes T_n^{\kappa*}(\vecrho)\right] e^{-2\pi j \vec{b}_{mn} \cdot (\vecrho - \vec{\rho_s})}\ K\ \ip d\Omega \label{eqn-ie}
\end{equation}

where $j=\sqrt{-1}$, the term $V_{mn}$ is the measured complex 4-vector
of polarization correlations on the baseline $\vec{b}_{mn}$ between
antennas $m$ and $n$ (by common convention referred to as visibilities
in this discipline), $\ip$ is the Stokes 4-vector of radio brightness
in unit direction $\vecrho$, vector $\vec{\rho_s}$ is the center of
the field $\Omega$, and $K$ is a constant $(4 \times 4)$ matrix that
maps the Stokes parameters $\{I,Q,U,V\}$ into the polarization
receptor basis (e.g. orthogonal circular or linear) of the visibility
polarization correlations. Direction-independent gains of instrumental
type $\kappa$ at antenna $m$ are denoted $G_m^{\kappa}$, with
associated direction-dependent gains $T_m^{\kappa}=f(\vecrho)$; both
are $(2 \times 2)$ Jones matrices in the polarization receptor basis,
and of arbitrary (e.g. pixel or functional) parametrization. The outer
matrix product is denoted by $\otimes$, and complex conjugation by an
asterisk.

\subsection{Polarization self-calibration}

For the interferometric polarization self-calibration and imaging
problem considered here as a representative problem for the study of
imaging fidelity assessment, the imaging equation~\ref{eqn-ie}
contains only direction-independent Jones matrices $G_m^{\kappa} =
\{P_m,\ \dm\}$, i.e. $T_m^{\kappa}$ is the null set $\{\emptyset\}$ (or equivalently, the $T_m^\kappa$ are identity matrices), and where
$P_m=\rm{diag}(e^{-j \amt}, e^{+j \amt})$ is the Jones matrix
containing the feed parallactic angle term $\amt$, known analytically,
and $\dm$ is an anti-diagonal matrix containing the antenna-based
instrumental polarization leakage terms for each nominally orthogonal
polarization receptor basis. As described in Paper I, our polarization
self-calibration method, as a joint iterative solution for $\dm$ and
$\ip$, is a subset of general self-calibration within the imaging
equation framework. For each cycle of (non-progressive) iterative
refinement of $\ip$, we solve for the instrumental polarization by
minimizing $\chi^2$, formed as the $L_2$ complex norm at the position
of the unknown $\dm$ in equation~\ref{eqn-ie}, by integrating the
imaging equation from both the right-hand and left-hand sides to that
position.

\subsection{Imaging fidelity assessment by bootstrap resampling}

Before considering bootstrap resampling as a technique for imaging
fidelity assessment we need to formulate the problem of
radio-interferometric calibration and imaging as a statistical
inference problem. General statistical inference and estimation in
signal processing is described by \citet{kay93}. The measured
visibility dataset, $V_{mn}^{obs}=V_{mn} + \mathcal{N},\
\forall(m,n,t)$, can be considered a single realization of a vector
over time $t$ of random variables $\mv$ drawn from a joint,
multi-variate parent probability density function (PDF) of parametric
form $\pvd$. Independent, identically-distributed (IID) thermal noise
in each visibility measurement is denoted by $\mathcal{N}$. We
consider the joint self-calibration solvers for $D_m$ and $\ip$ as
statistical point estimators, denoted $\hdm$ and $\hip$
respectively. In this standard statistical inference framework, the
problem of imaging fidelity assessment is that of determining the
sampling distribution $\fip$ of $\hip$. As described in Paper I, this
problem is neither tractable analytically, nor is $\pvd$ (or therefore
$\fip$) known a priori.

Statistical resampling offers an alternative inference method for
$\fip$ that does not require analytic tractability or detailed
knowledge of $\pvd$. These methods are computationally expensive
however, but contemporary advances in HPC capabilities now make them
viable approaches. The single realization of $\mv$, comprising the
observed visibility dataset, can be used to construct an empirical
distribution function $\tpvd$. Resamples $\mv^*$ drawn from $\tpvd$,
conditional on the observed data $V_{mn}^{obs}$, and under statistical
conditions where the bootstrap is applicable, mirror the statistical
relationship between $\mv$ and the unknown parent distribution
$\pvd$. The imaging estimator $\hip$, is here chosen to represent the
estimator for the restored image. Acting on the resampled visibility
datasets, this estimator yields images $S^*$; their statistical
distribution relative to $S^*_{xy_0}$, which is obtained from the $\hip$ acting
on the observed (template) realization, provides a bootstrap estimate
of $\fip$ and hence an assessment of imaging fidelity by our current
definition. Here the subscript $xy$ denotes the use of a pixel basis for the images.

Radio-interferometric data have long-range statistical dependence and
thus require bootstrap innovations developed for dependent data
\citep{lah03}. In Paper I, we demonstrated the successful use of two
such methods, namely the model-based and subsample bootstraps, for a
single simulated test case in polarization self-calibration and
imaging. We present an expanded evaluation in the following sections.

\section{Simulation methods}

\subsection{Run codes}

The primary goal of this study is to evaluate the statistical
performance of the bootstrap resampling techniques advanced in Paper I
over a larger domain subset. We expand the previous
study along two axes, namely the expected level of image fidelity and
the range of source brightness distribution morphologies
considered. We retain the same interferometric polarization and
imaging problem used in Paper I as a representative test problem in
this discipline. The details of our polarization self-calibration
heuristic is described in detail in Paper I; broadly summarized, this
is a joint solver for $\ip$ and $\dm$ over ten non-progressive
self-calibration iterations, starting from an unpolarized
unit-brightness point source brightness distribution and zero
instrumental polarization for each antenna in the array. In the
current work, we also retain the the simulated Very Long Baseline
Array (VLBA\footnote{The National Radio Astronomy Observatory is a
facility of the National Science Foundation operated under agreement
by Associated Universities, Inc.}) configuration and instrumental
polarization terms enumerated in Table 1 of Paper I.

\begin{figure}
\plotone{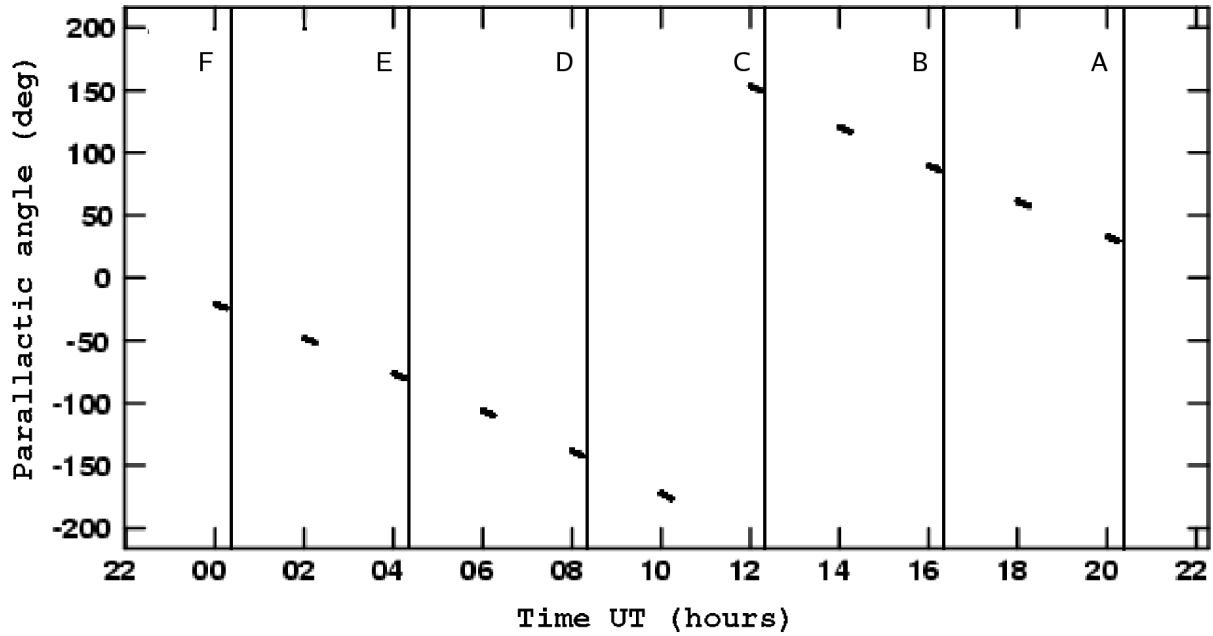}
\caption{Parallactic angle variation at the central antenna (Pie Town) for the
data used in the Monte Carlo simulations for run codes A-F, as
enumerated in Table~\ref{tbl-par}. Vertical lines denote the schedule
truncation for each run code, as used to vary the parallactic angle
range in the simulated data.\label{fig-par}}
\end{figure}

To vary the expected image fidelity, we adjust the range of
parallactic angle for the simulated array configuration by truncating
the simulated observation duration. The VLBA is comprised of antennas
with azimuth-elevation antenna mounts which, unlike equatorial mounts,
produce non-zero, time-variable feed parallactic angle variations
\citep{tho01}. These parallactic angle ranges are labeled as run codes
A through F, in order of decreasing parallactic angle range, and are
enumerated in Table~\ref{tbl-par}. A graphical representation is
provided in Figure~\ref{fig-par}. Run code A is the case of
nearly-complete parallactic angle coverage considered as the sole case
studied in Paper I. Interferometric polarization calibration of
linearly-polarized calibrators has a higher degree of systematic error
for reduced parallactic angle coverage; this results from the
increasing degeneracy introduced between the polarization basis
functions $e^{\pm j \amt}$ and the unknown full-polarization source
brightness distribution $\ip$ for low parallactic angle range
\citep{mor64,con69,cot84,rob84}.

\begin{figure}
\plotone{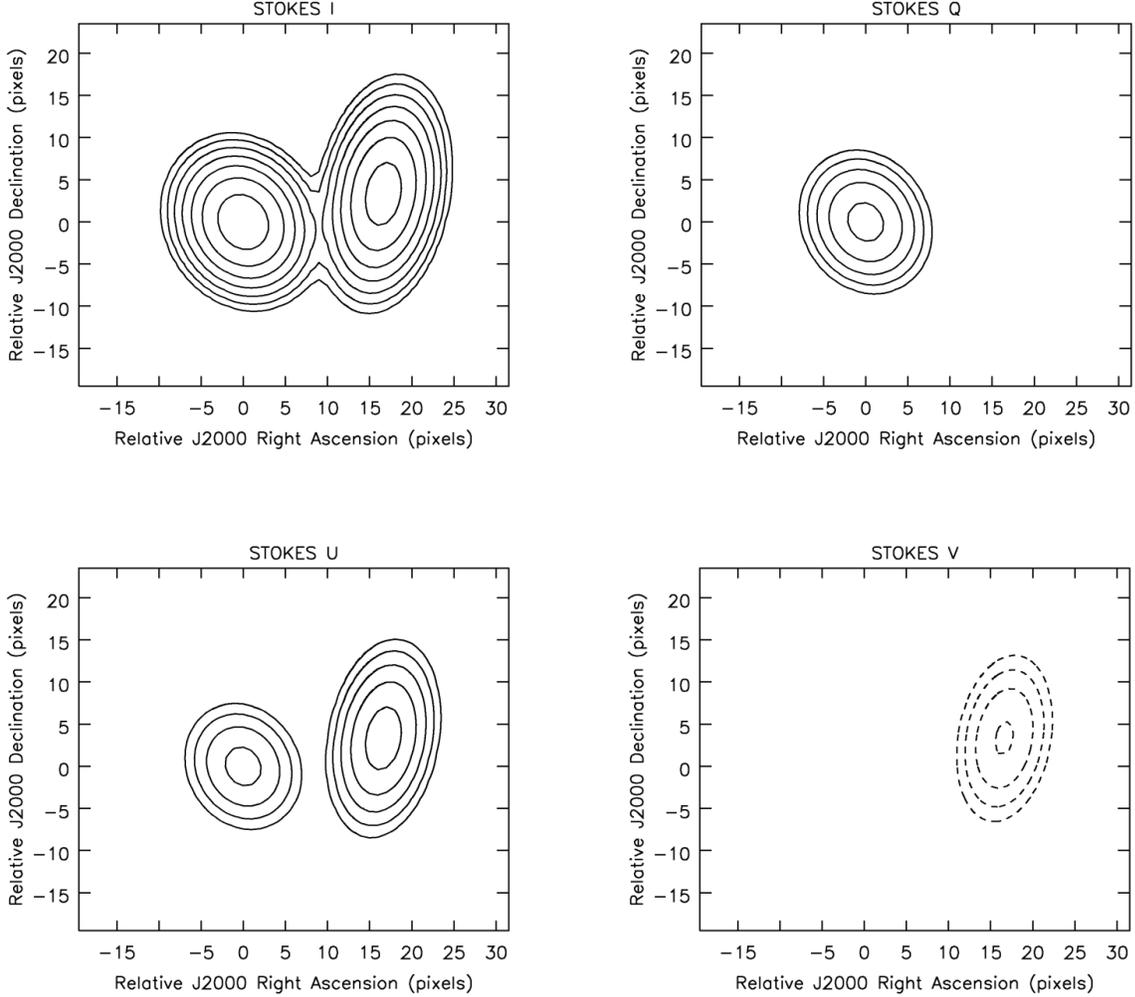}

\caption{Simulation source model, convolved with a circular restoring
beam of 156.007 $\mu as$, plotted in Stokes $\{I,Q,U,V\}$, using
contour levels of $\{-64,\ -32,\ -16,\ -8,\ -4,\ -2,\ -1,\ 1,\ 2,\ 4,\
8,\ 16,\ 32,\ 64\} \times 4.434 \times 10^{-3}$ Jy per beam. The peak
brightness in Stokes $\{I,Q,U,V\}$ is $4.434 \times 10^{-1}$, $8.867
\times 10^{-2}$, $9.591 \times 10^{-2}$, and $-3.836\times 10^{-2}$ Jy
per beam respectively. The source model consists of two polarized
Gaussian components, denoted here by subscripts one and two, with
parameters: $(\Delta\alpha_1=0,\ \Delta\delta_1=0),\
b_{maj_1}=0.2\ \rm{mas},\ b_{min_1}=0.15\ \rm{mas},\ \rm{P.A.}_1=30\
\rm{deg},\ I_1=1\ \rm{Jy},\ Q_1=0.2\ \rm{Jy},\ U_1=0.1\ \rm{Jy},\
V_1=0\ \rm{Jy},\ (\Delta\alpha_2=-0.5,\ \Delta\delta_2=+0.1)\
\rm{mas},\ b_{maj_2}=0.3\ \rm{mas},\ b_{min_2}=0.1\ \rm{mas},\
\rm{P.A.}_2=-10\ \rm{deg},\ I_2=1\ \rm{Jy},\ Q_2=0\ \rm{Jy},\
U_2=0.25\ \rm{Jy},\ V_2=-0.1\ \rm{Jy}$. This Figure is reproduced 
from Figure 1 and Table 2 in Paper I.\label{fig-true-ad}}

\end{figure}

To vary source polarization morphology, we decrease the angular
separation between the two simulated Gaussian components used in Paper
I (as parametrized in Table 2 of Paper I), so increasing their degree
of spatial overlap and hence polarization complexity. We reproduce the
source model in Stokes $\{I,Q,U,V\}$ at the original component
separation used in Paper I in Figure~\ref{fig-true-ad}, and include
the quantitative component parameter values in the caption to that
Figure. The component separations used in the current study of source
morphology variation are tabulated in Table~\ref{tbl-sep} in this
paper as run codes X, Y, and Z. For these run codes,however, we retain
the same complete parallactic angle range used in Paper I, i.e. the
same parallactic angle range as run code A in Table~\ref{tbl-par}.

\subsection{Monte Carlo reference simulations}

For each run code \{A-F,\ X-Z\}, we assess the statistical performance
of each bootstrap resampling method by inter-comparison with results
obtained by direct Monte Carlo simulation.  As in Paper I, an ensemble
of $N_s = 256$ visibility datasets were generated for each run code by
direct Monte Carlo simulation. Additive IID thermal noise
contributions $\mathcal{N}$ were drawn as a phasor from the complex
normal distribution $\mathcal{CN}(0,\sigma_{th}^2)$; this distribution is defined by \citet{kay93}.  The
joint polarization self-calibration and imaging estimator for $\hdm$
and $\hip$, was applied to each visibility dataset realization within the Monte Carlo ensemble generated for each run code. The bias and mean-squared error (MSE) of
the sampling distribution $\fip$ of $\hip$ were then computed for each
restored ensemble image, in a pixel basis $S_{xy}$, relative to the
true source model brightness distribution $\breve{S_xy}$ used to
generate the simulated data, the latter convolved with a common median
restoring beam matched separately to the spatial resolution of each
run code. The variance was computed as
$N_s^{-1}\sum (S_{xy}^2) - {\overline{S_{xy}}}^2$, where
$\overline{S_{xy}}$ is the sample mean across the ensemble, and the
MSE per pixel relative to the true model was computed as $N_s^{-1}\sum
{(S_{xy} - {\breve{S_xy}})}^2$. Each statistic was computed per image
pixel, per self-calibration iteration number, and per Stokes parameter
$\{I,Q,U,V\}$. These Monte Carlo estimates of the variance, bias,
mean, and MSE of the sampling distribution $\fip$ of the imaging
estimator $\hip$ provide an estimate of truth against which the
statistical performance of the bootstrap resampling methods can be
assessed.

\subsection{Bootstrap simulations}

Both the model-based and subsample bootstrap methods described in
Paper I were used in the current study. These bootstrap methods are
applicable to data with long-term statistical dependence, as is the
case in radio interferometry. The parameters for each bootstrap
method, both the model-based (M1-M4) and subsample (S1-S4) bootstrap
are reproduced here from Paper I as Table~\ref{tbl-mcode} and
Table~\ref{tbl-scode} respectively. The reader is referred to Paper I
for specific implementation details for these bootstrap methods. For
each run code \{A-F,\ X-Z\}, a bootstrap ensemble (of size 256 realizations) was generated for
each separate bootstrap method \{M1-M4,\ S1-S4\} by resampling from a
single template realization (here chosen to be either the 127th or
128th of $N_s=256$) taken from the matching Monte Carlo ensemble for that run
code. For each resulting bootstrap ensemble, the bias and MSE of the sampling
distribution of the imaging estimator, $\hip,$ were computed from the
restored images obtained across the ensemble, $S^*_{xy}$, relative to the
restored image, $S^*_{xy_0}$, obtained for the template visibility dataset
realization. This technique can be used directly with real data
obtained by physical observation, unlike Monte Carlo simulation, which
presupposes knowledge of the unknown true source model.

The primary purpose of this study is to assess if a single observed
visibility dataset realization can be used, through resampling, to
estimate imaging fidelity, quantified here as the variance, bias, and
MSE of the sampling distribution $\fip$ of a regularized imaging
estimator $\hip$. We assess the statistical performance of each
bootstrap method in the only practical way, namely against the
sampling distribution properties obtained by direct Monte Carlo
simulation. Conclusions regarding the statistical applicability of
bootstrap resampling for imaging fidelity obtained in this way
translate directly to real observations if the general validity of the
approach can be assessed numerically over a sufficiently complete
domain subset. As part of that more complete evaluation, for each run
code \{A-F,\ X-Z\}, we have completed a Monte Carlo simulation
(denoted by code MC) and a set of resampling simulations for bootstrap
codes \{M1-M4,S1-S4\}.

\subsection{HPC implementation}

As in Paper I, a modified version of the
AIPS++\footnote{http://aips2.nrao.edu} package was used to perform
polarization self-calibration and imaging, but for the work described
in this paper, compiled within the build framework for
radio-interferometric analysis software described by \citet{kem08}. In
addition to the changes to the code base to support polarization
self-calibration and statistical resampling, we also replaced the
default $\chi^2$ Newton solver for the imaging equation in the
original package with a conjugate-gradient solver as implemented in
the OptSolve++ package\footnote{OptSolve++ is distributed by Tech-X
corporation (http://www.techxhome.com)}.

The computations were parallelized over visibility dataset realization
within each bootstrap or Monte Carlo ensemble and run on a 32-bit
Intel Xeon Linux cluster operated as a national computational resource
by the National Center for Supercomputing
Applications\footnote{www.ncsa.uiuc.edu}. The cluster has a peak
single-precision floating point performance per processor of
approximately 12 Gflops. Each run was typically distributed over 64
processors and ran for $\sim$ 3-4 hours resulting in an integrated
compute cost of $O$(Tflops). We note as before, however, that the
problem is highly parallelizable and scalable over realization and has
good load balancing characteristics; only the template realization and
final statistical processing need to be performed in serial segments
of the code.

\section{Simulation results}

\begin{figure}
\plotone{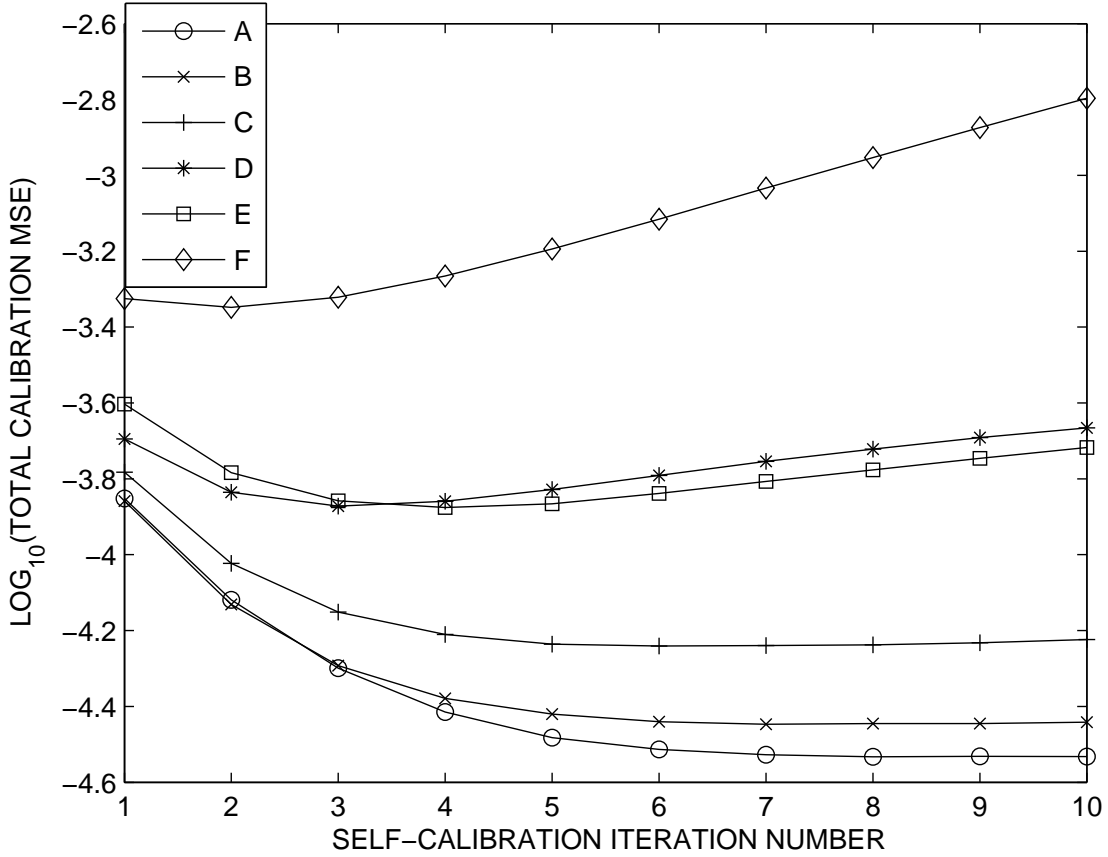}

\caption{Total mean-squared error in the off-diagonal $\dm$ matrix
elements, plotted for each run code \{A-F\} against polarization
self-calibration iteration number. As noted in Paper I, the total
calibration MSE is computed here as: ${{1}\over{2N}} \sum_{p \in
(R,L)} \sum_m^{N} (d_m^p - \breve{d}_{m}^p){(d_m^p -
\breve{d}_{m}^p)}^*$, where $d^p_m$ is the instrumental polarization
determined by the solver, and $\breve{d}_{m}^p$ is the true value used
to generate the simulated data. The decreasing parallactic angle
ranges for run codes \{A-F\} are listed in
Table~\ref{tbl-par}. \label{fig-calmse}}

\end{figure}

The performance of the polarization self-calibration algorithm in
direct Monte Carlo simulations over run codes \{A-F\} in shown in
Figure~\ref{fig-calmse}; these run codes have parallactic angle ranges
depicted in Figure~\ref{fig-par}. Here the MSE of the instrumental
polarization estimator $\hdm$, averaged over antenna and receptor
polarization, with reference to the known truth used in the
simulations, is separately plotted for each run code against
self-calibration iteration number. The detailed computation of the MSE
for $\hdm$ is described in the caption to Figure~\ref{fig-calmse}.

As described in preceding sections, the primary purpose of the current
work is to assess the statistical performance of the bootstrap
resampling techniques advanced in Paper I, over an expanded domain
subset in this discipline. This has been achieved by increasing the
range of imaging fidelity and source morphologies considered,
enumerated by run codes \{A-F,\ X-Z\} in Tables~\ref{tbl-par} and
~\ref{tbl-sep}. For each run code, a direct Monte Carlo simulation
(MC) was performed and bootstrap resampling methods \{M1-M4,\ S1-S4\}
subsequently applied to a single visibility dataset drawn from each
Monte Carlo ensemble.

\begin{figure}[h] 
\advance\leftskip-1cm
\advance\rightskip-1cm
 $\begin{array}{c@{\hspace{2mm}}c@{\hspace{2mm}}c} 
\includegraphics[width=57mm, height=57mm, trim = 0mm 5mm 10mm 5mm, clip]{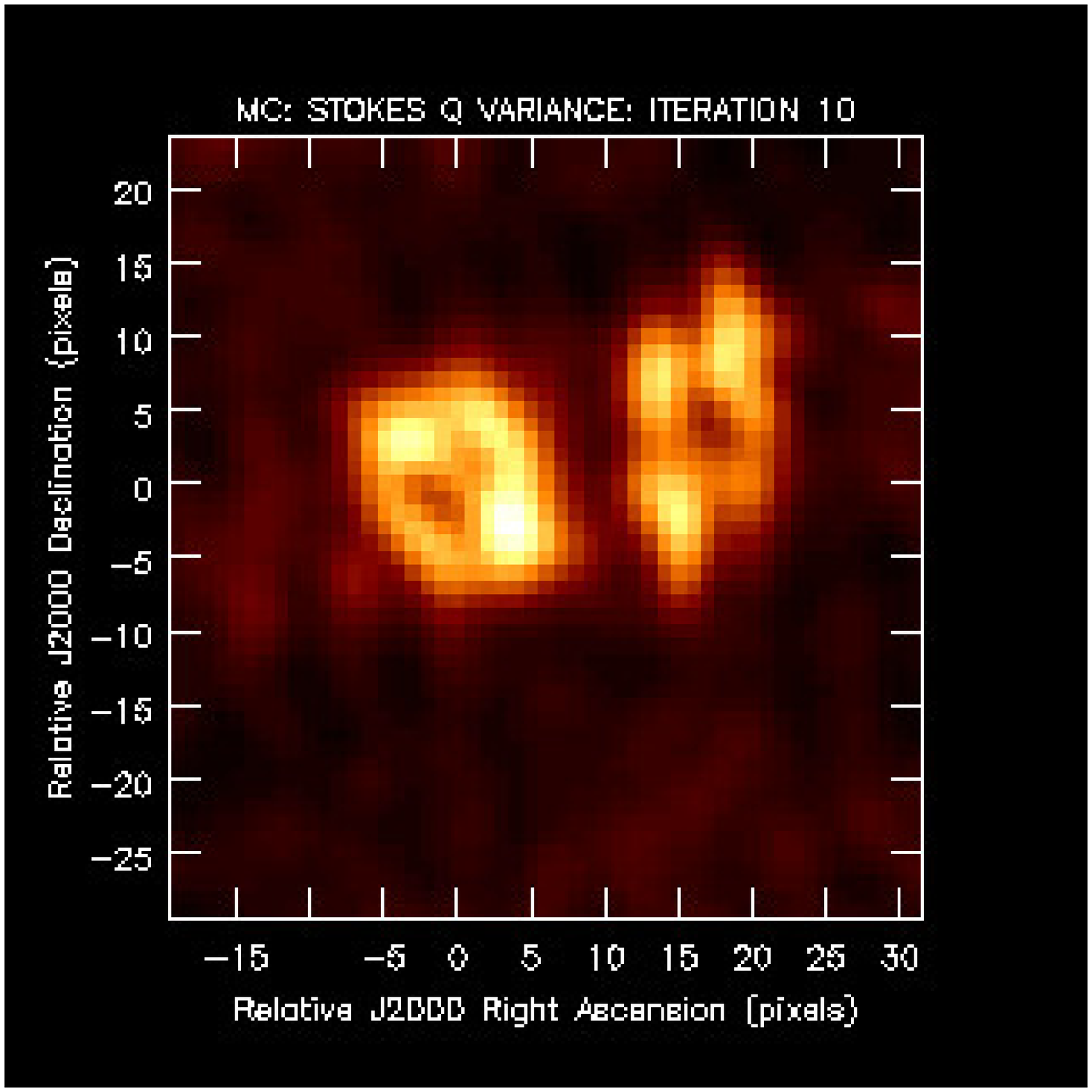} & 
\includegraphics[width=57mm, height=57mm, trim = 0mm 5mm 10mm 5mm, clip]{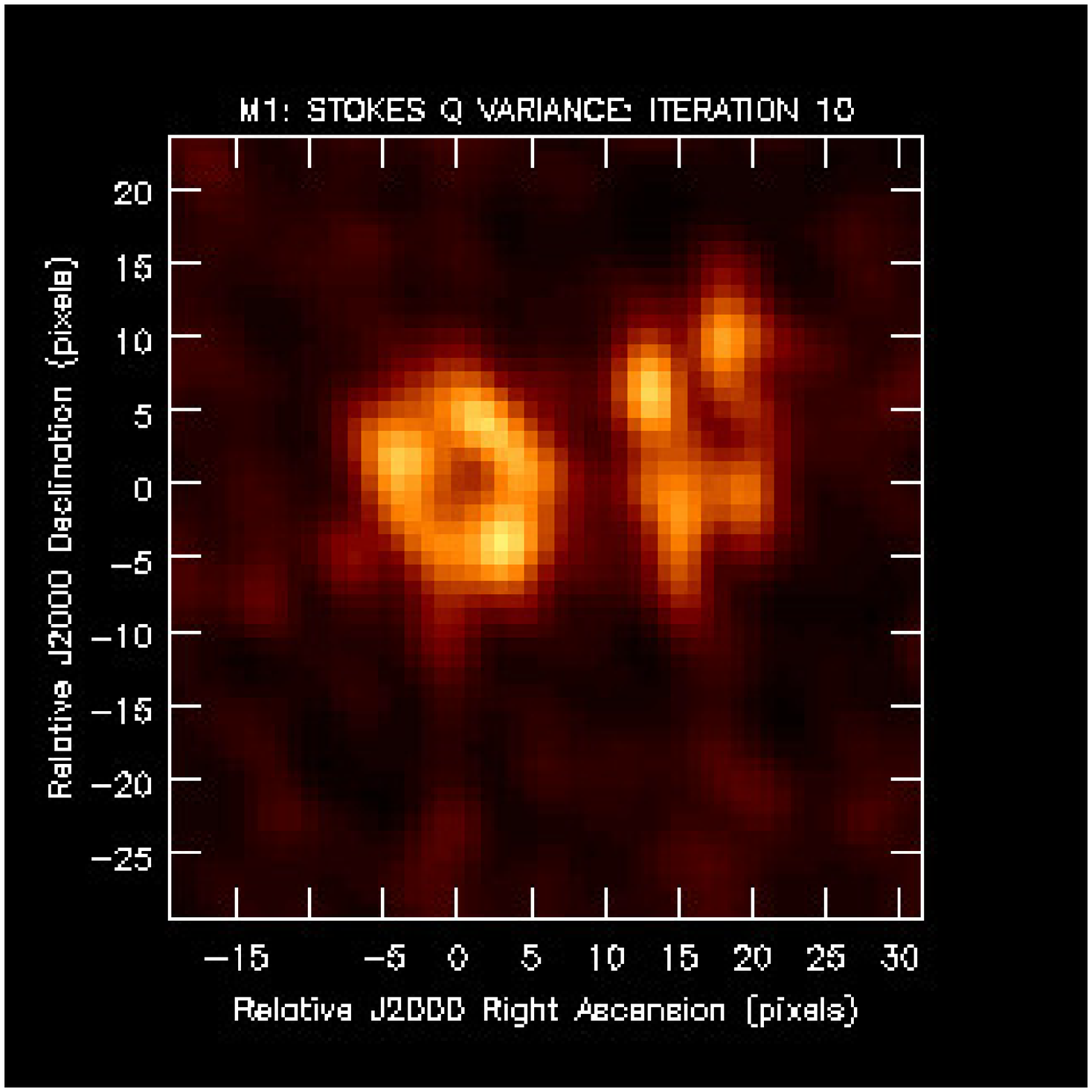} & 
\includegraphics[width=57mm, height=57mm, trim = 0mm 5mm 10mm 5mm, clip]{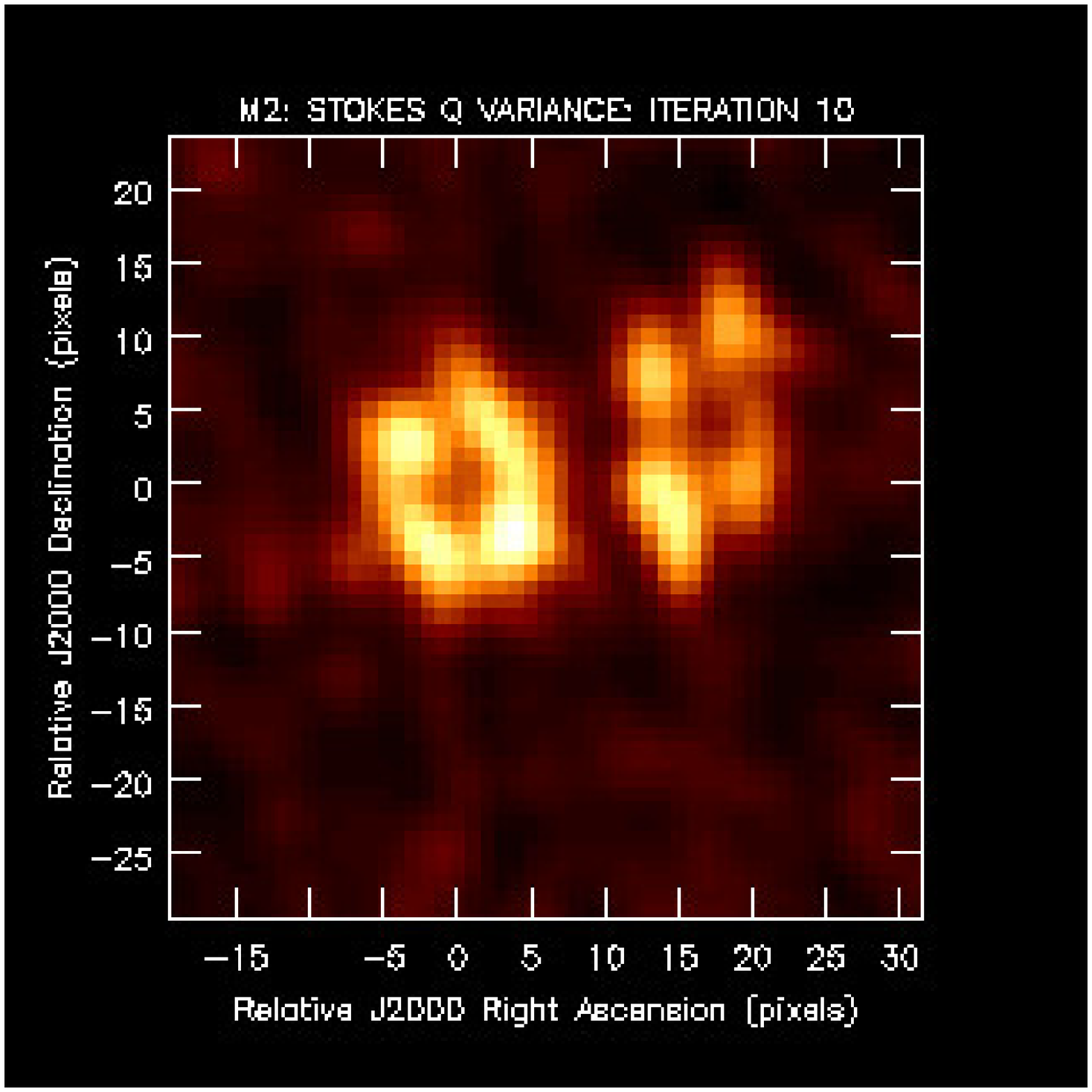} \\ 
\includegraphics[width=57mm, height=57mm, trim = 0mm 5mm 10mm 5mm, clip]{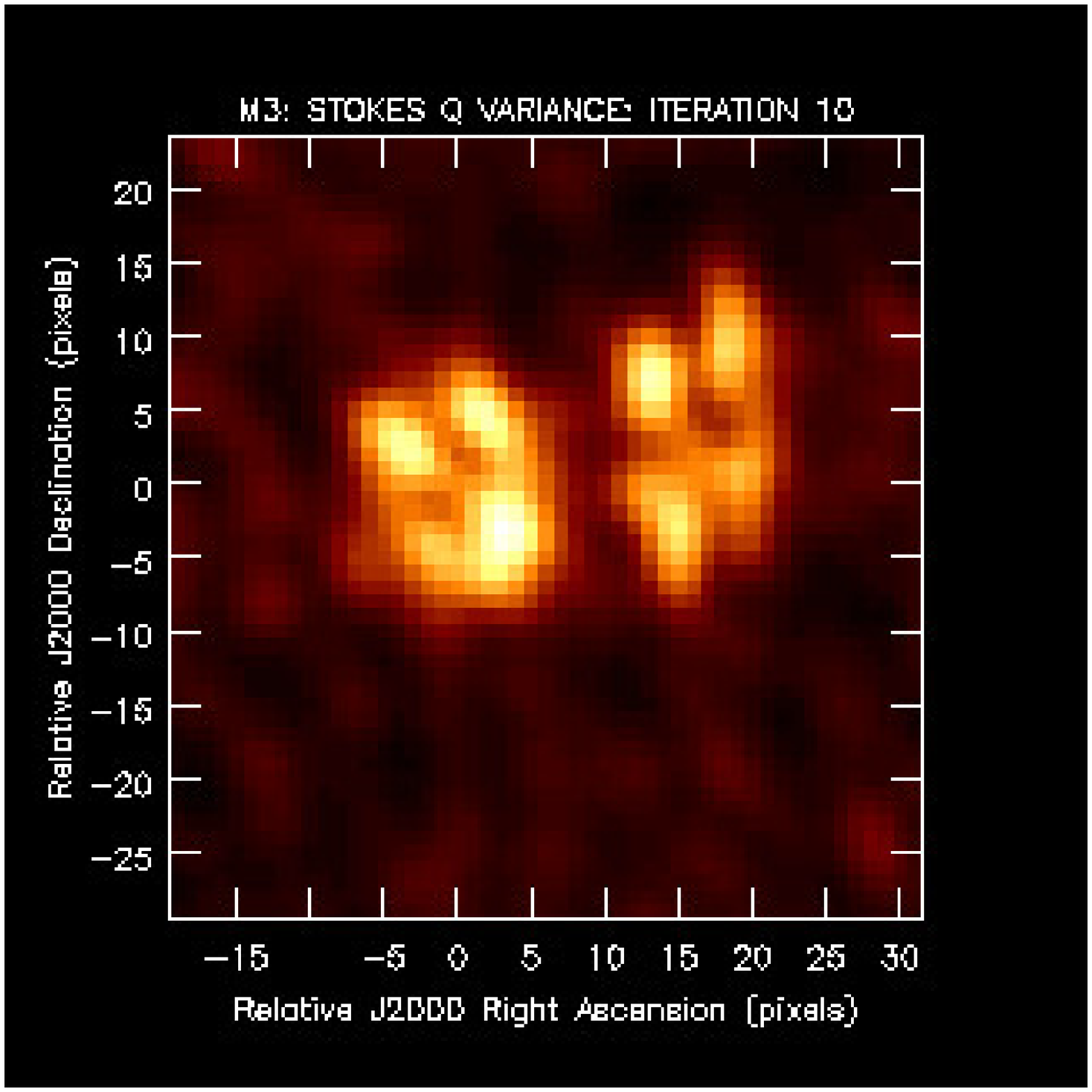} & 
\includegraphics[width=57mm, height=57mm, trim = 0mm 5mm 10mm 5mm, clip]{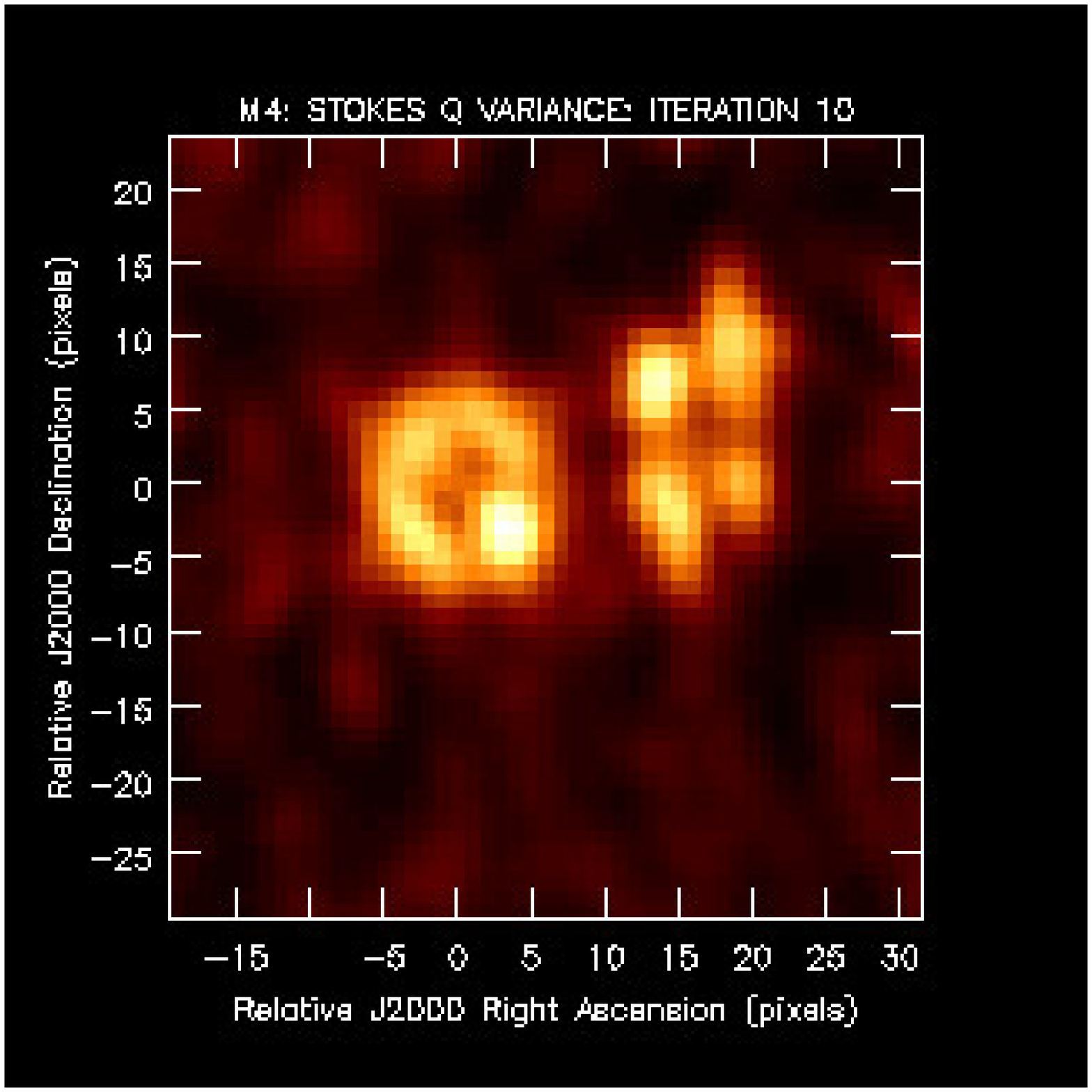} & 
\includegraphics[width=57mm, height=57mm, trim = 0mm 5mm 10mm 5mm, clip]{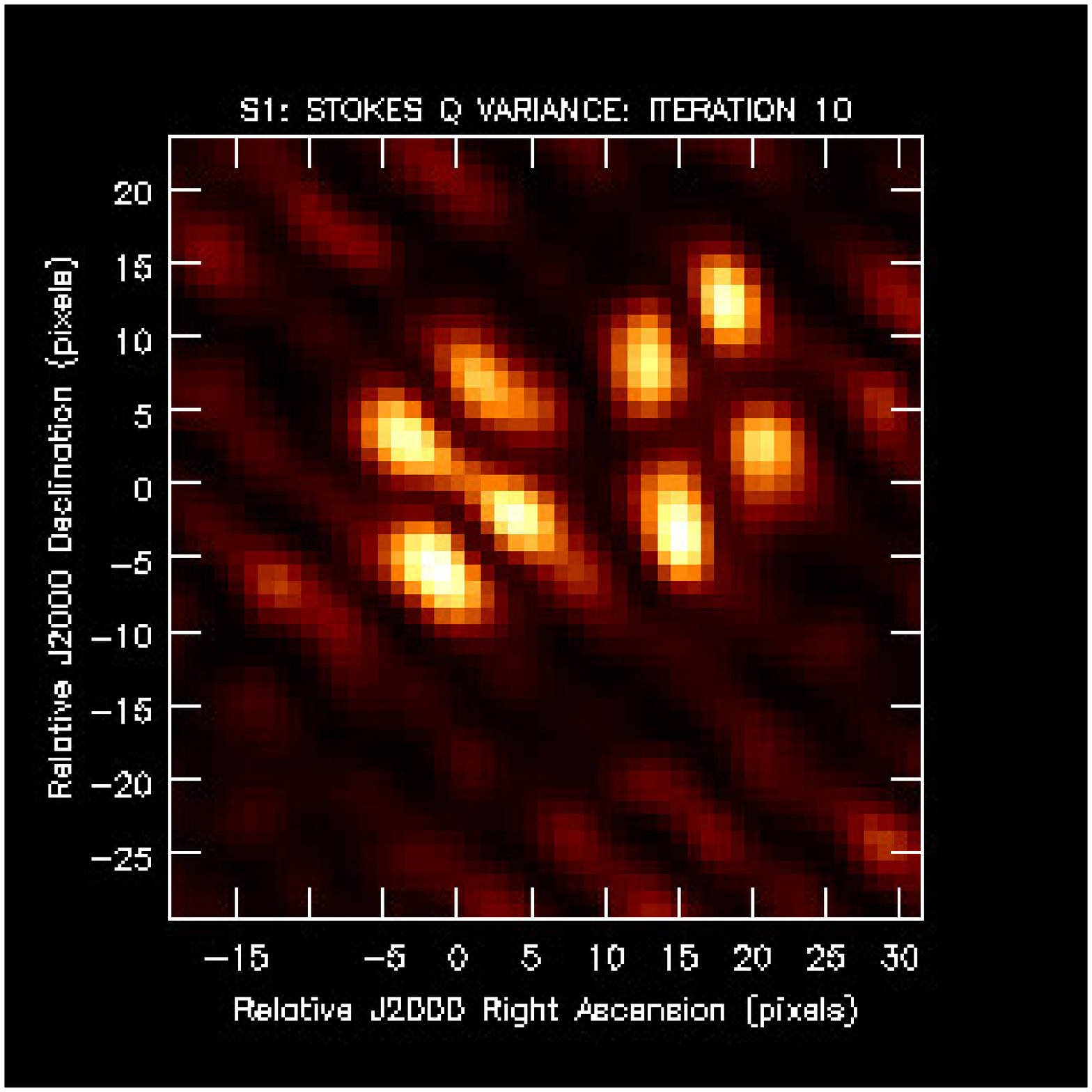} \\ 
\includegraphics[width=57mm, height=57mm, trim = 0mm 5mm 10mm 5mm, clip]{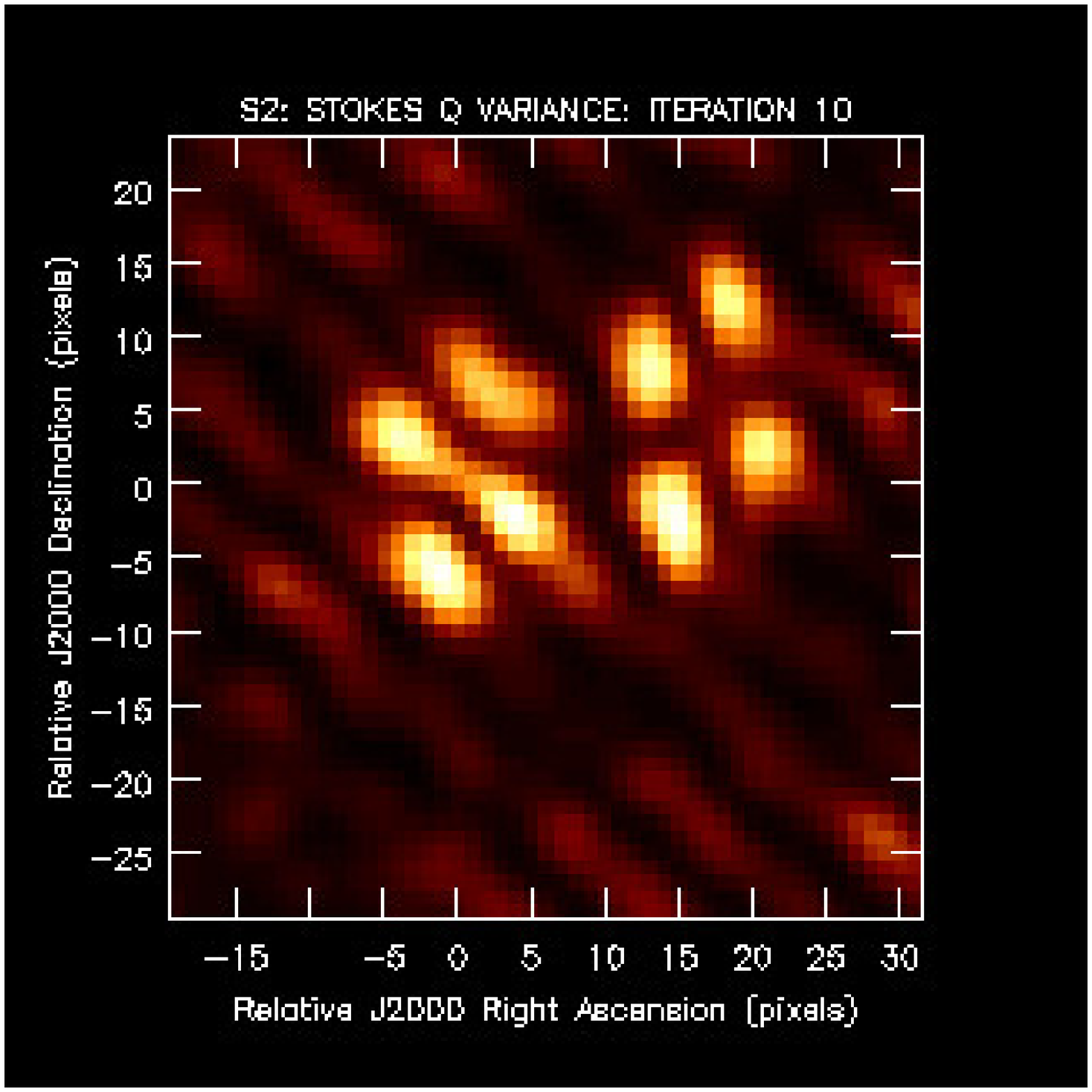} & 
\includegraphics[width=57mm, height=57mm, trim = 0mm 5mm 10mm 5mm, clip]{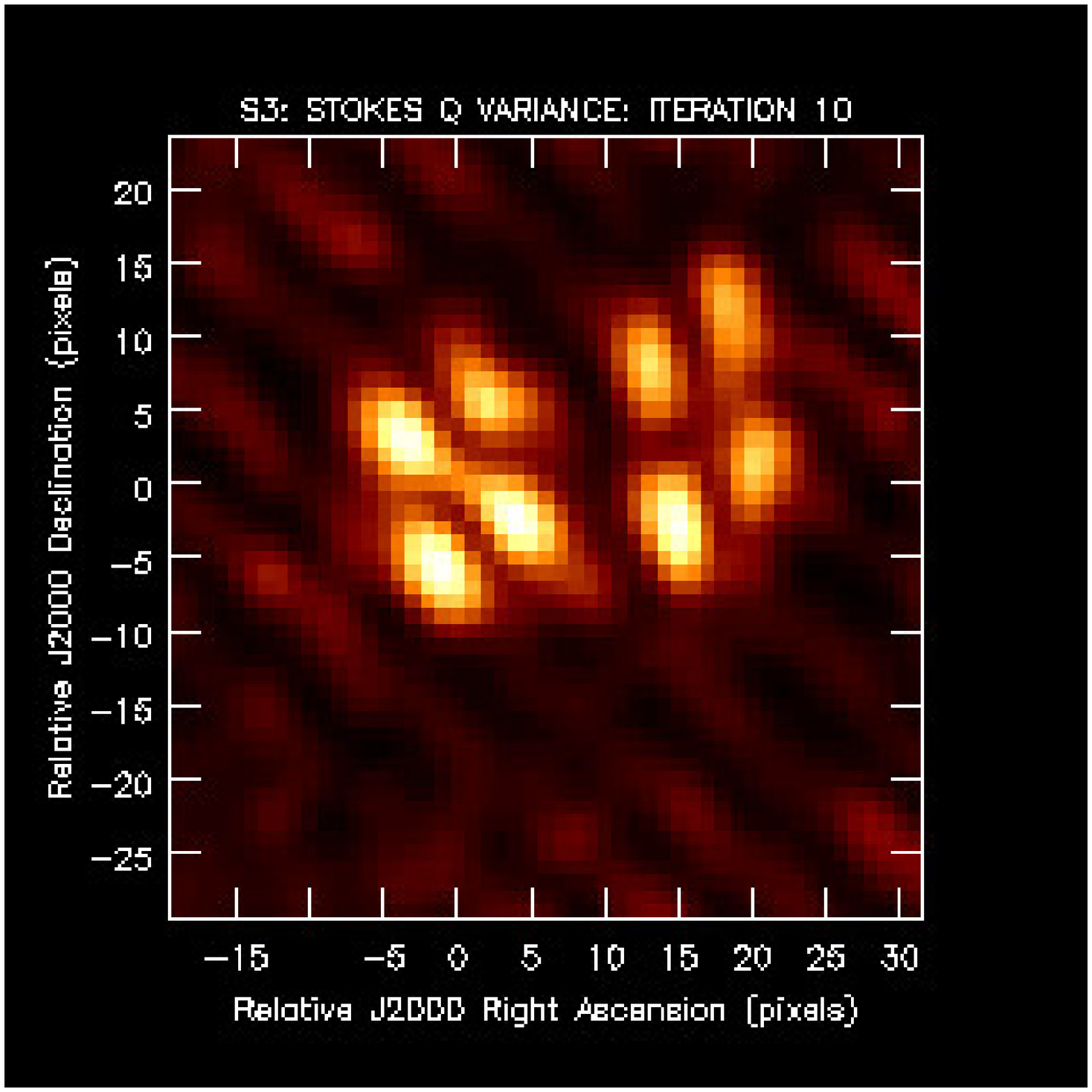} & 
\includegraphics[width=57mm, height=57mm, trim = 0mm 5mm 10mm 5mm, clip]{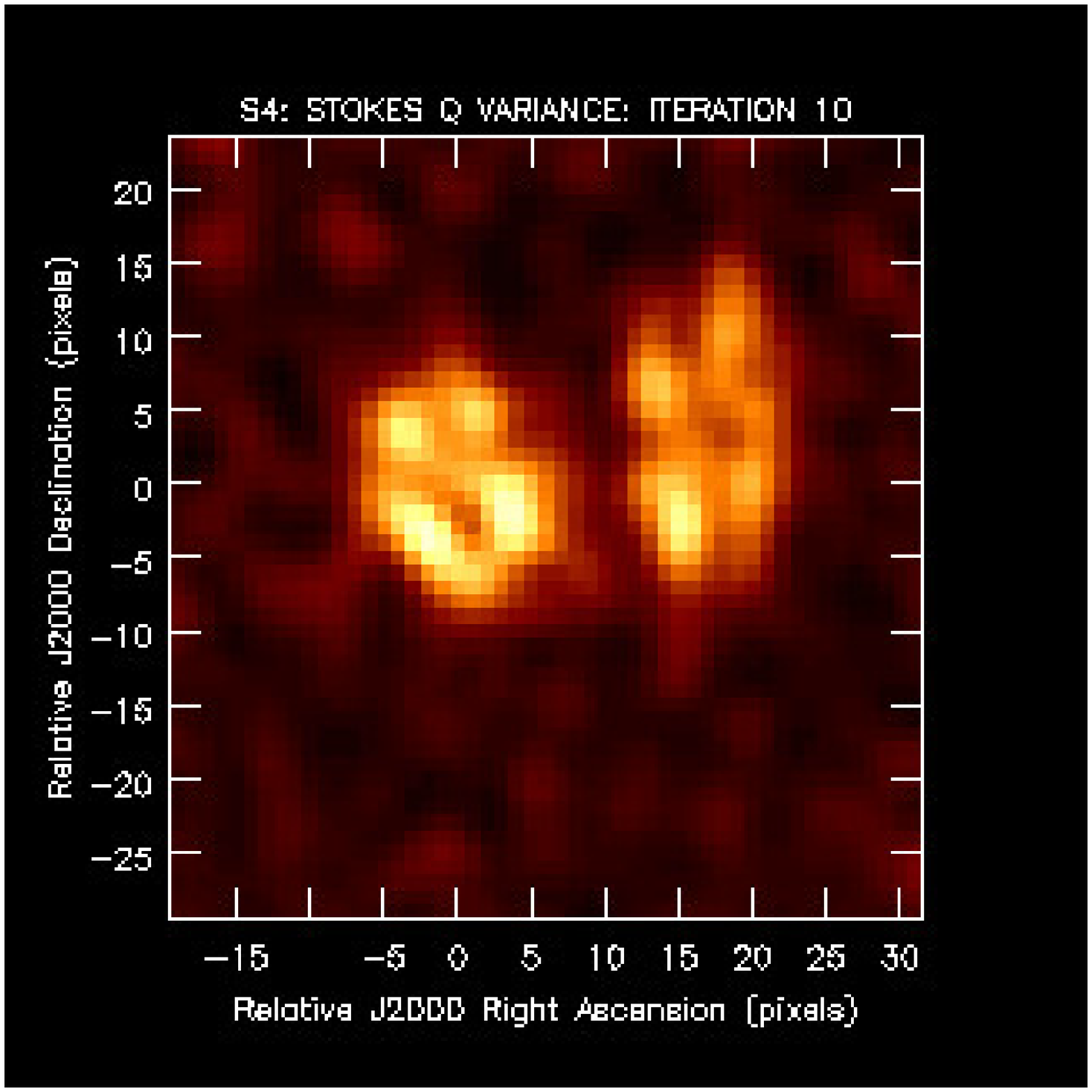} \\ 
\end{array}$ 

\caption{The imaging estimator Stokes $Q$ variance for the final
iteration of polarization self-calibration for run code B, as measured
by direct Monte Carlo simulation (MC), model-based bootstrap
resampling (M1-M4), and subsample resampling (S1-S4) methods({\it in
sequential order from top left to bottom right}). The parameters for
these bootstrap resampling codes are listed in Table~\ref{tbl-mcode}
and Table~\ref{tbl-scode} respectively. The run code parameters are
given in Table~\ref{tbl-par} and Table~\ref{tbl-sep}. The same default
color mapping is used for each bootstrap image variance cube, where
increasing color brightness denotes increasing image variance. This
figure is included to allow morphological comparison of bootstrap
variance images; a quantitative assessment of bootstrap performance is
given in Table~\ref{tbl-vmse}. See captions to Figures~\ref{fig-qvar-abc}, \ref{fig-qvar-def}, and \ref{fig-qvar-xyz} for quantitative variance ranges used in the associated contour plots.}

\label{fig-col-qvar-b} 
\end{figure}

\begin{figure}[h] 
\advance\leftskip-1cm
\advance\rightskip-1cm
 $\begin{array}{c@{\hspace{2mm}}c@{\hspace{2mm}}c} 
\includegraphics[width=57mm, height=57mm, trim = 0mm 5mm 10mm 5mm, clip]{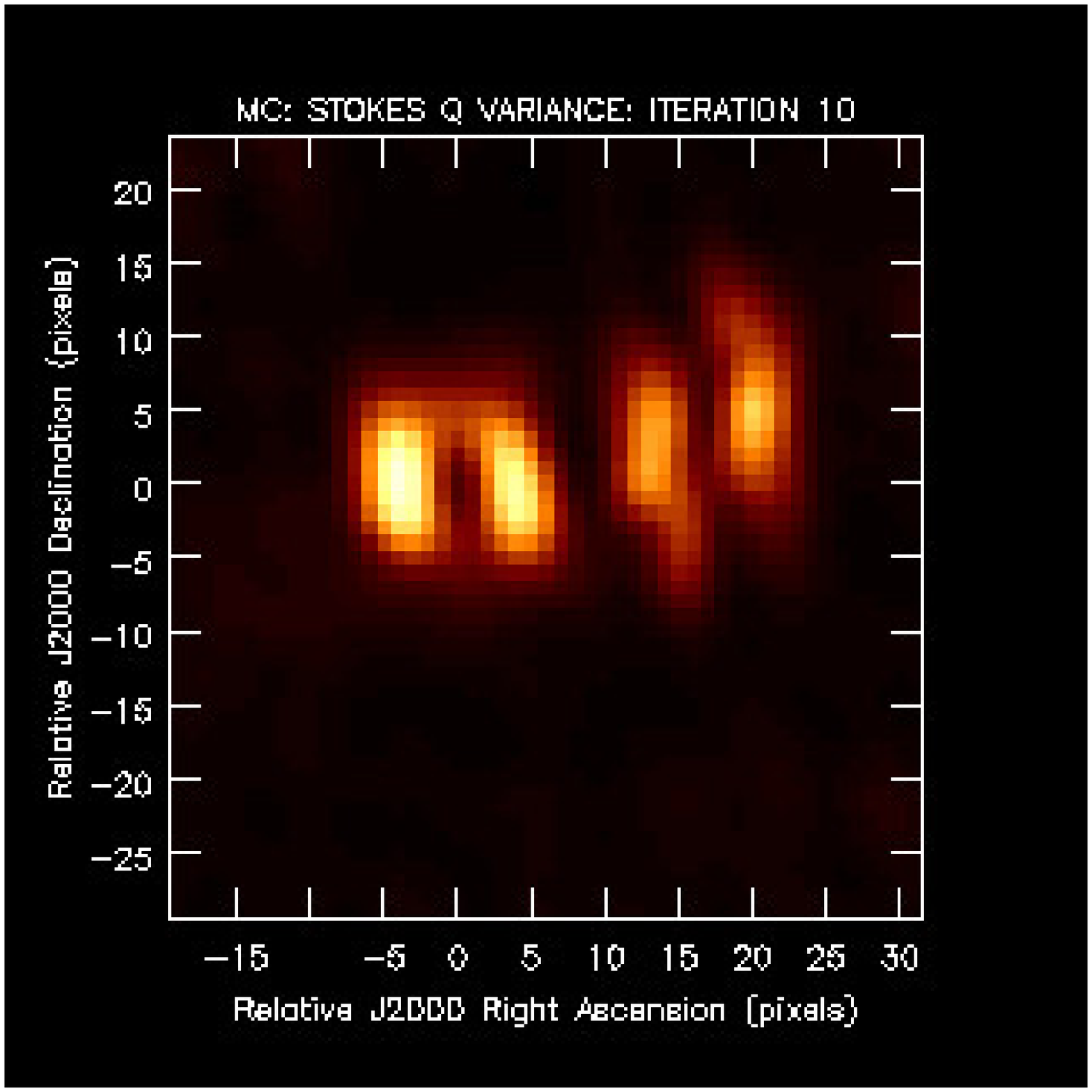} & 
\includegraphics[width=57mm, height=57mm, trim = 0mm 5mm 10mm 5mm, clip]{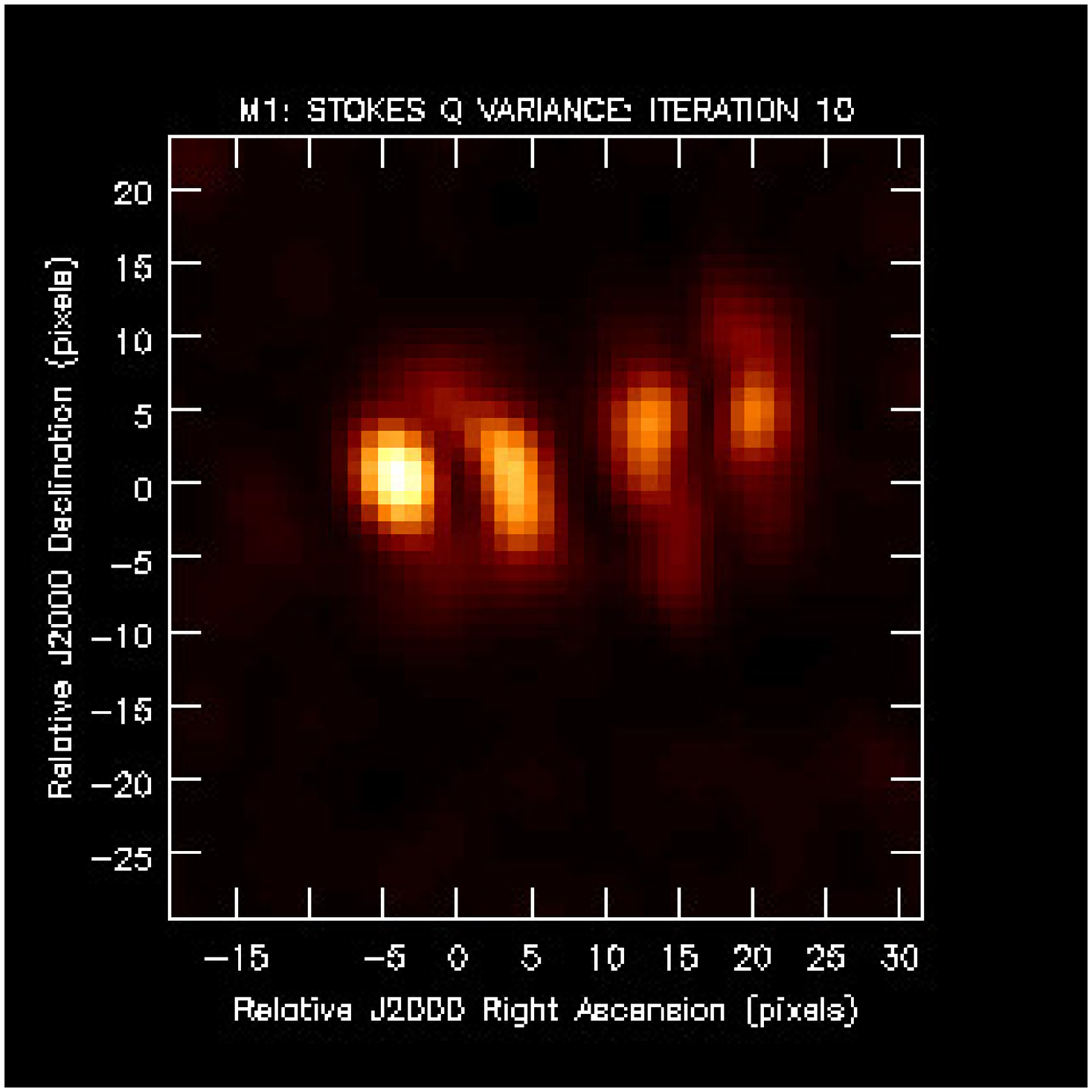} & 
\includegraphics[width=57mm, height=57mm, trim = 0mm 5mm 10mm 5mm, clip]{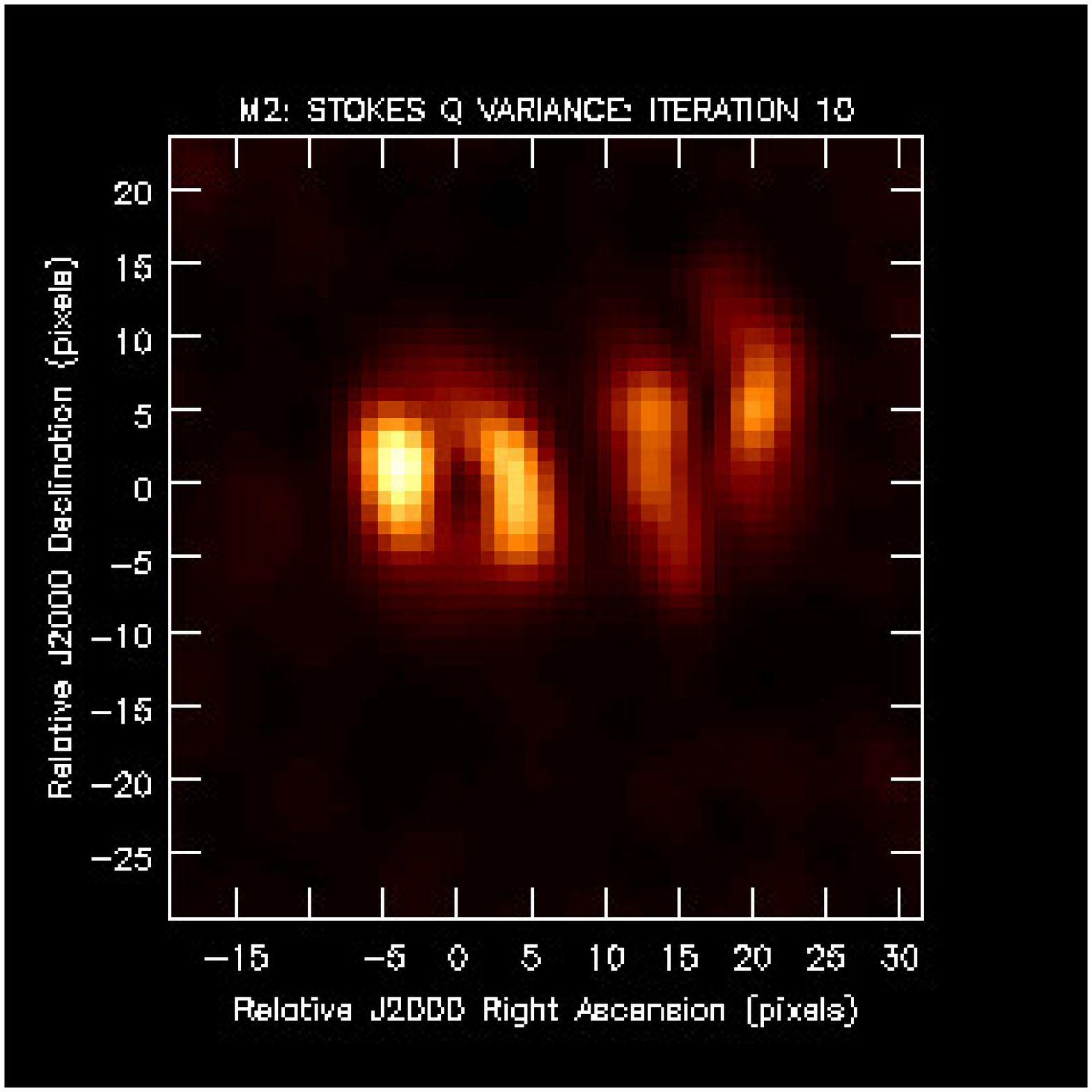} \\ 
\includegraphics[width=57mm, height=57mm, trim = 0mm 5mm 10mm 5mm, clip]{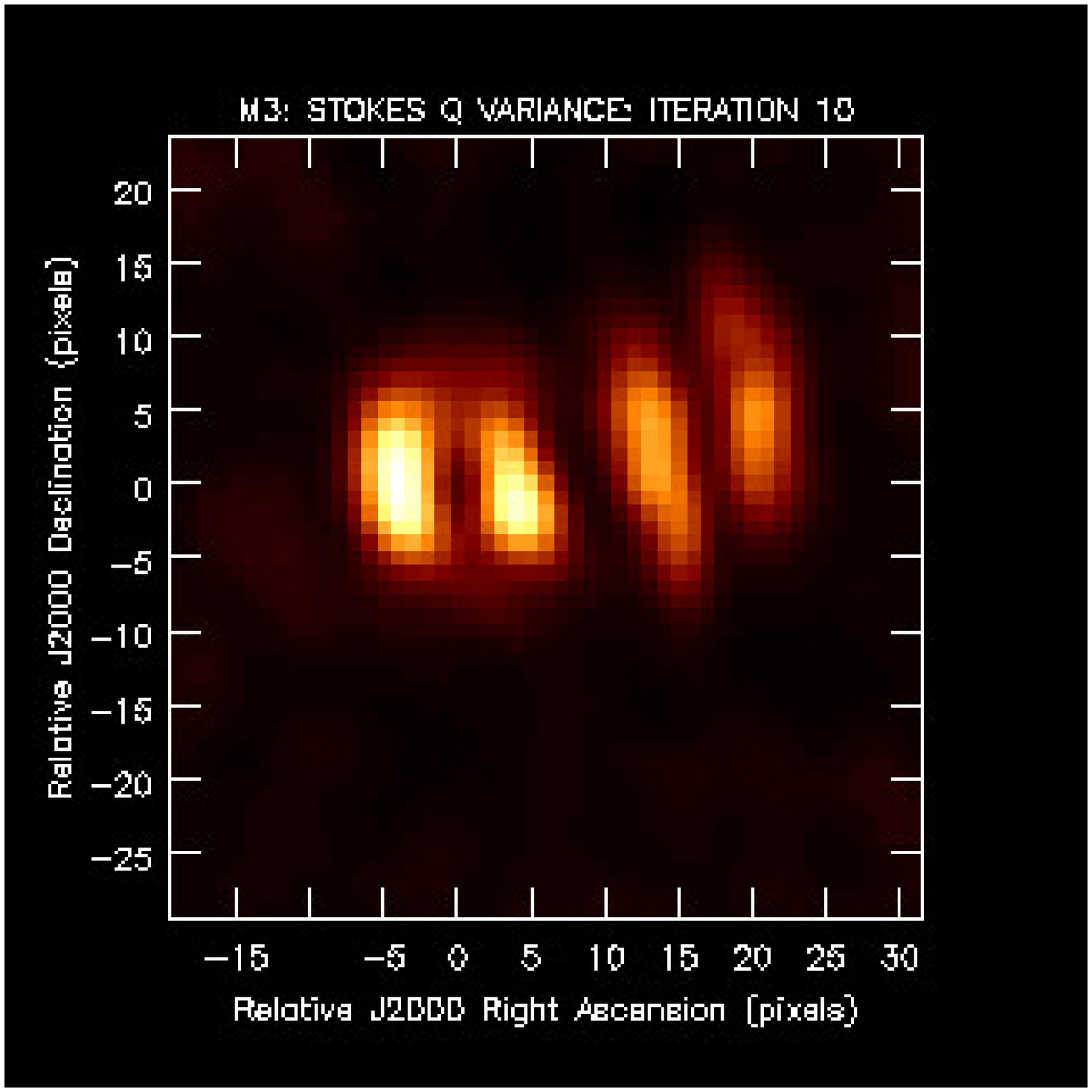} & 
\includegraphics[width=57mm, height=57mm, trim = 0mm 5mm 10mm 5mm, clip]{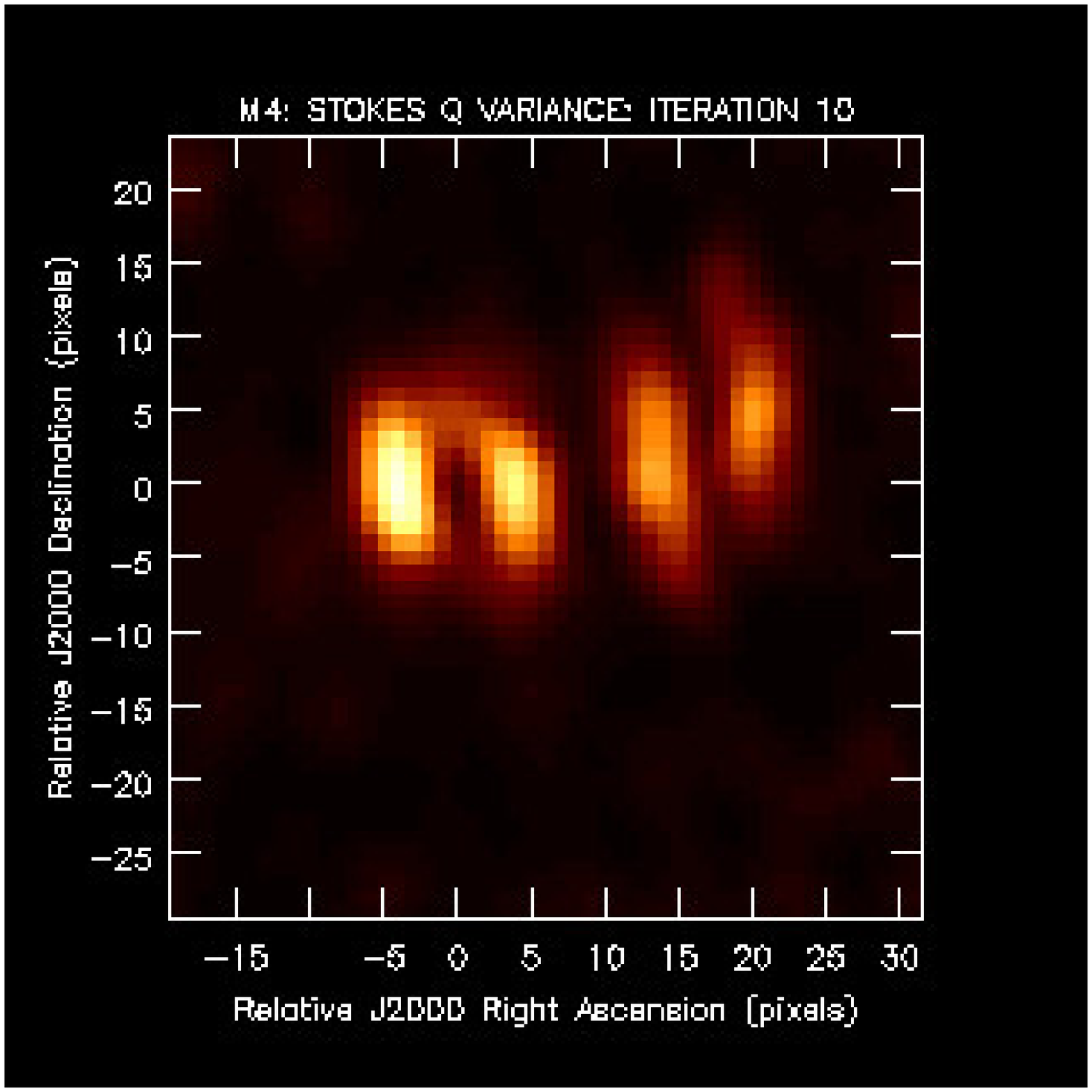} & 
\includegraphics[width=57mm, height=57mm, trim = 0mm 5mm 10mm 5mm, clip]{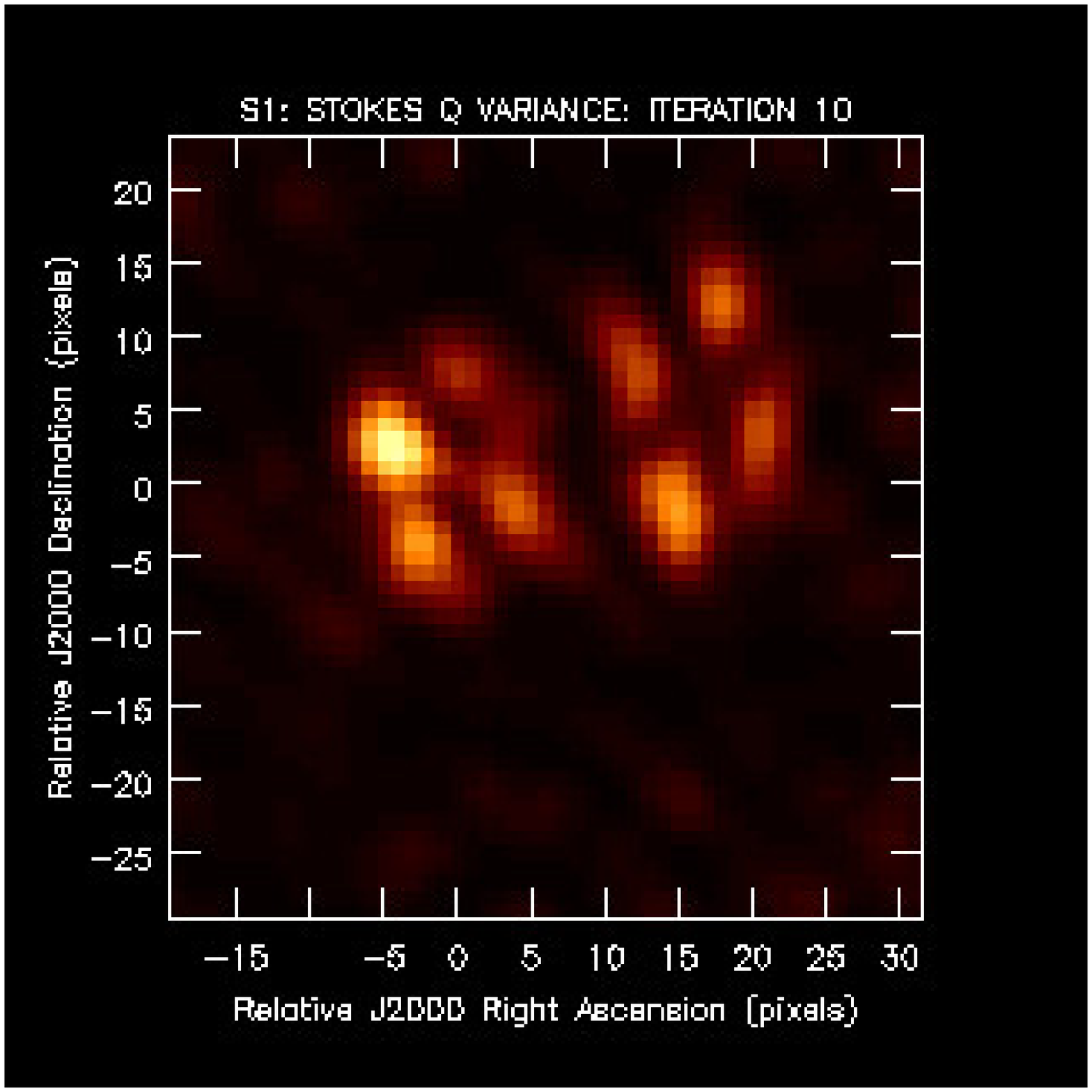} \\ 
\includegraphics[width=57mm, height=57mm, trim = 0mm 5mm 10mm 5mm, clip]{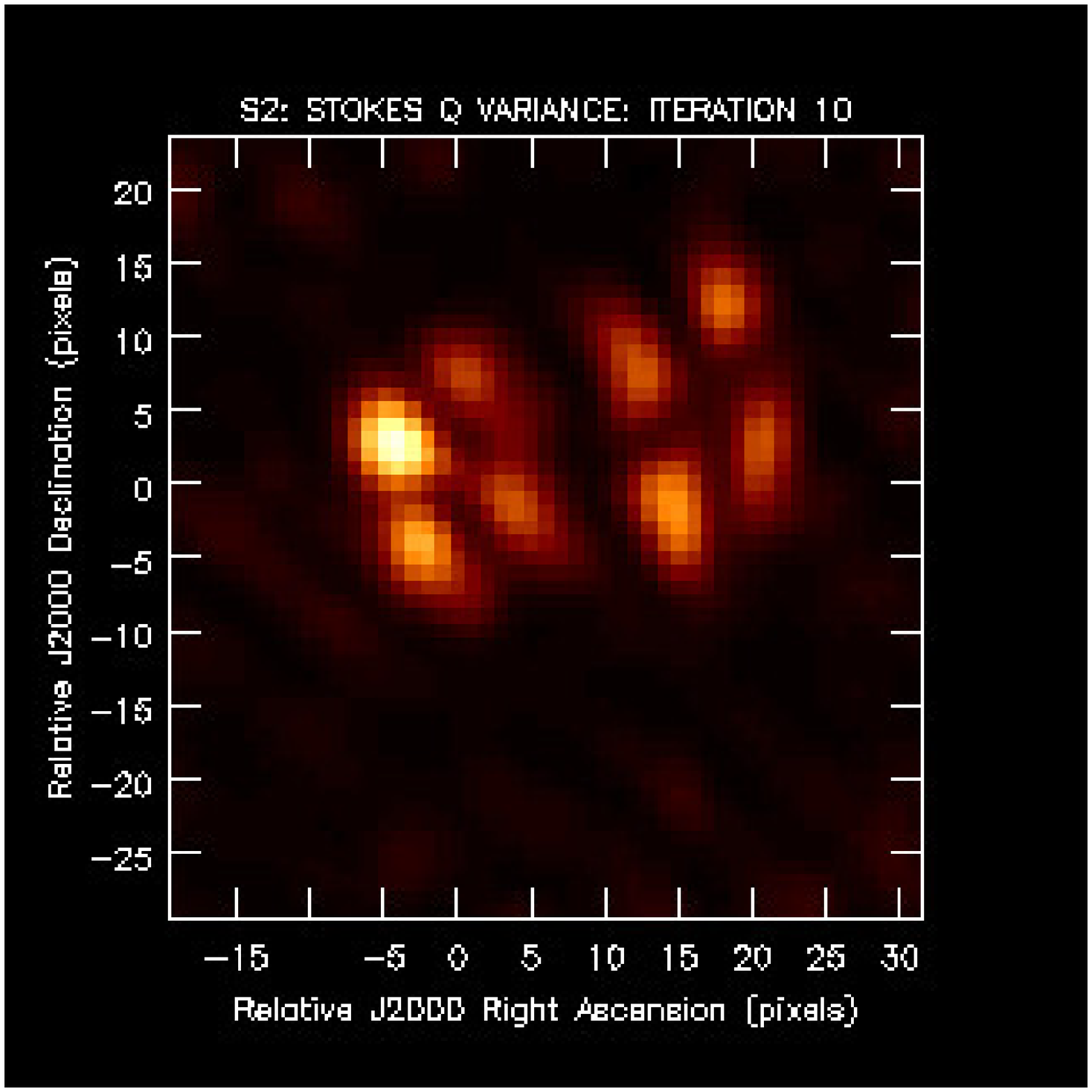} & 
\includegraphics[width=57mm, height=57mm, trim = 0mm 5mm 10mm 5mm, clip]{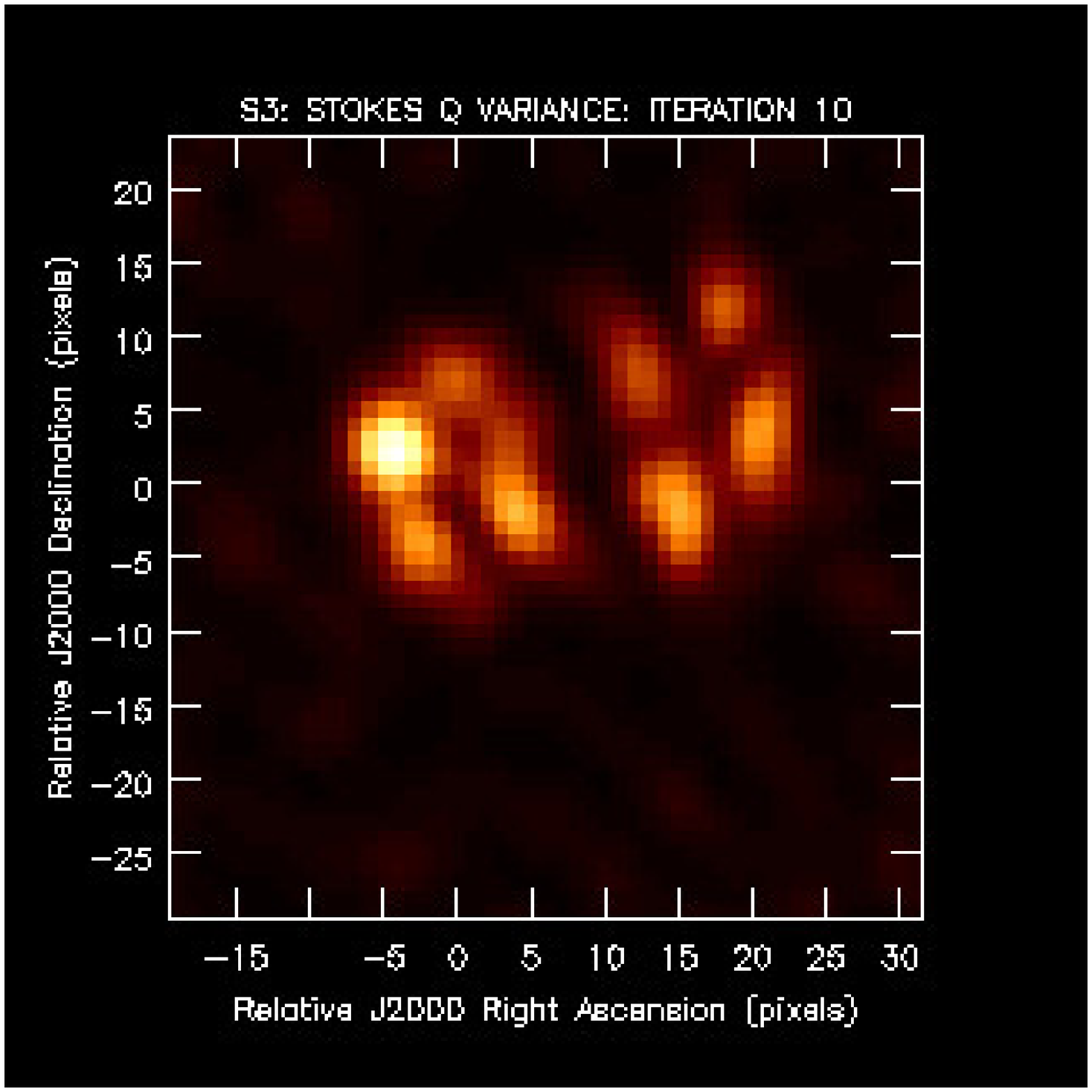} & 
\includegraphics[width=57mm, height=57mm, trim = 0mm 5mm 10mm 5mm, clip]{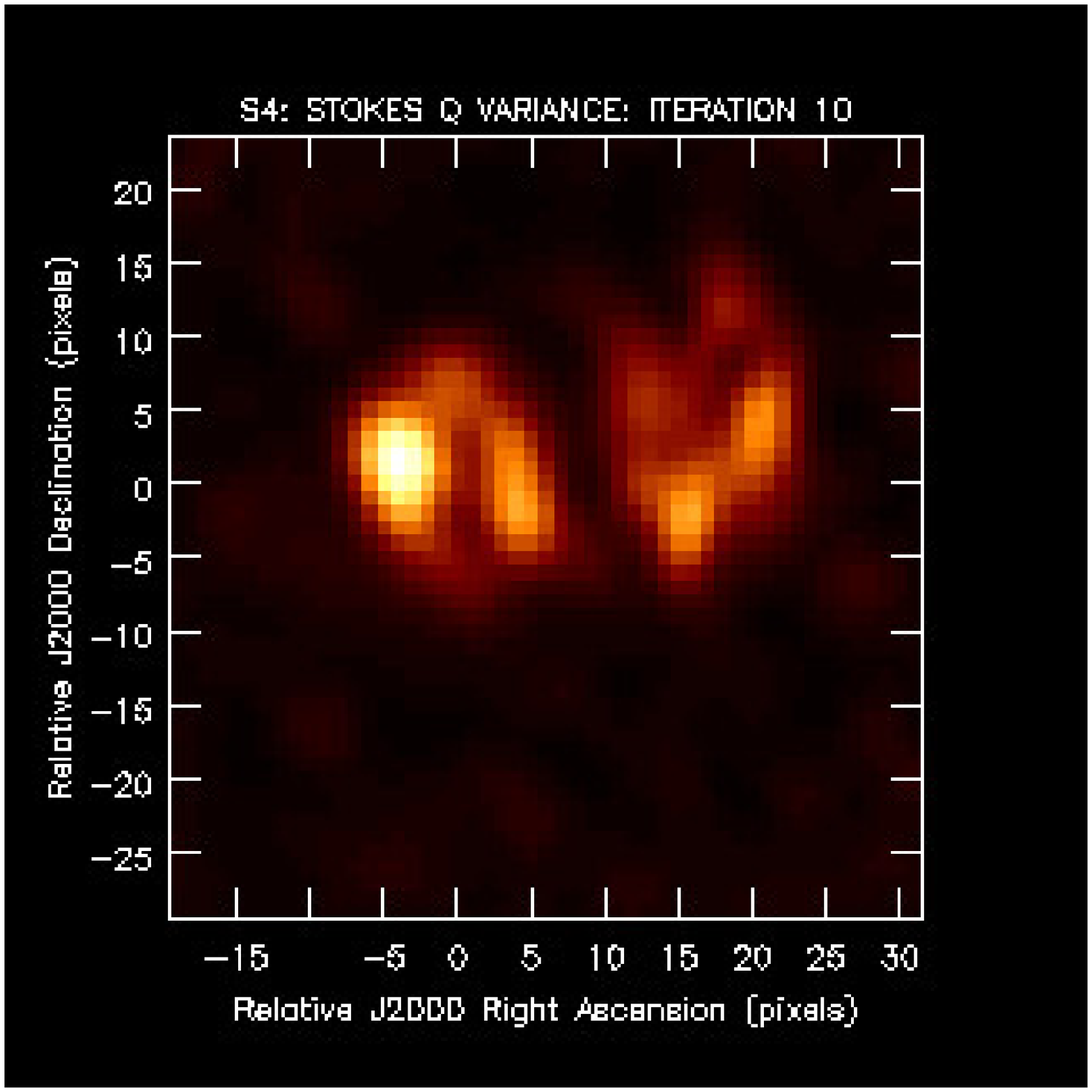} \\ 
\end{array}$ 

\caption{Same as Figure~\ref{fig-col-qvar-b}, but for run code D.}

\label{fig-col-qvar-d} 
\end{figure}

\begin{figure}[h] 
\advance\leftskip-1cm
\advance\rightskip-1cm
 $\begin{array}{c@{\hspace{2mm}}c@{\hspace{2mm}}c} 
\includegraphics[width=57mm, height=57mm, trim = 0mm 5mm 10mm 5mm, clip]{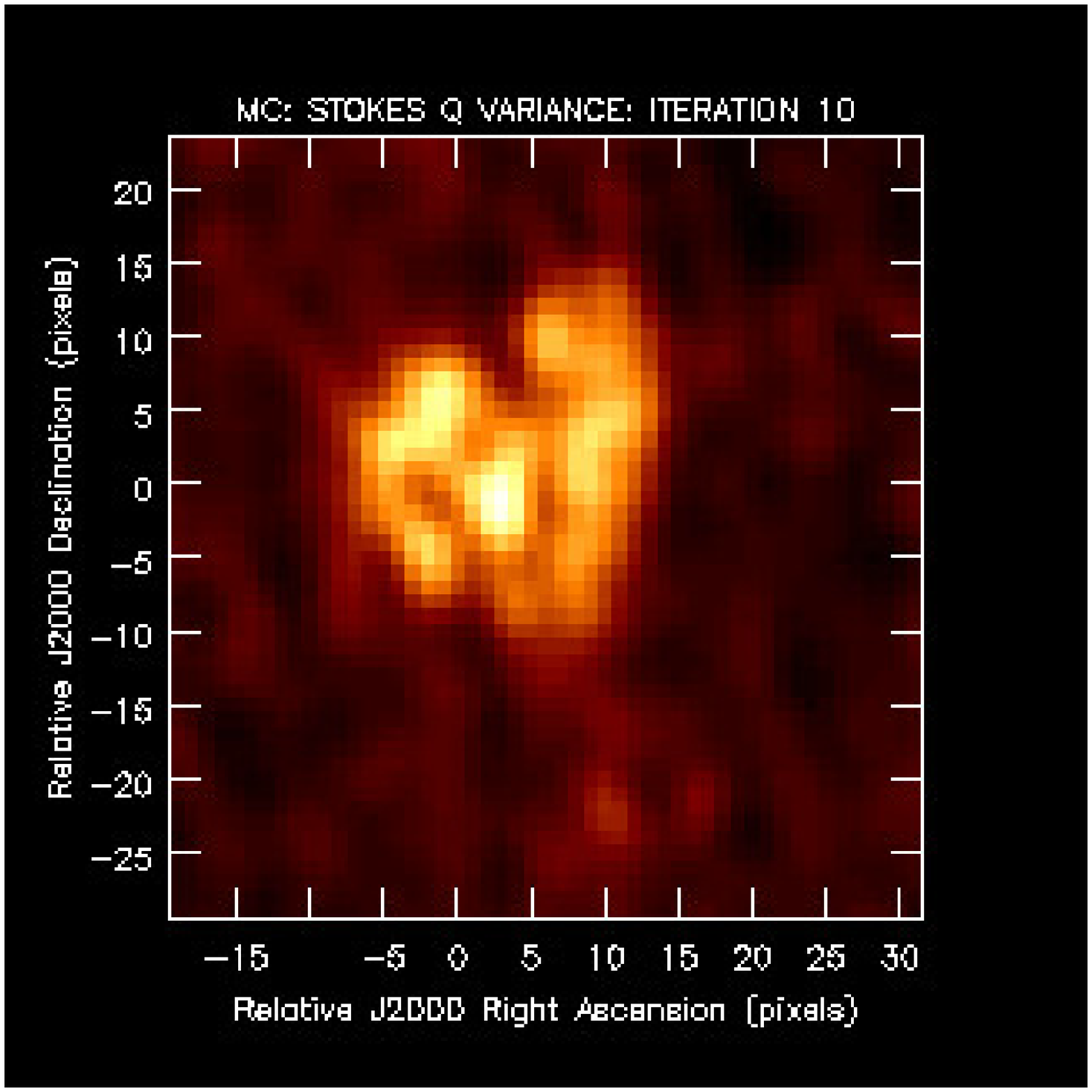} & 
\includegraphics[width=57mm, height=57mm, trim = 0mm 5mm 10mm 5mm, clip]{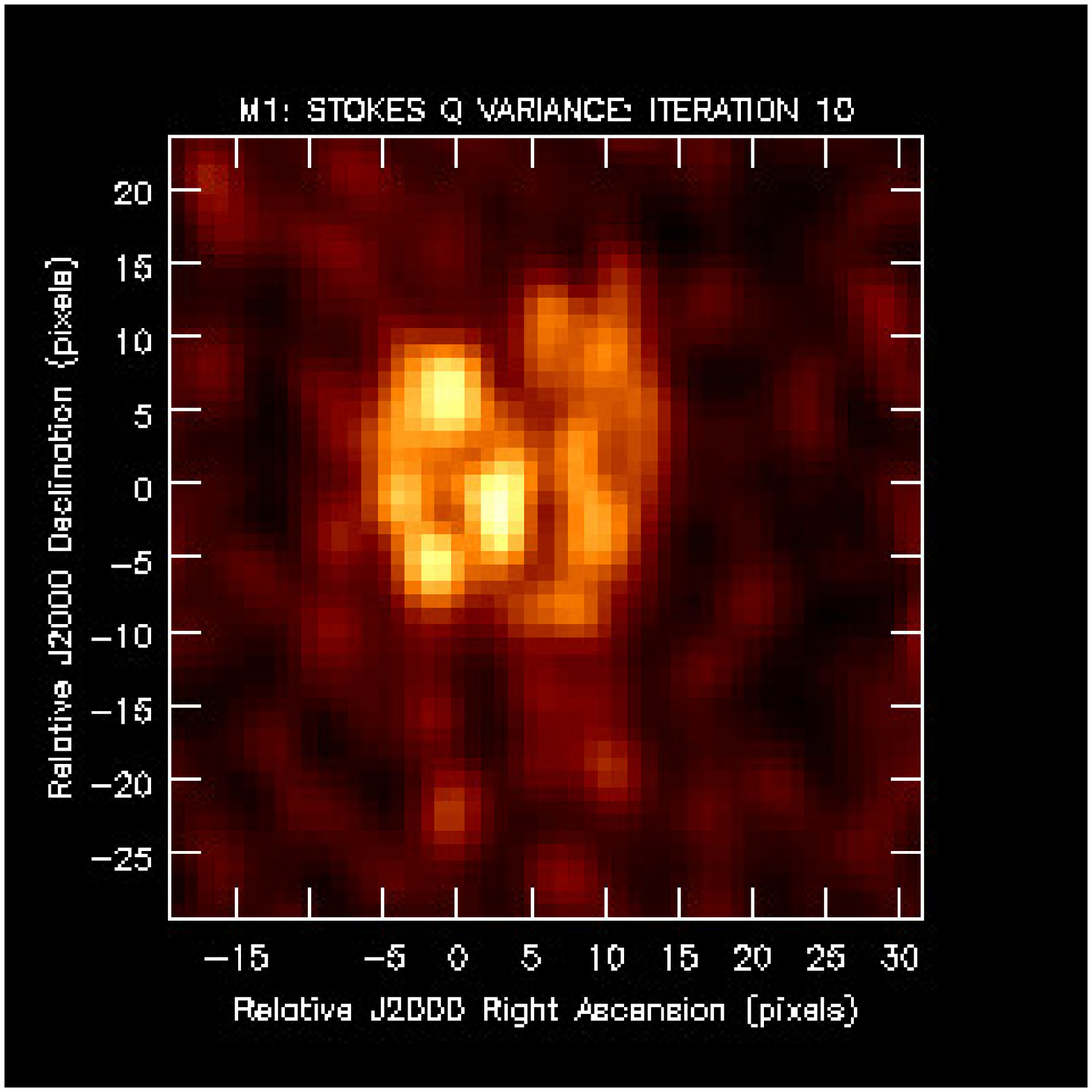} & 
\includegraphics[width=57mm, height=57mm, trim = 0mm 5mm 10mm 5mm, clip]{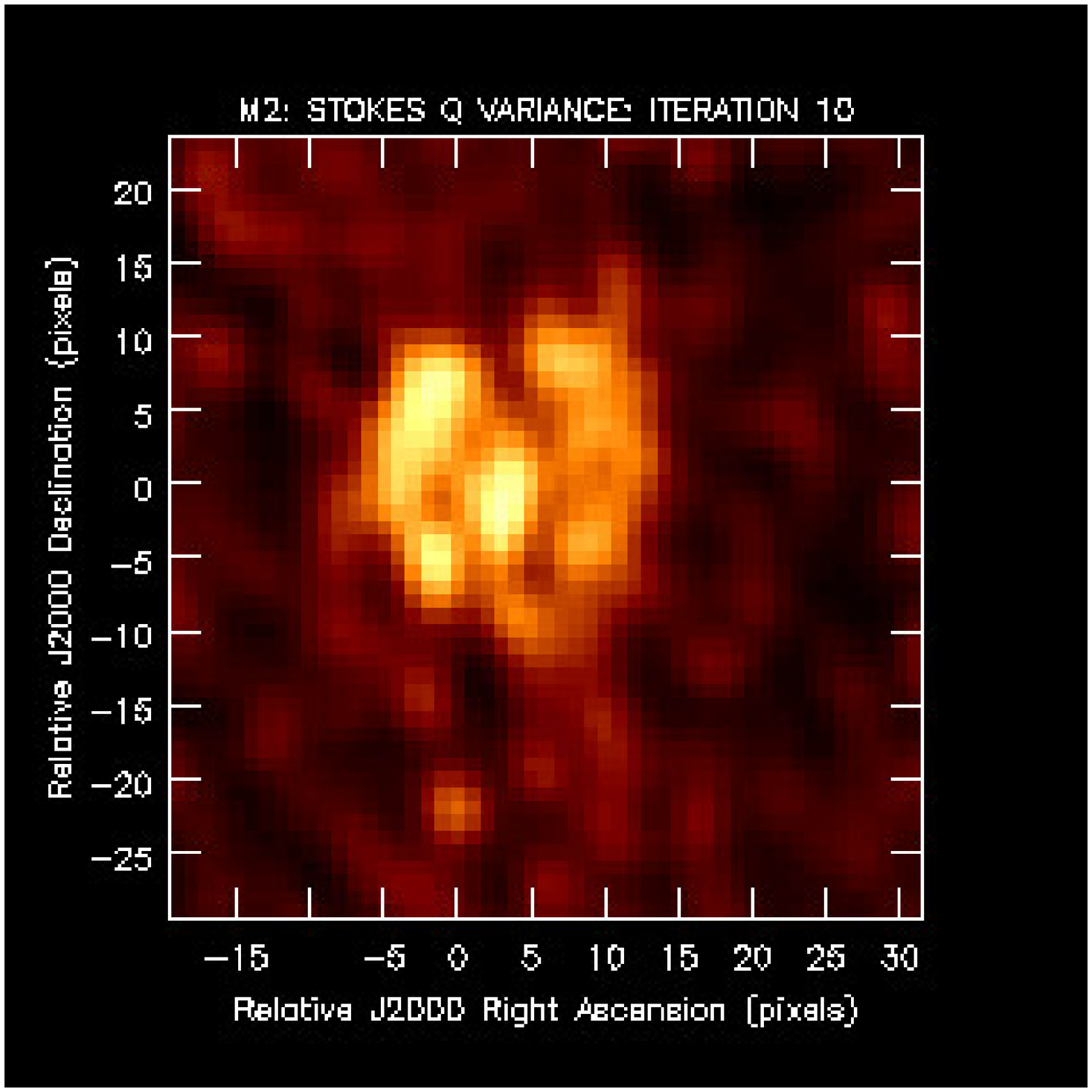} \\ 
\includegraphics[width=57mm, height=57mm, trim = 0mm 5mm 10mm 5mm, clip]{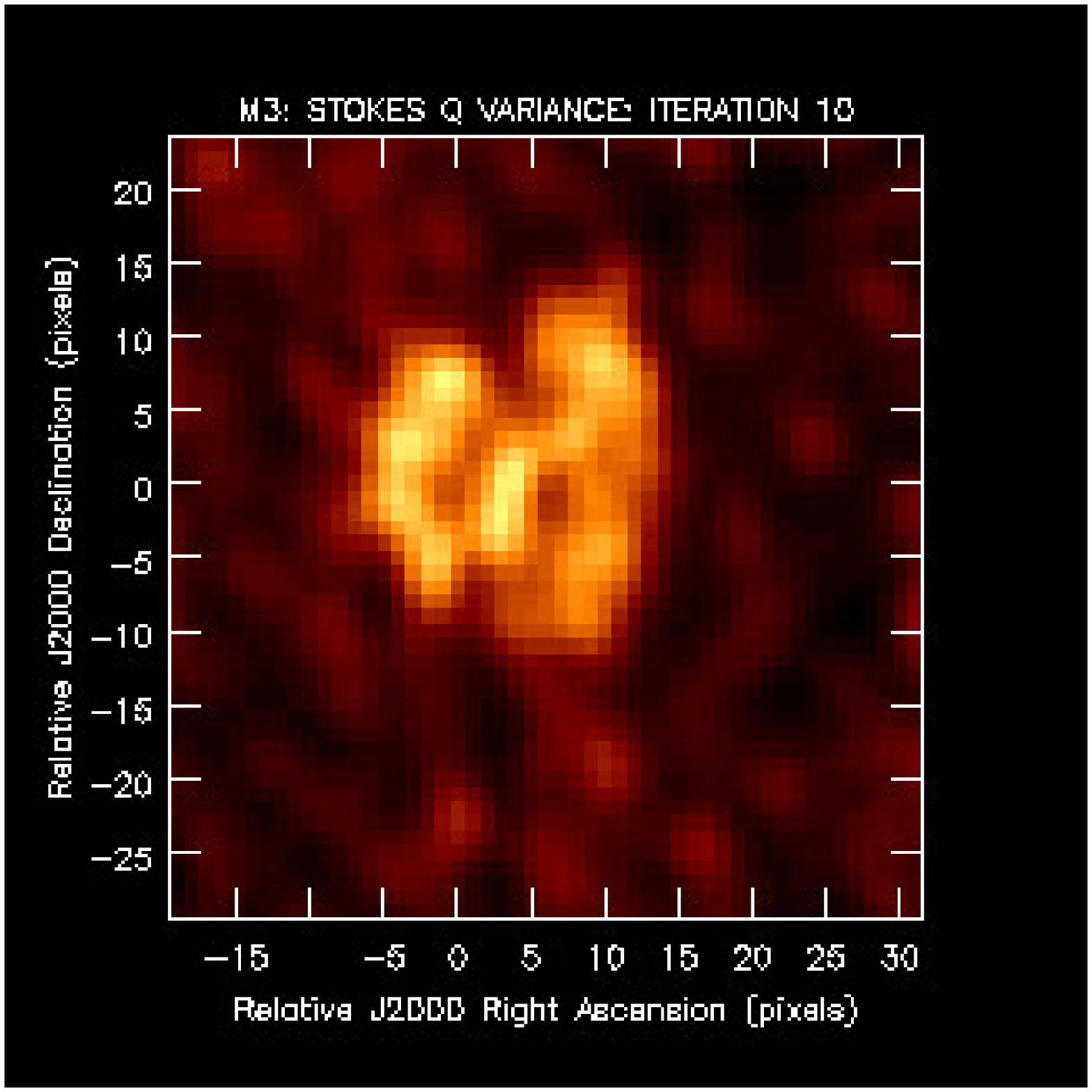} & 
\includegraphics[width=57mm, height=57mm, trim = 0mm 5mm 10mm 5mm, clip]{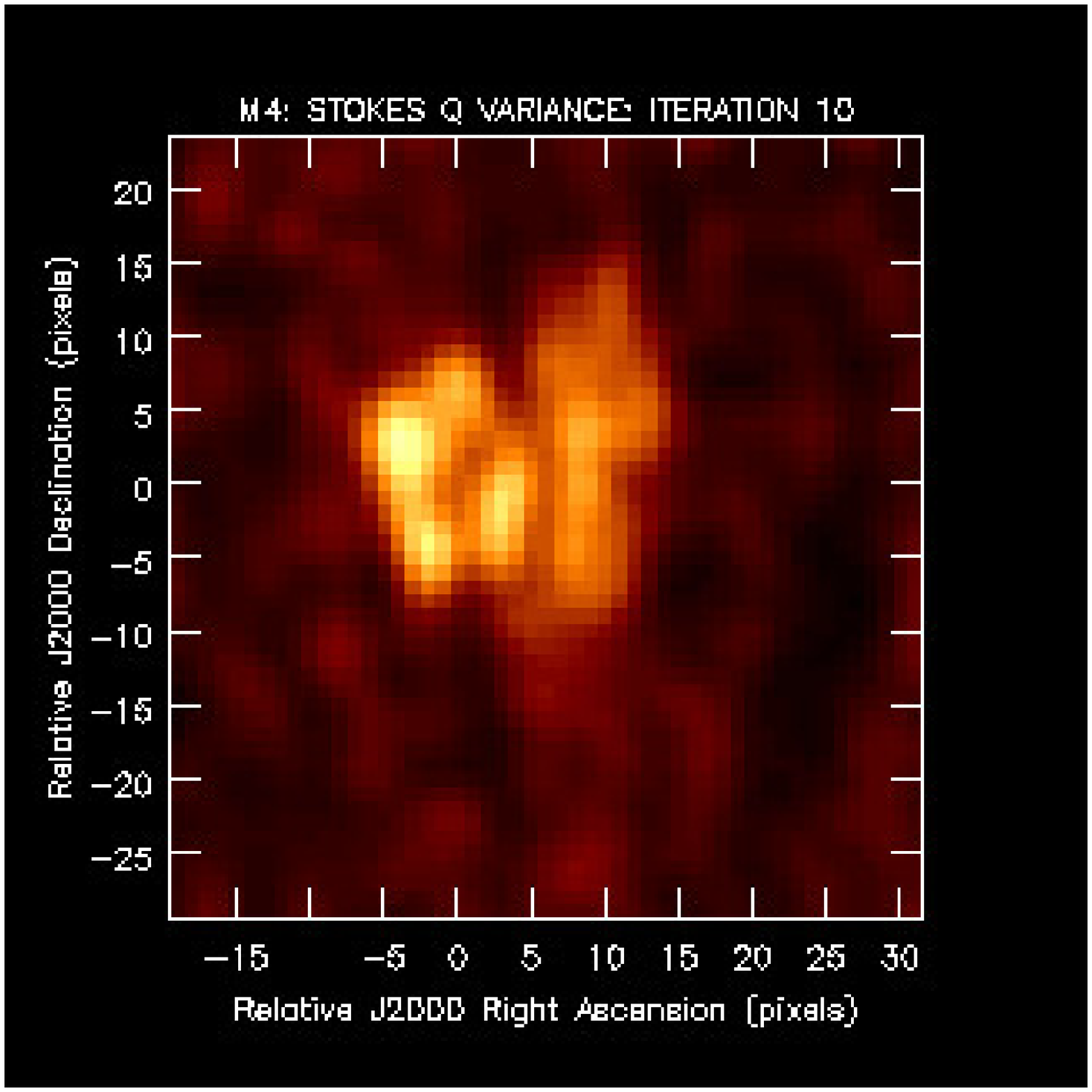} & 
\includegraphics[width=57mm, height=57mm, trim = 0mm 5mm 10mm 5mm, clip]{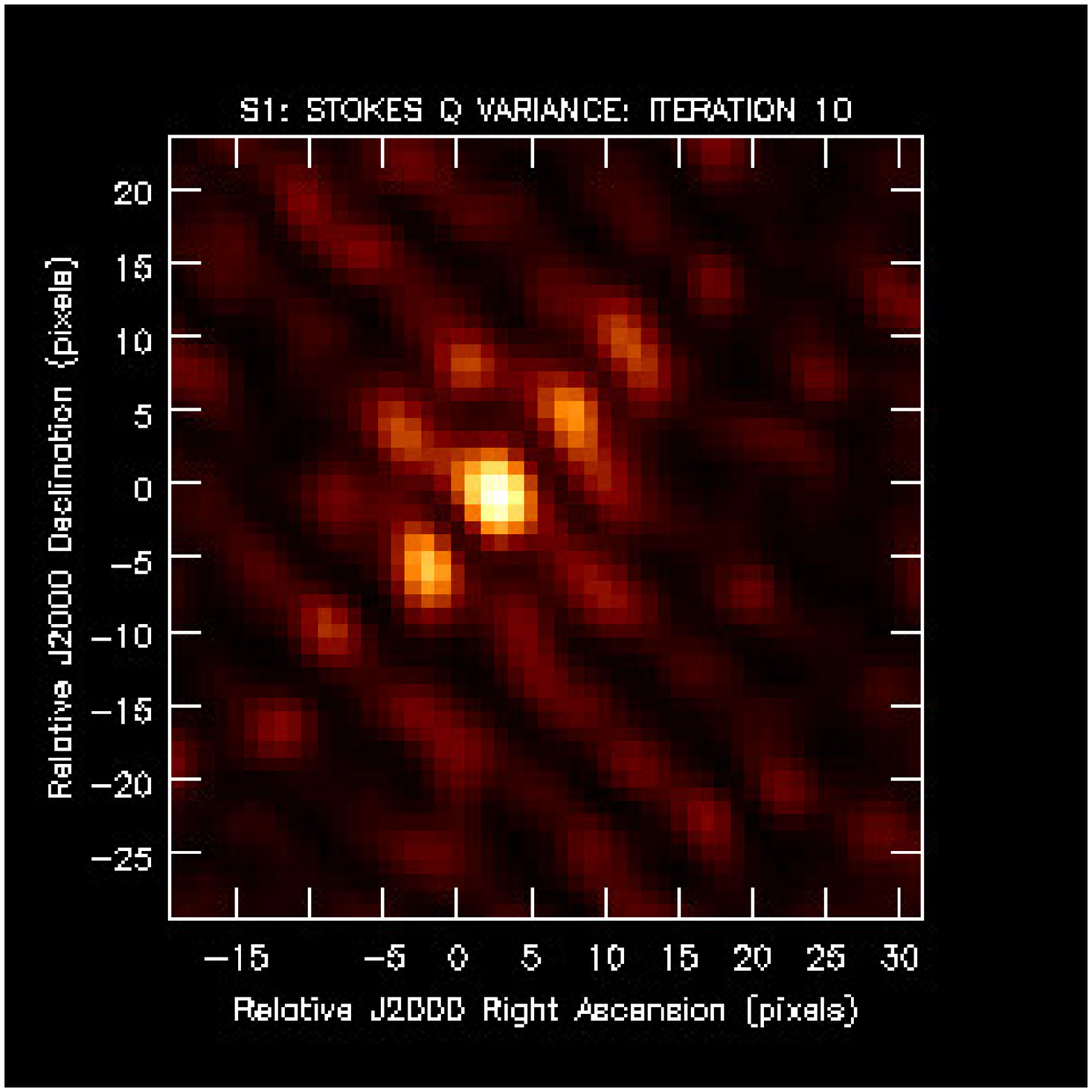} \\ 
\includegraphics[width=57mm, height=57mm, trim = 0mm 5mm 10mm 5mm, clip]{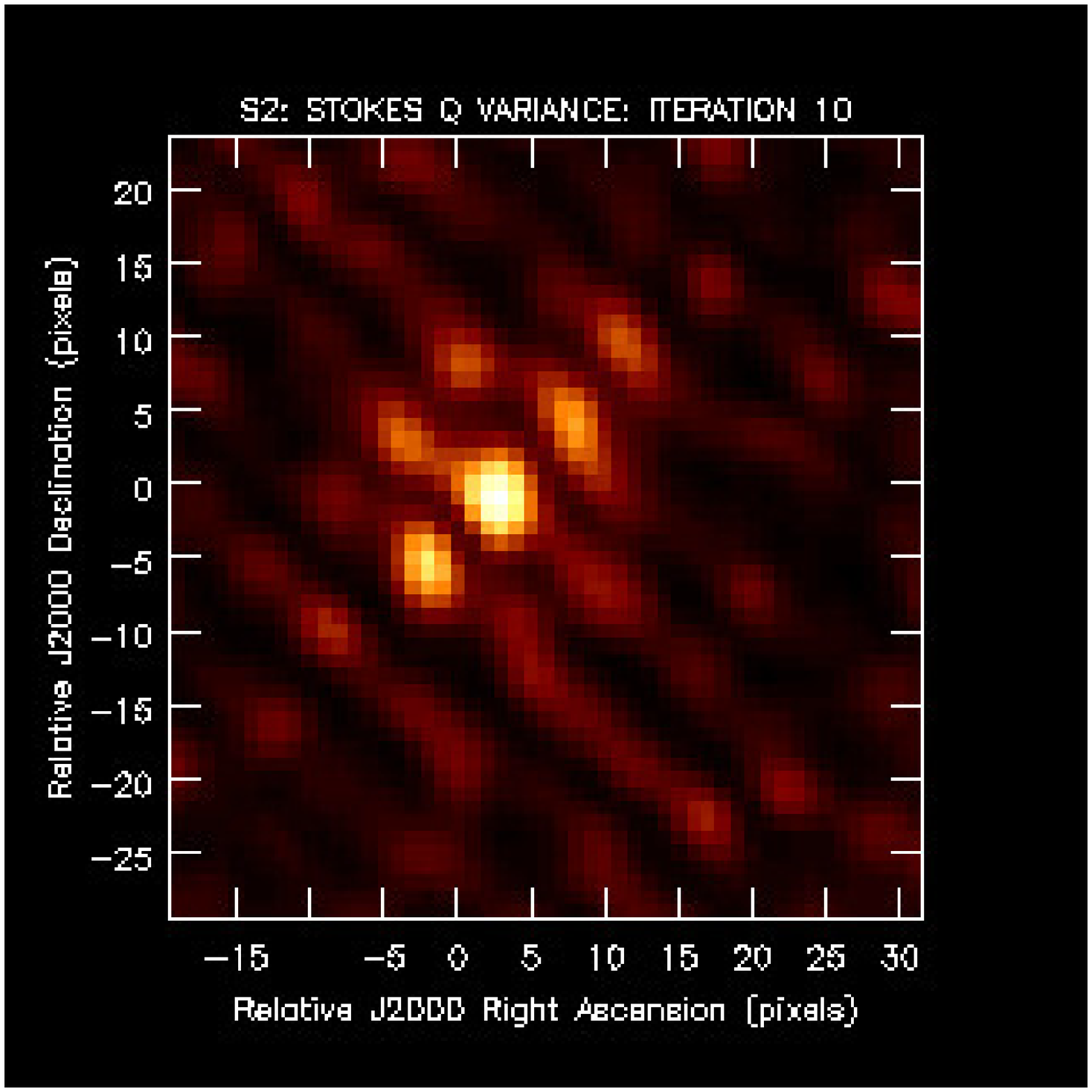} & 
\includegraphics[width=57mm, height=57mm, trim = 0mm 5mm 10mm 5mm, clip]{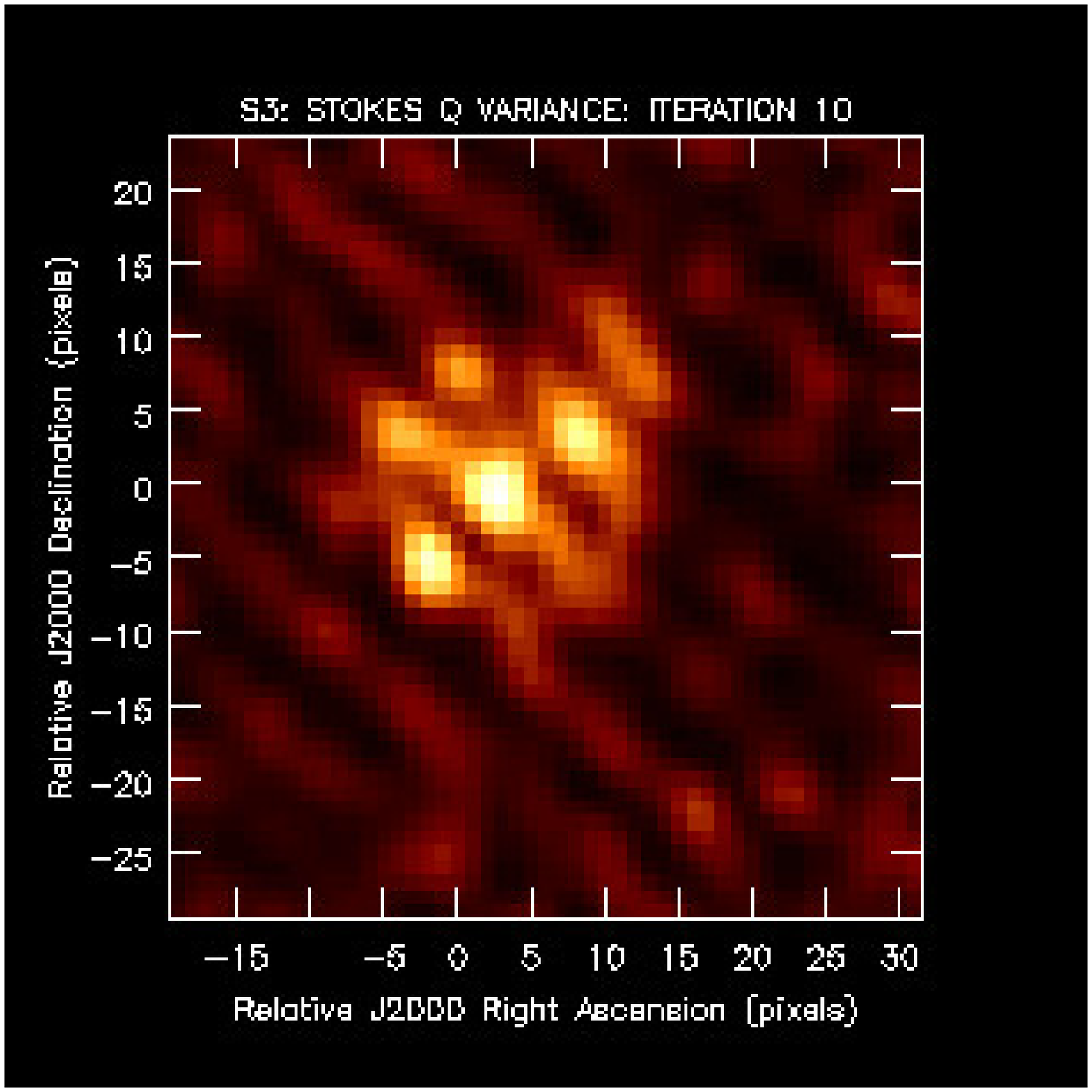} & 
\includegraphics[width=57mm, height=57mm, trim = 0mm 5mm 10mm 5mm, clip]{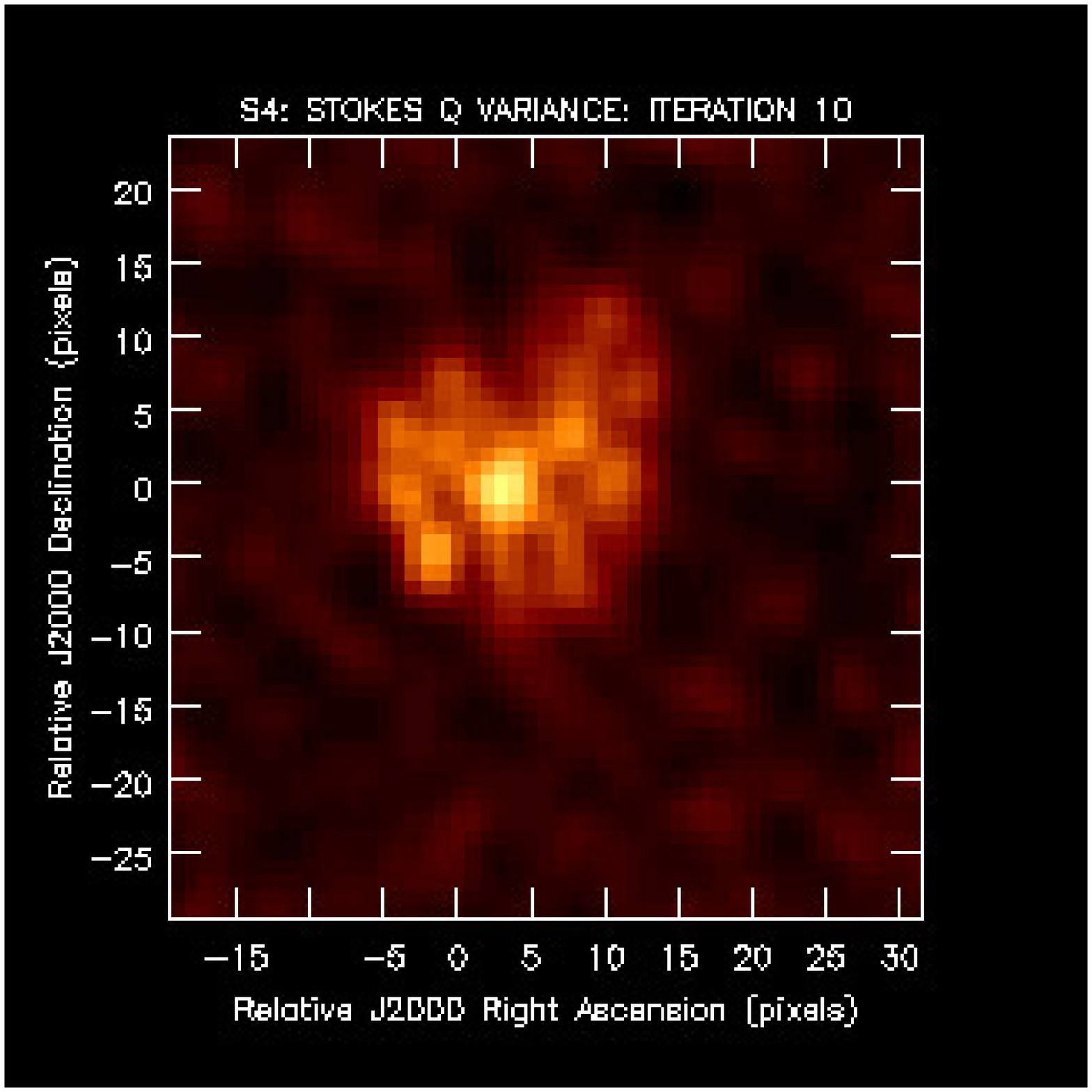} \\ 
\end{array}$ 

\caption{Same as Figure~\ref{fig-col-qvar-d}, but for run code X.}

\label{fig-col-qvar-x} 
\end{figure}

\begin{figure}[h] 
\advance\leftskip-1cm
\advance\rightskip-1cm
 $\begin{array}{c@{\hspace{2mm}}c@{\hspace{2mm}}c} 
\includegraphics[width=57mm, height=57mm, trim = 0mm 5mm 10mm 5mm, clip]{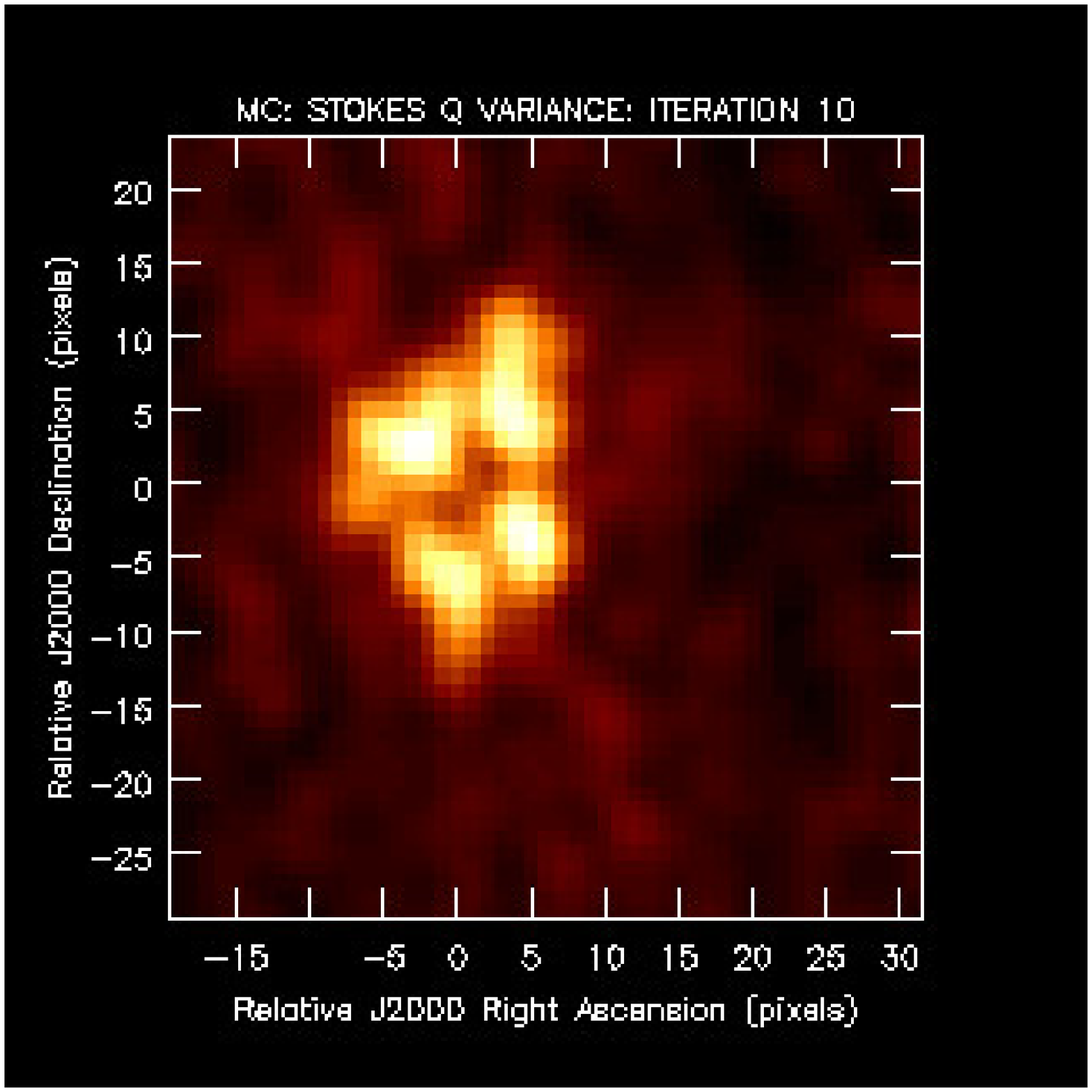} & 
\includegraphics[width=57mm, height=57mm, trim = 0mm 5mm 10mm 5mm, clip]{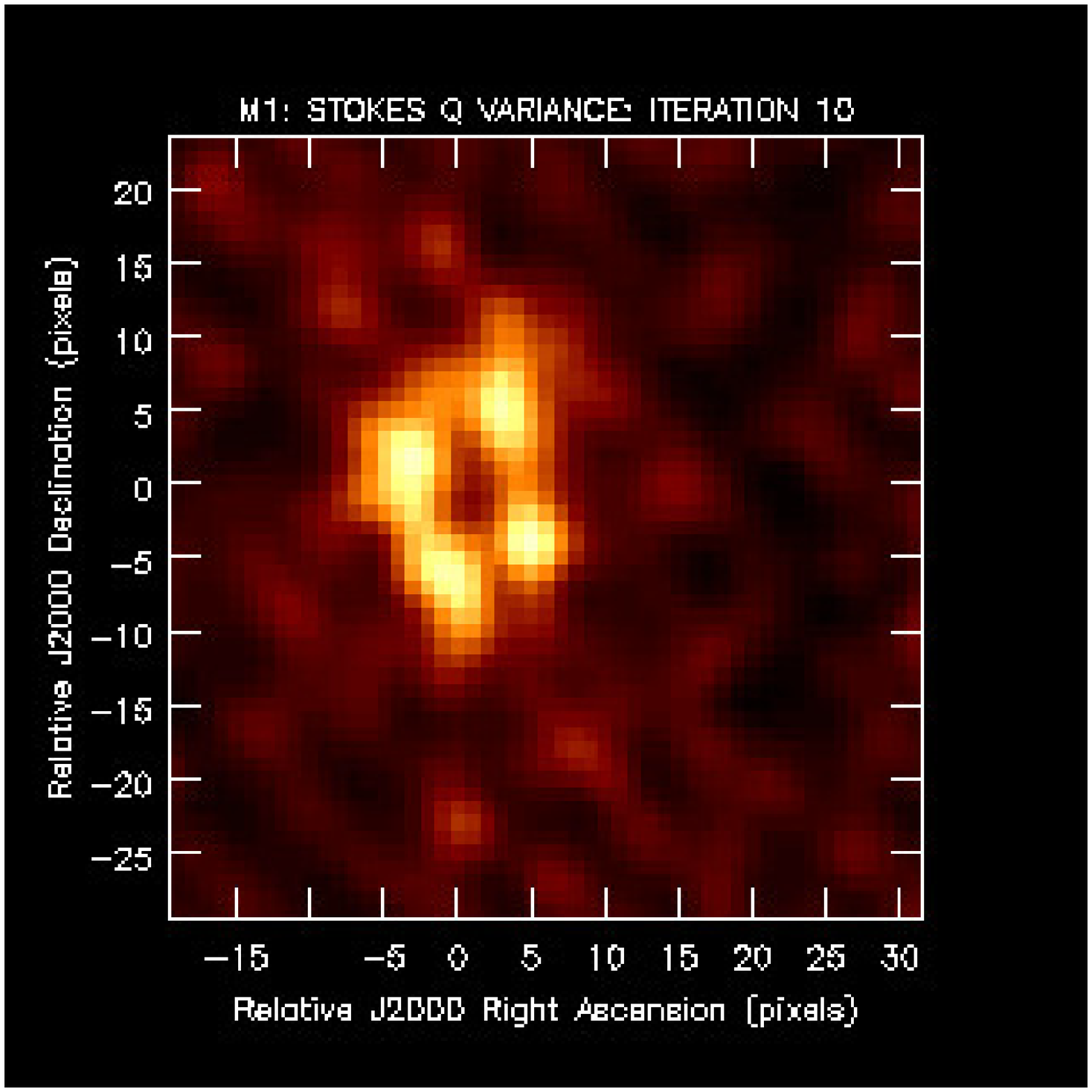} & 
\includegraphics[width=57mm, height=57mm, trim = 0mm 5mm 10mm 5mm, clip]{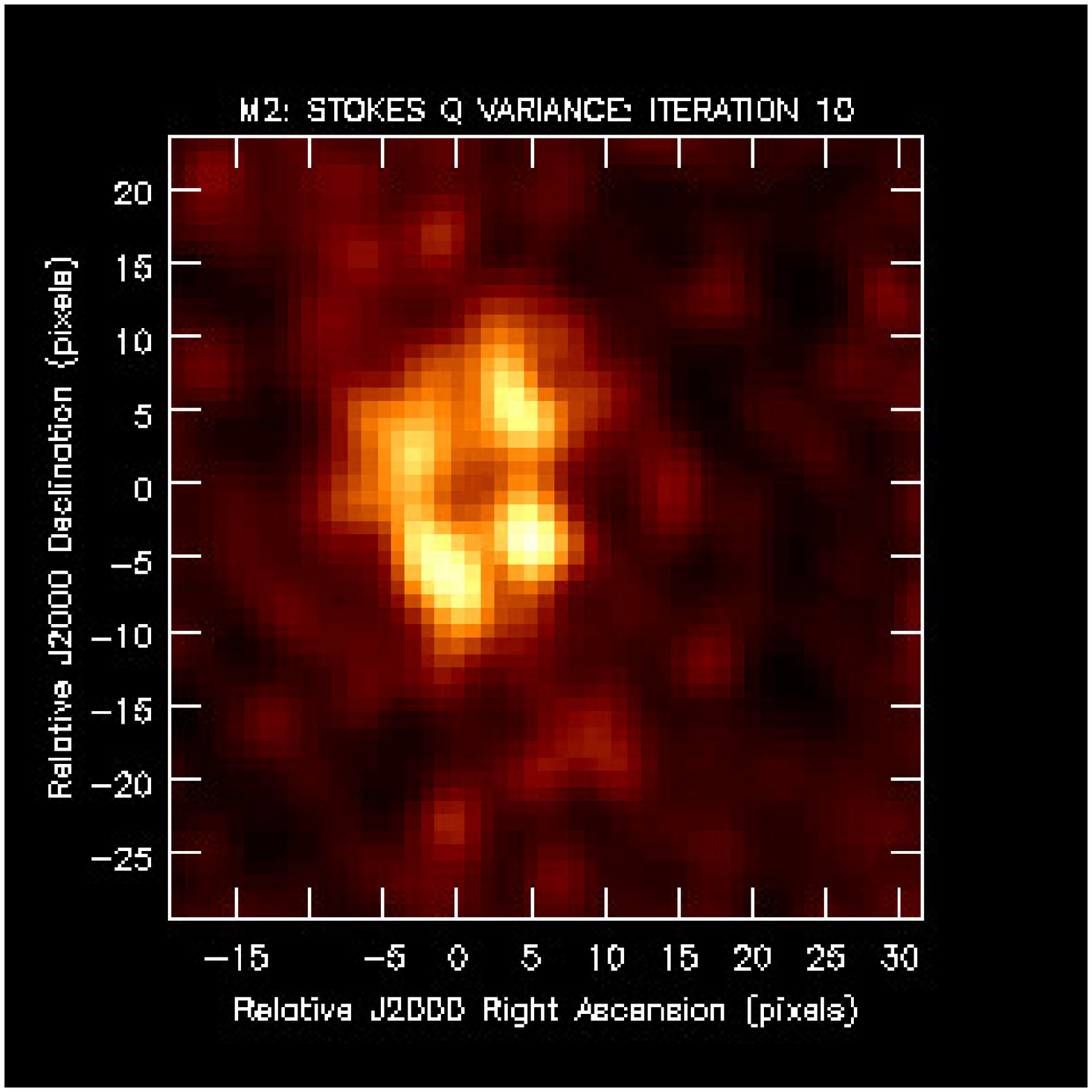} \\ 
\includegraphics[width=57mm, height=57mm, trim = 0mm 5mm 10mm 5mm, clip]{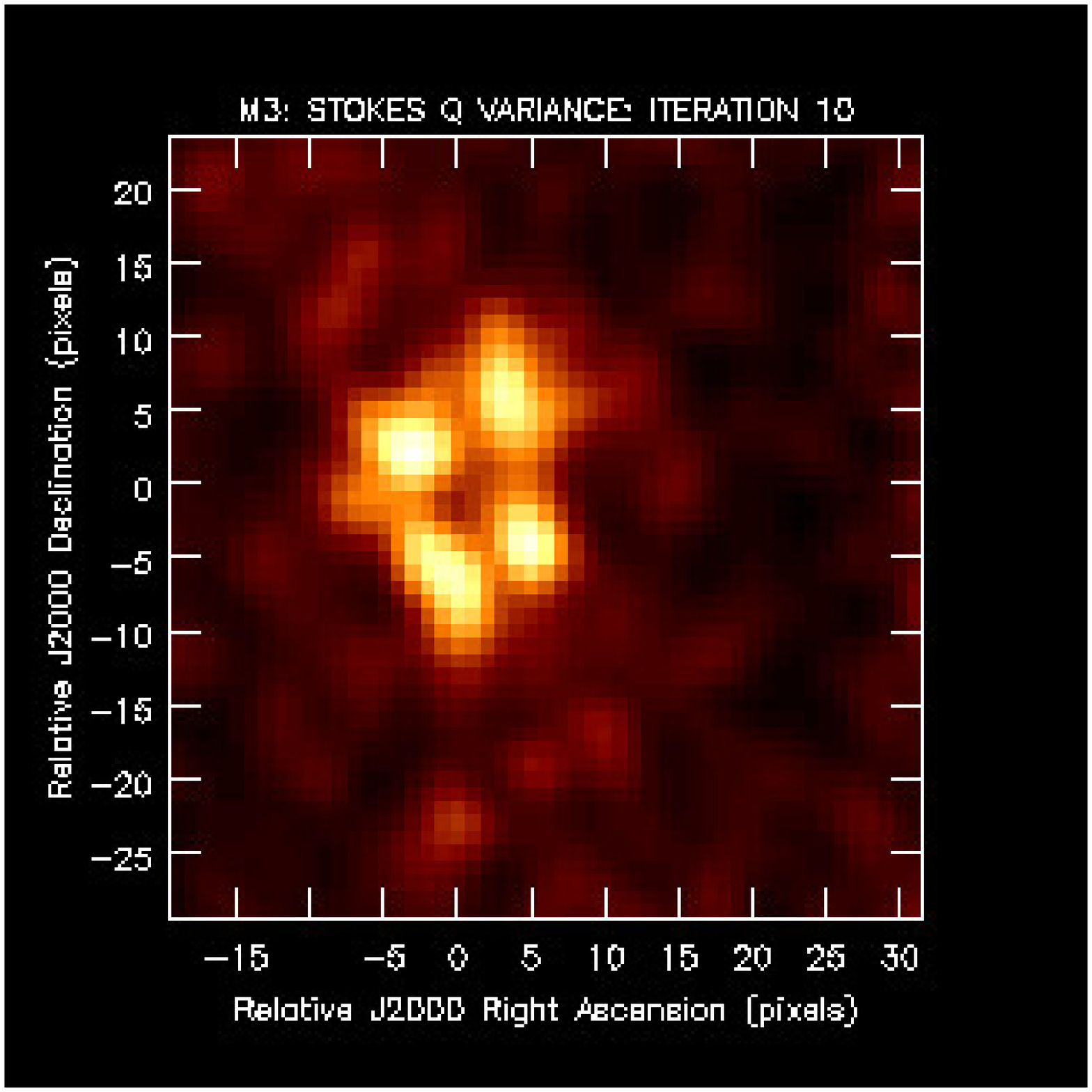} & 
\includegraphics[width=57mm, height=57mm, trim = 0mm 5mm 10mm 5mm, clip]{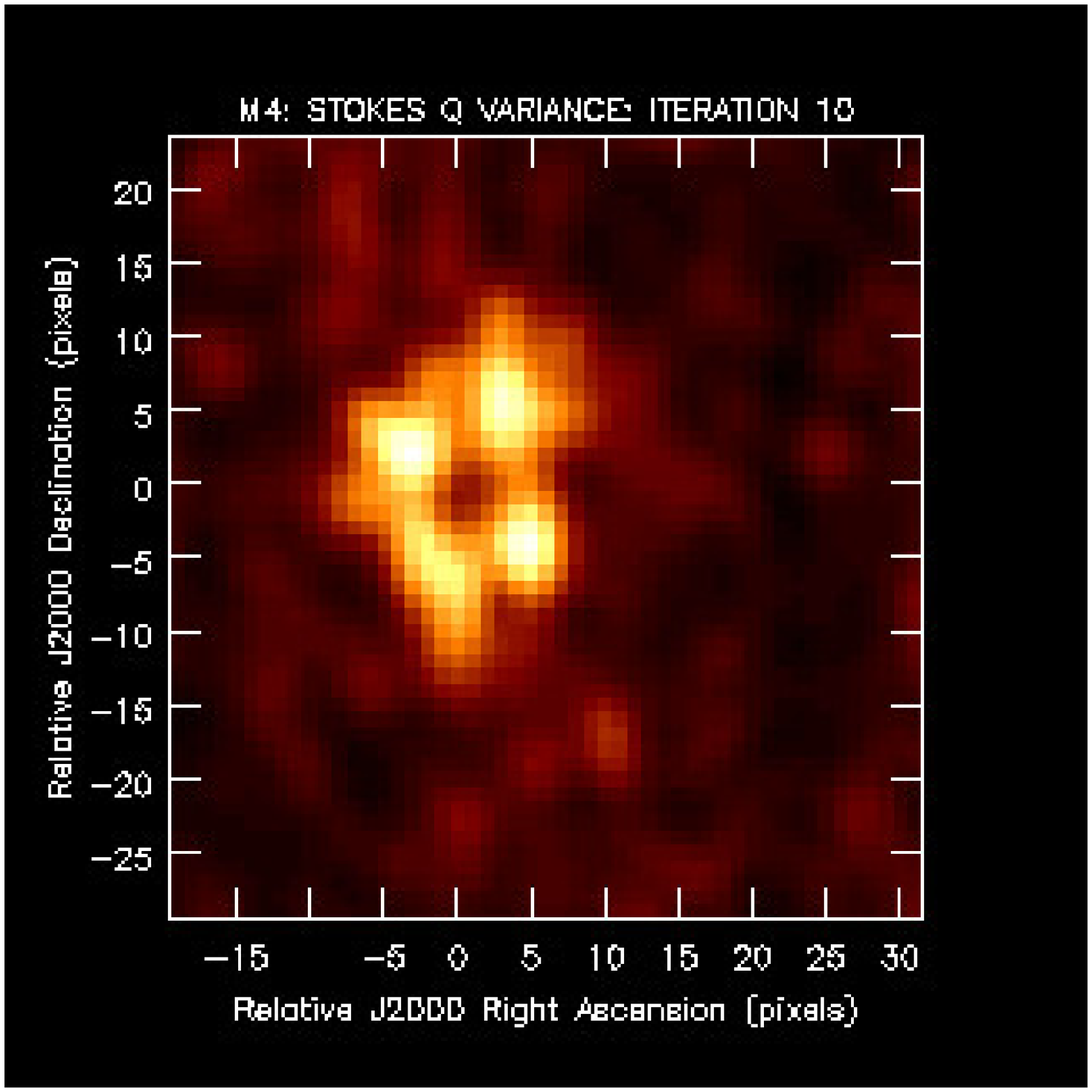} & 
\includegraphics[width=57mm, height=57mm, trim = 0mm 5mm 10mm 5mm, clip]{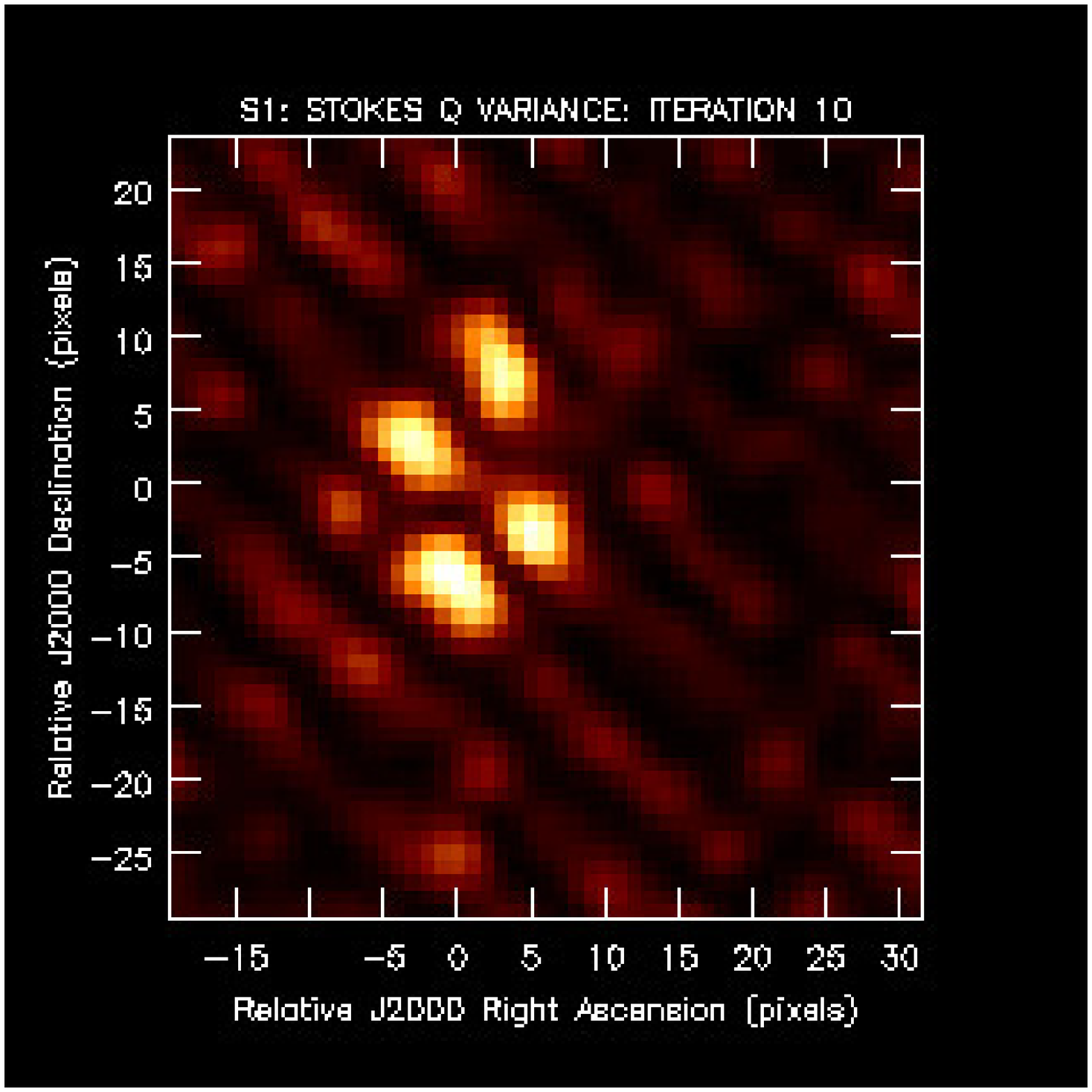} \\ 
\includegraphics[width=57mm, height=57mm, trim = 0mm 5mm 10mm 5mm, clip]{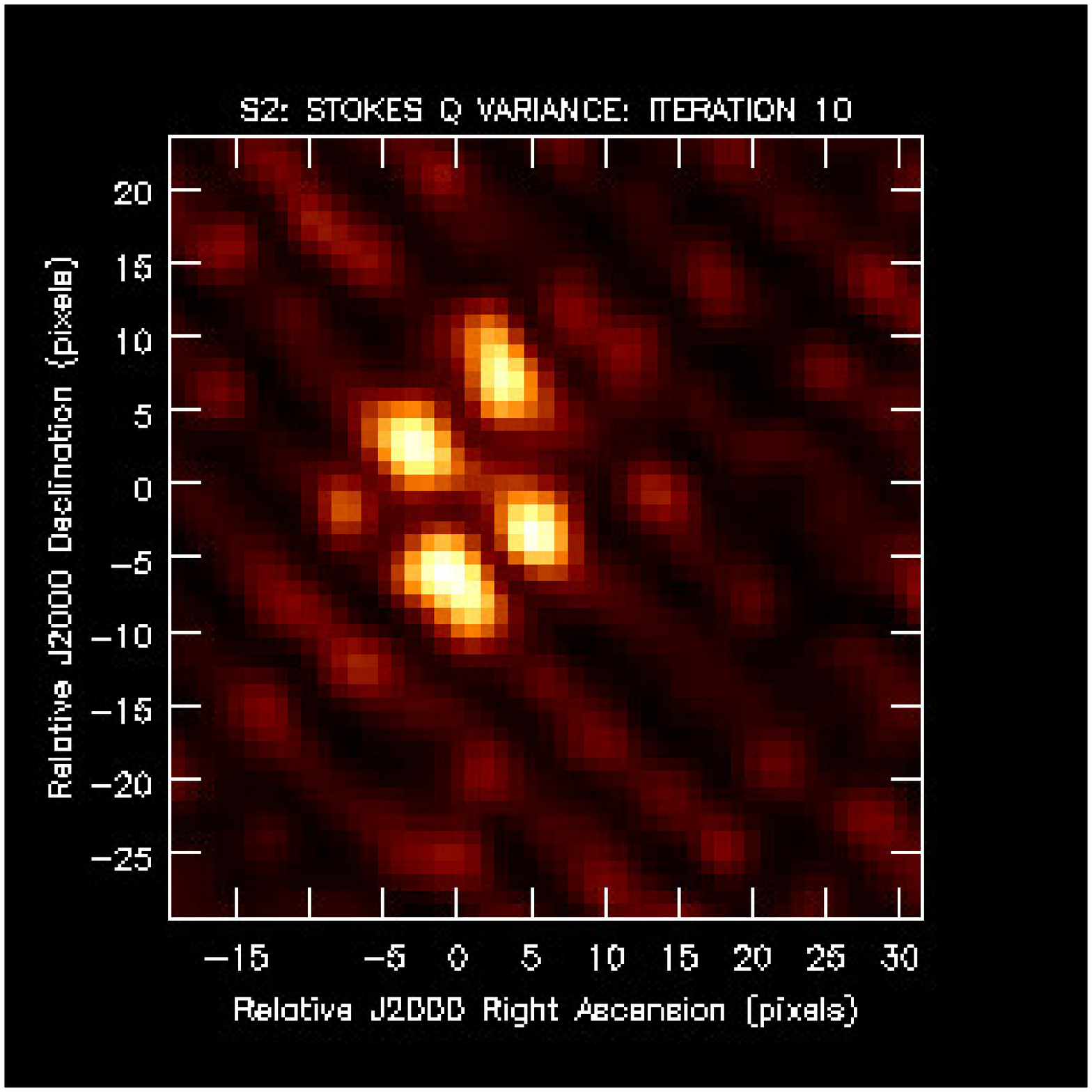} & 
\includegraphics[width=57mm, height=57mm, trim = 0mm 5mm 10mm 5mm, clip]{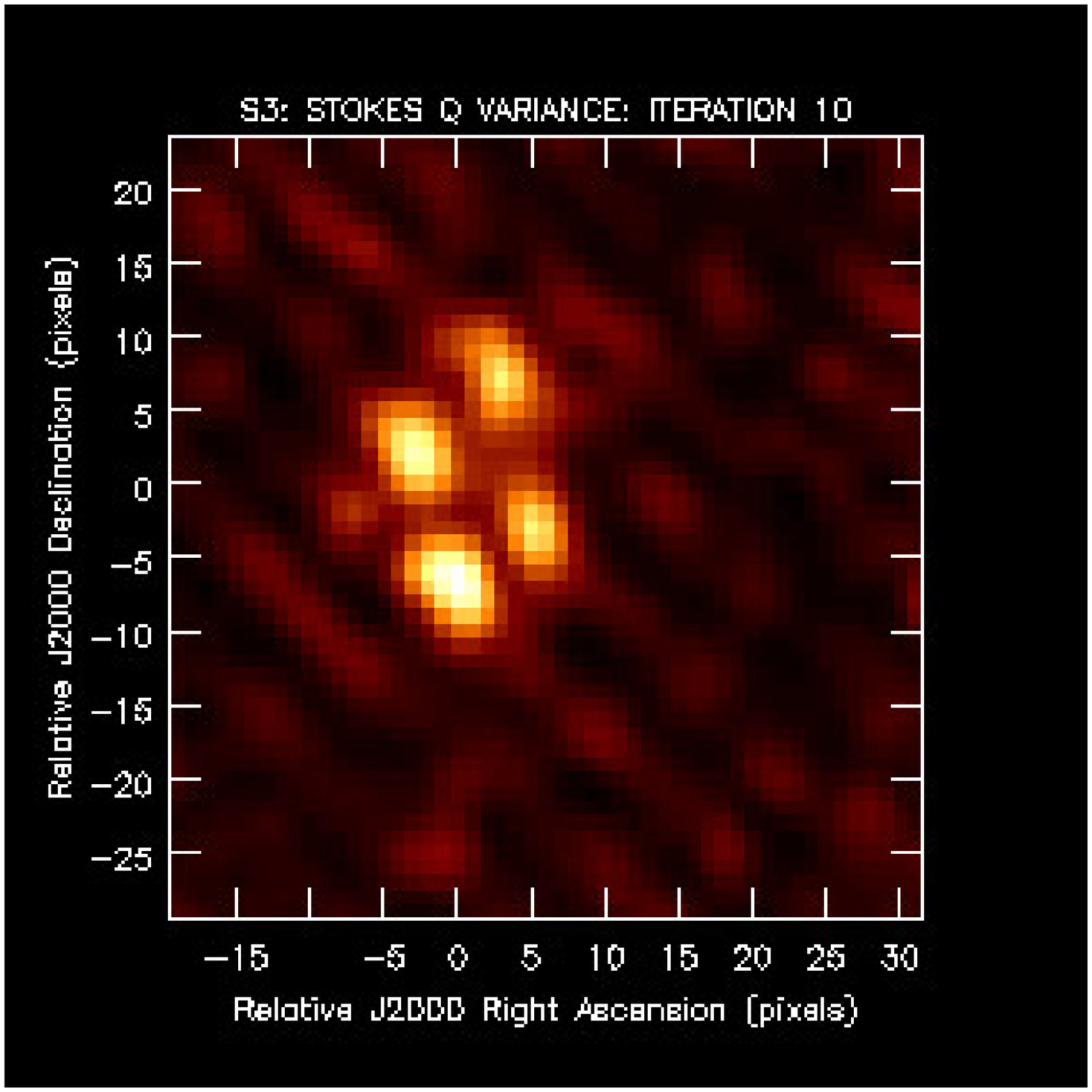} & 
\includegraphics[width=57mm, height=57mm, trim = 0mm 5mm 10mm 5mm, clip]{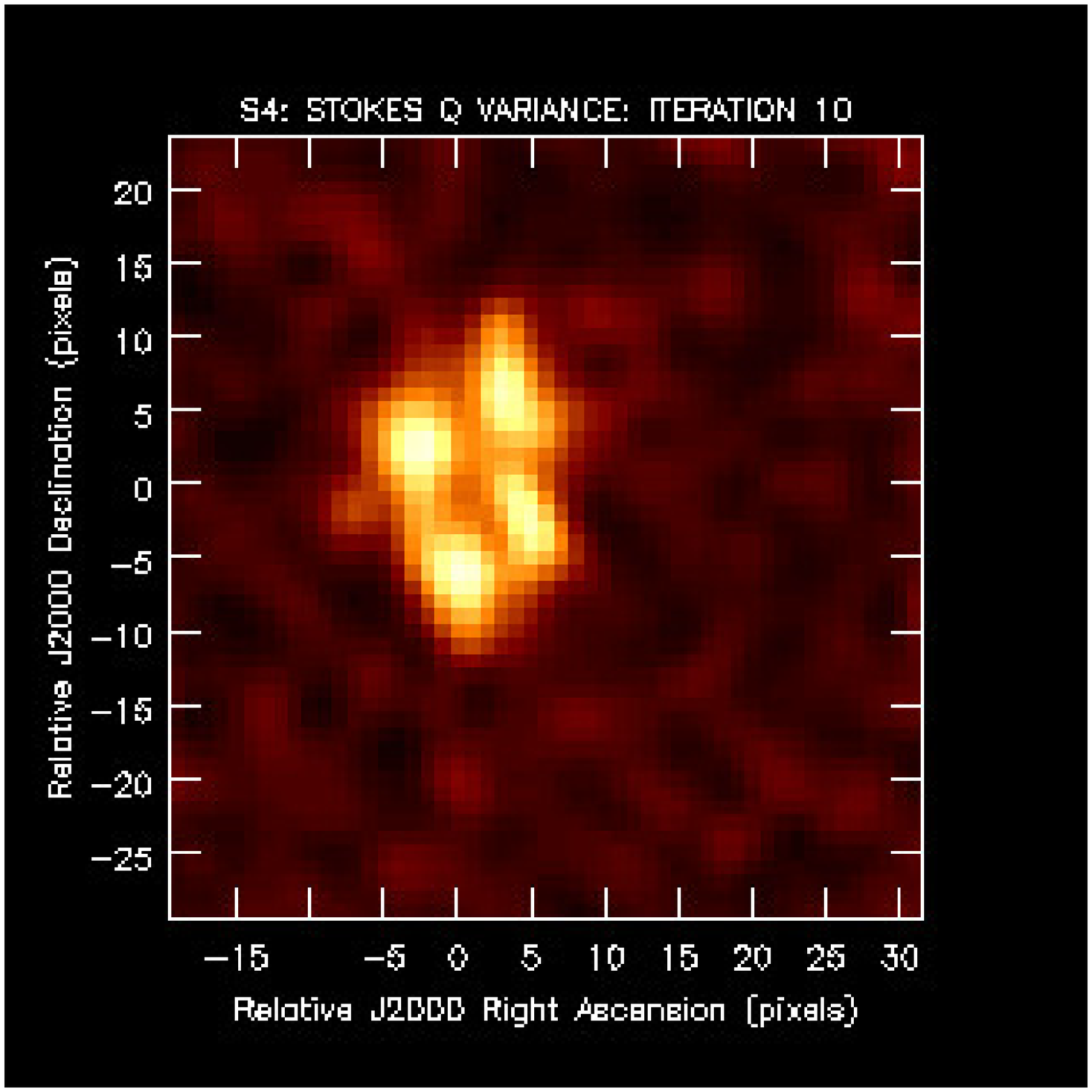} \\ 
\end{array}$ 

\caption{Same as Figure~\ref{fig-col-qvar-x}, but for run code Z.}

\label{fig-col-qvar-z} 
\end{figure}

For a qualitative assessment of the statistical performance of the
bootstrap methods relative to the Monte Carlo reference, we plot in
Figures~\ref{fig-col-qvar-b} through \ref{fig-col-qvar-z} the
pixel-based Stokes $Q$ variance images for the final iteration of
polarization self-calibration for a subset of run codes, chosen here
to be \{B,D,X,Z\}; these run codes are representative of typical
bootstrap statistical performance observed. We choose Stokes \textit{Q} here as both Stokes \{$Q,U$\} show residual  calibration errors most clearly for the underlying polarization calibration algorithm. In each of these figures,
the reference Monte Carlo variance image is shown in the upper left
corner; the remaining images, in sequential order, are the variance
images obtained by bootstrap resampling methods \{M1-M4,\ S1-S4\}. A
default color mapping was used in generating the raster images in
these figures; for each image the same color palette was mapped to the
range of pixel data values in the image variance cube resulting from
each run; each cube has axes $(x,y)$, Stokes parameter $\{I,Q,U,V\}$,
and self-calibration iteration number. The variance images contain
contributions from both calibration and deconvolution errors, as
discussed in the introduction.

For a quantitative assessment of bootstrap statistical performance we
use a similar goodness-of-fit statistic $v_{MSE}$, to that defined in
Paper I to compare the variance image obtained using a given bootstrap
resampling method, $var_{xy}$, to the the reference variance image
obtained by Monte Carlo simulation, $var_{xy}^{MC}$. In the current
work, the statistic is computed over the inner quarter of the image,
all Stokes parameters $\{I,Q,U,V\}$, and all self-calibration iteration
numbers $k=1-10$, as:

\begin{equation}
v_{MSE} = \frac{1}{N} \sum_{IQUV} \sum_{j=1}^{10} \sum_{\Omega_{\frac{1}{4}}} \left ( \frac{var_{xy}}{v_f} - var_{xy}^{MC} \right )^2 \label{eqn-vmse}
\end{equation}

were $v_f$ is the variance scaling factor \citep{dav97}.

The value of $v_f$ was determined numerically for each bootstrap run
in a Newton minimization of the goodness-of-fit statistic $v_{MSE}$,
as defined in equation~\ref{eqn-vmse}. The resulting fitted values of
$v_f$ are listed in Table~\ref{tbl-vf}, and the associated minima in
$v_{MSE}$ given in Table~\ref{tbl-vmse}, both cross-tabulated by run
code \{A-F,\ X-Z\} and bootstrap code \{M1-M4,\ S1-S4\}. The values of
$v_{MSE}$ in Table~\ref{tbl-vmse} are normalized per row by the
minimum value obtained across bootstrap codes \{M1-M4,\ S1-S4\} for
each run code \{A-F,\ X-Z\}. The minimum $v_{MSE}$ is given in the
right-most column of the table. The optimal bootstrap methods are
therefore indicated in this table by unit relative $v_{MSE}$. Note
that the values of $v_{MSE}$ given in Table 5 of Paper I were
incorrectly scaled in absolute magnitude and need to be multiplied by
a factor $1.502 \times 10^{-5}$ for comparison with the $v_{MSE}$
values in the current work.

\begin{figure}[h] 
\advance\leftskip-1cm
\advance\rightskip-1cm
 $\begin{array}{c@{\hspace{2mm}}c@{\hspace{2mm}}c} 
\includegraphics[width=57mm, height=57mm, trim = 0mm 5mm 10mm 5mm, clip]{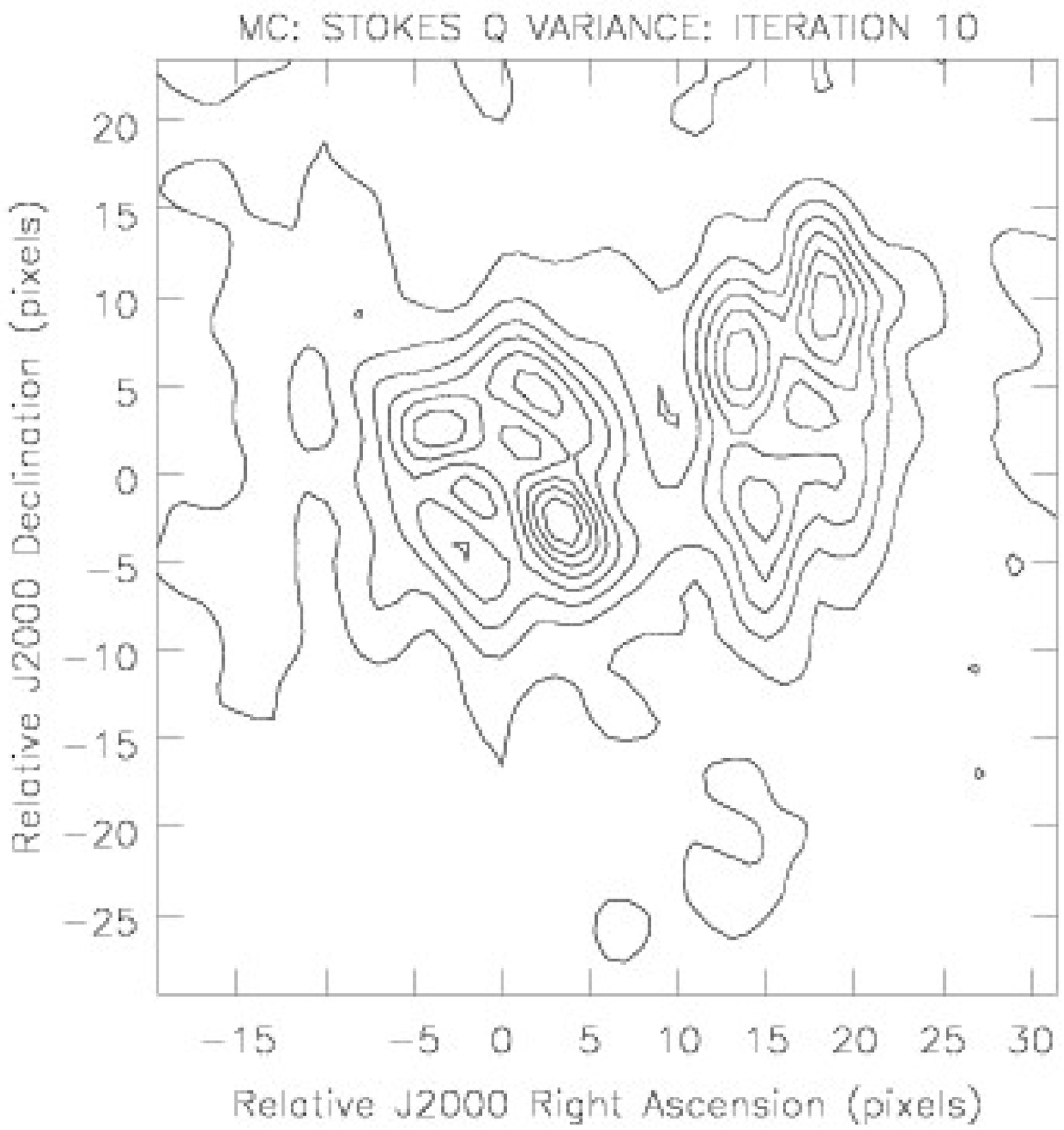} & 
\includegraphics[width=57mm, height=57mm, trim = 0mm 5mm 10mm 5mm, clip]{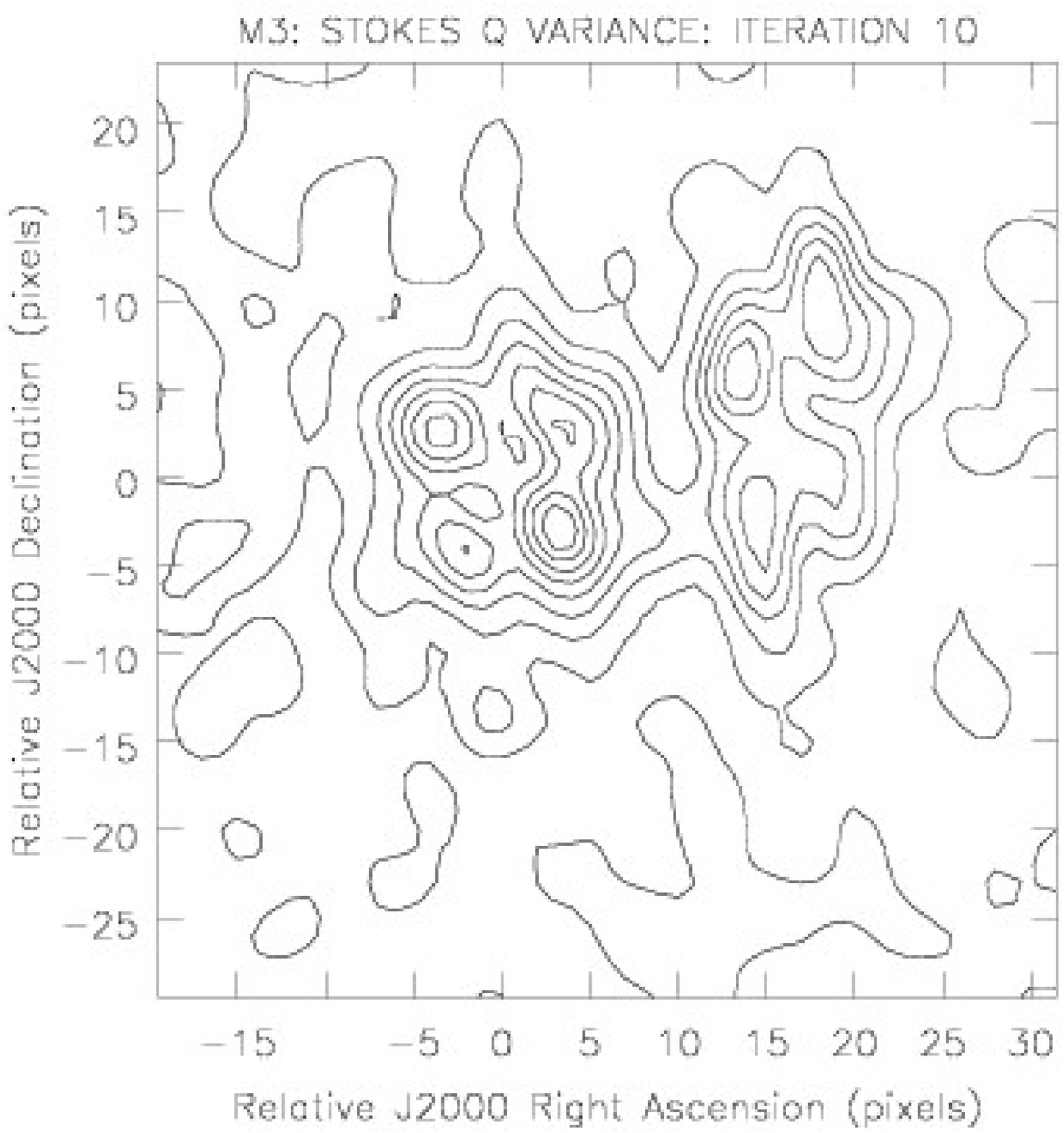} & 
\includegraphics[width=57mm, height=57mm, trim = 0mm 5mm 10mm 5mm, clip]{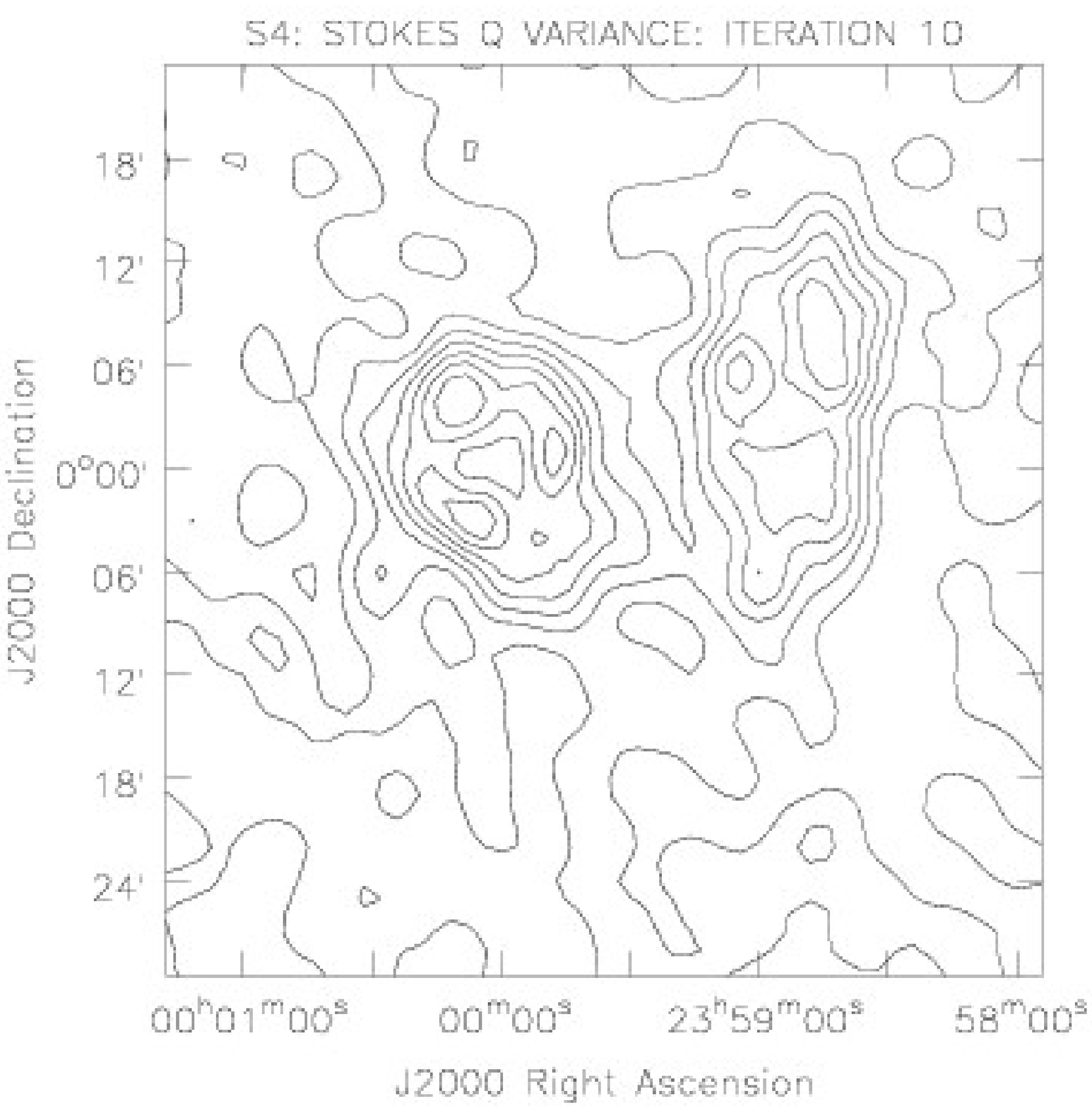} \\ 
\includegraphics[width=57mm, height=57mm, trim = 0mm 5mm 10mm 5mm, clip]{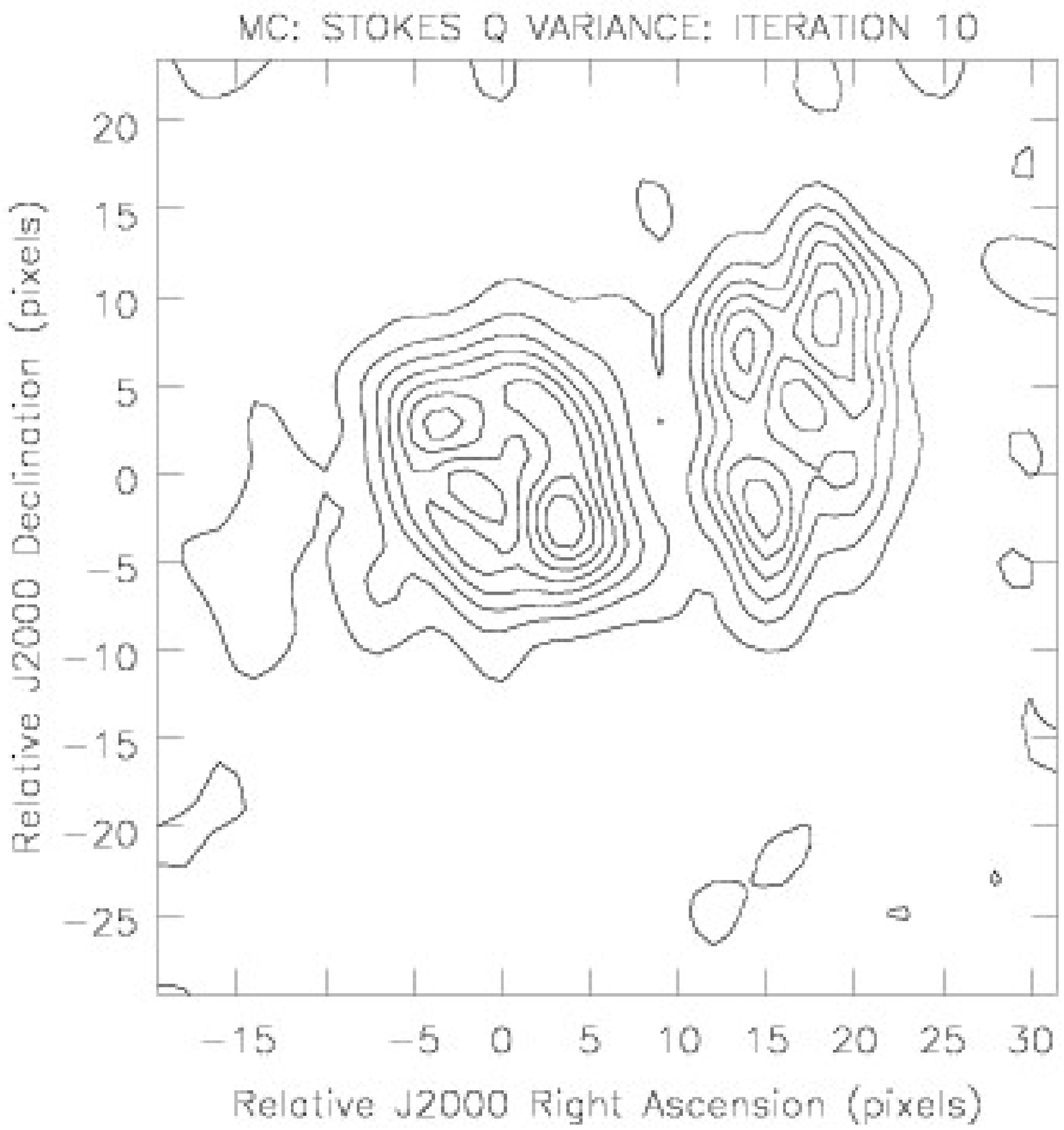} & 
\includegraphics[width=57mm, height=57mm, trim = 0mm 5mm 10mm 5mm, clip]{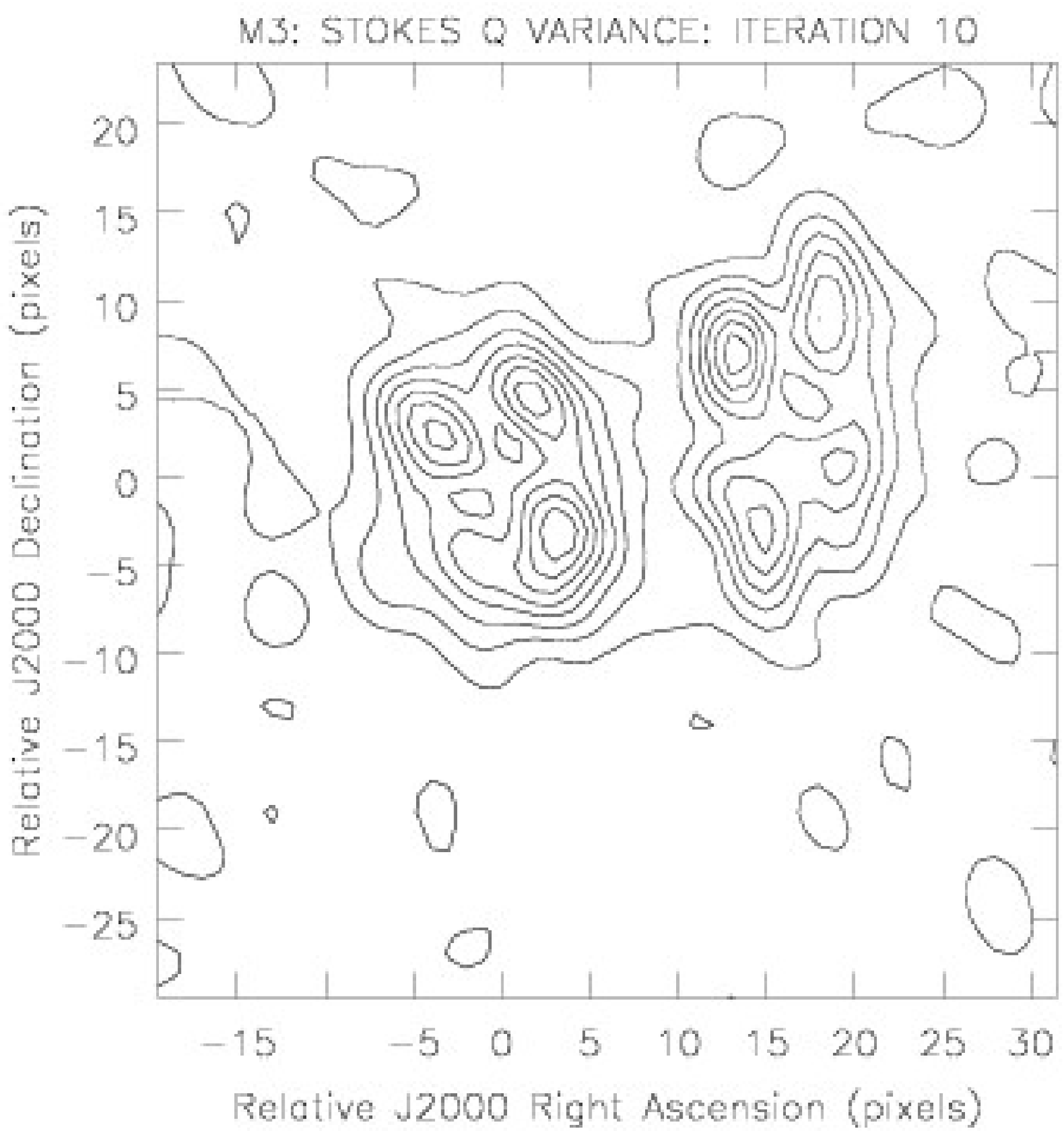} & 
\includegraphics[width=57mm, height=57mm, trim = 0mm 5mm 10mm 5mm, clip]{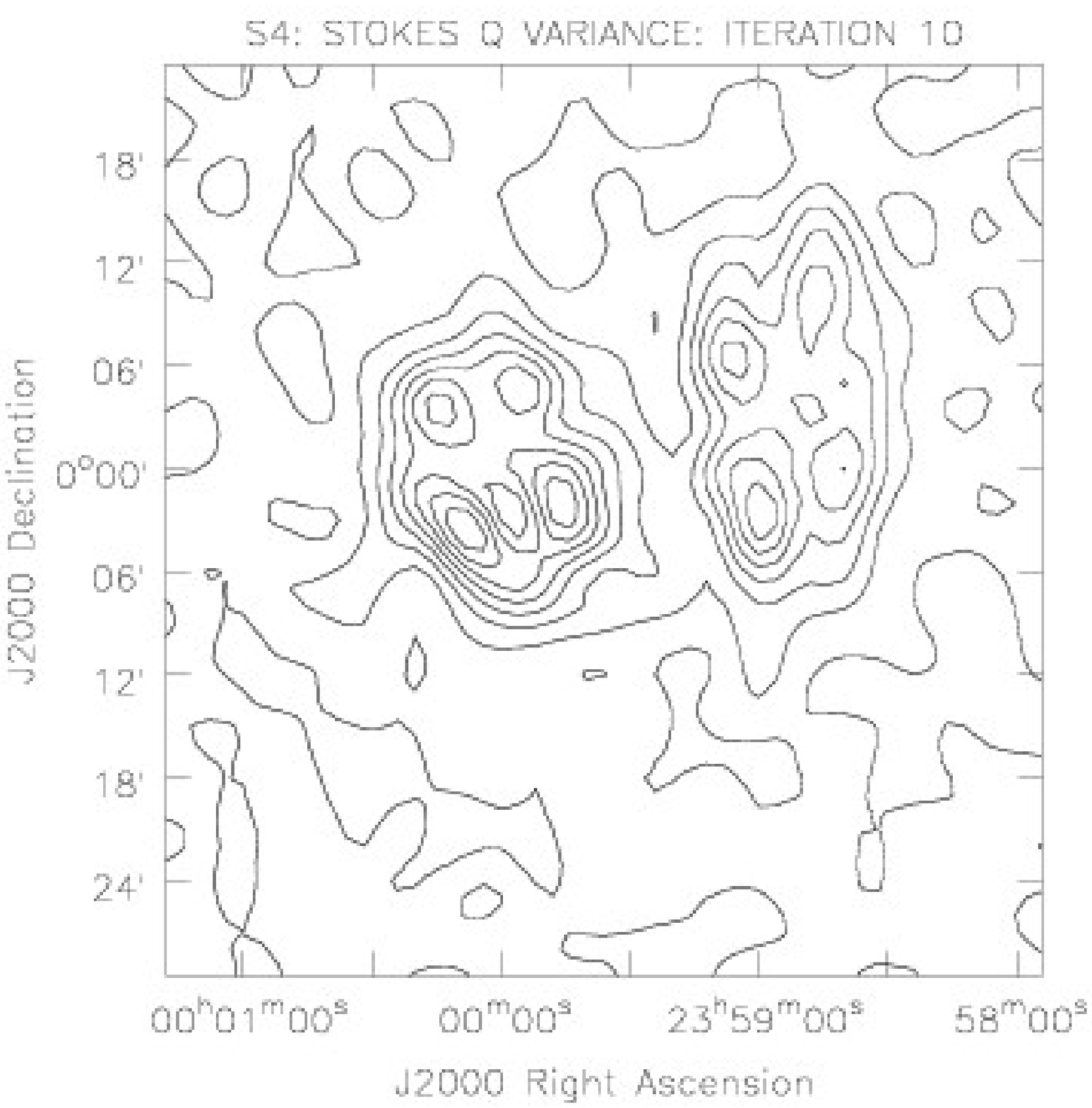} \\ 
\includegraphics[width=57mm, height=57mm, trim = 0mm 5mm 10mm 5mm, clip]{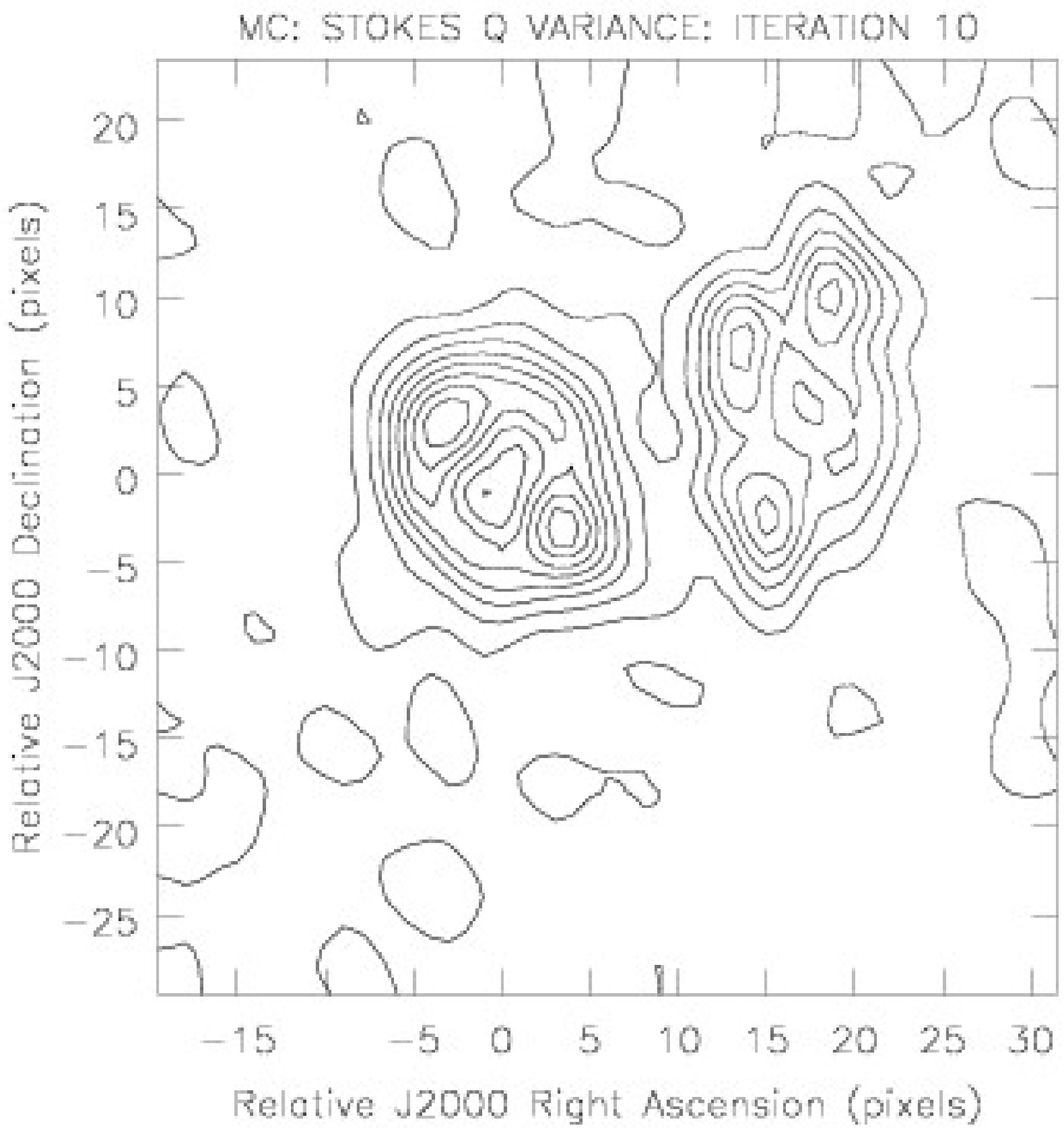} & 
\includegraphics[width=57mm, height=57mm, trim = 0mm 5mm 10mm 5mm, clip]{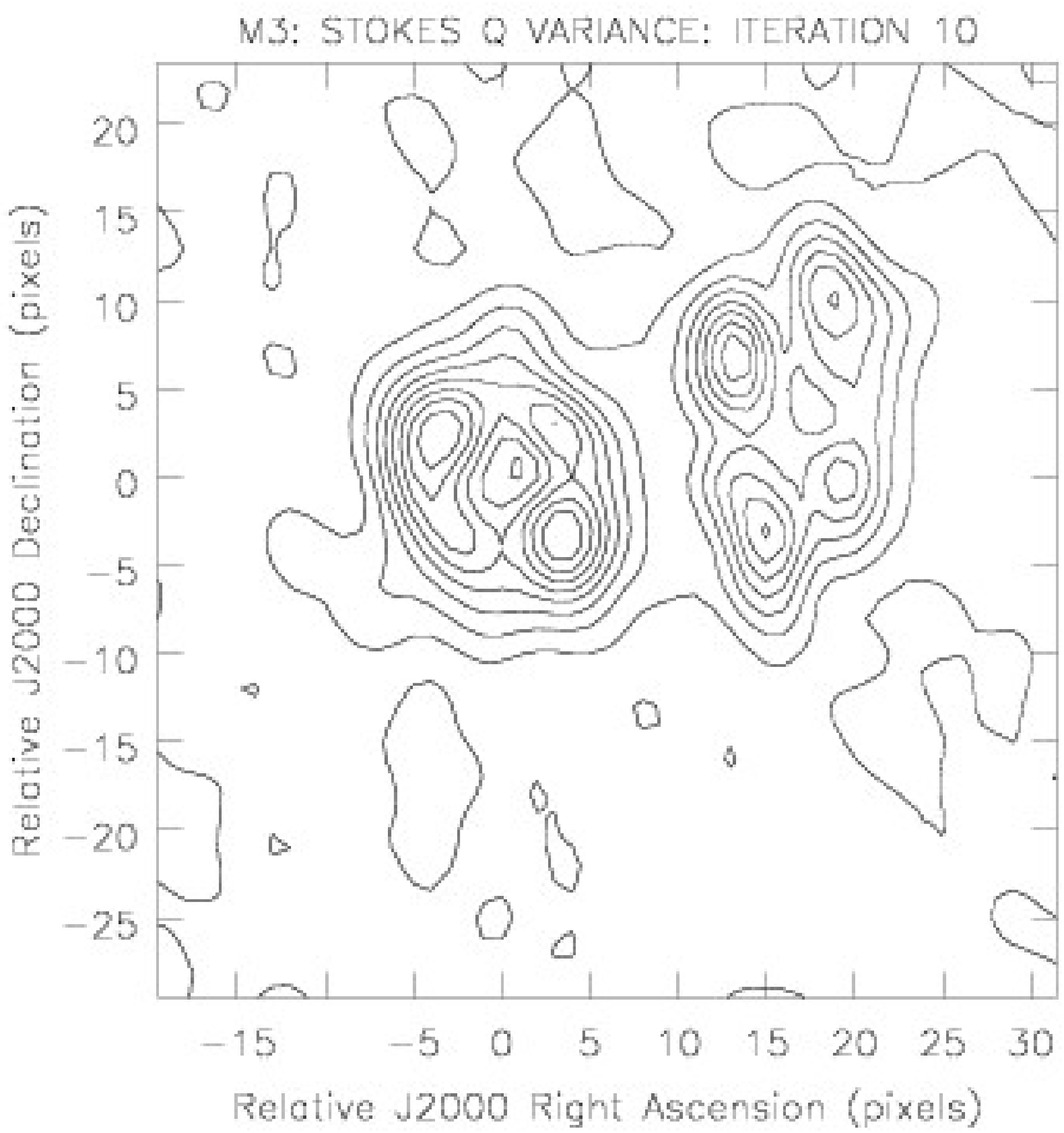} & 
\includegraphics[width=57mm, height=57mm, trim = 0mm 5mm 10mm 5mm, clip]{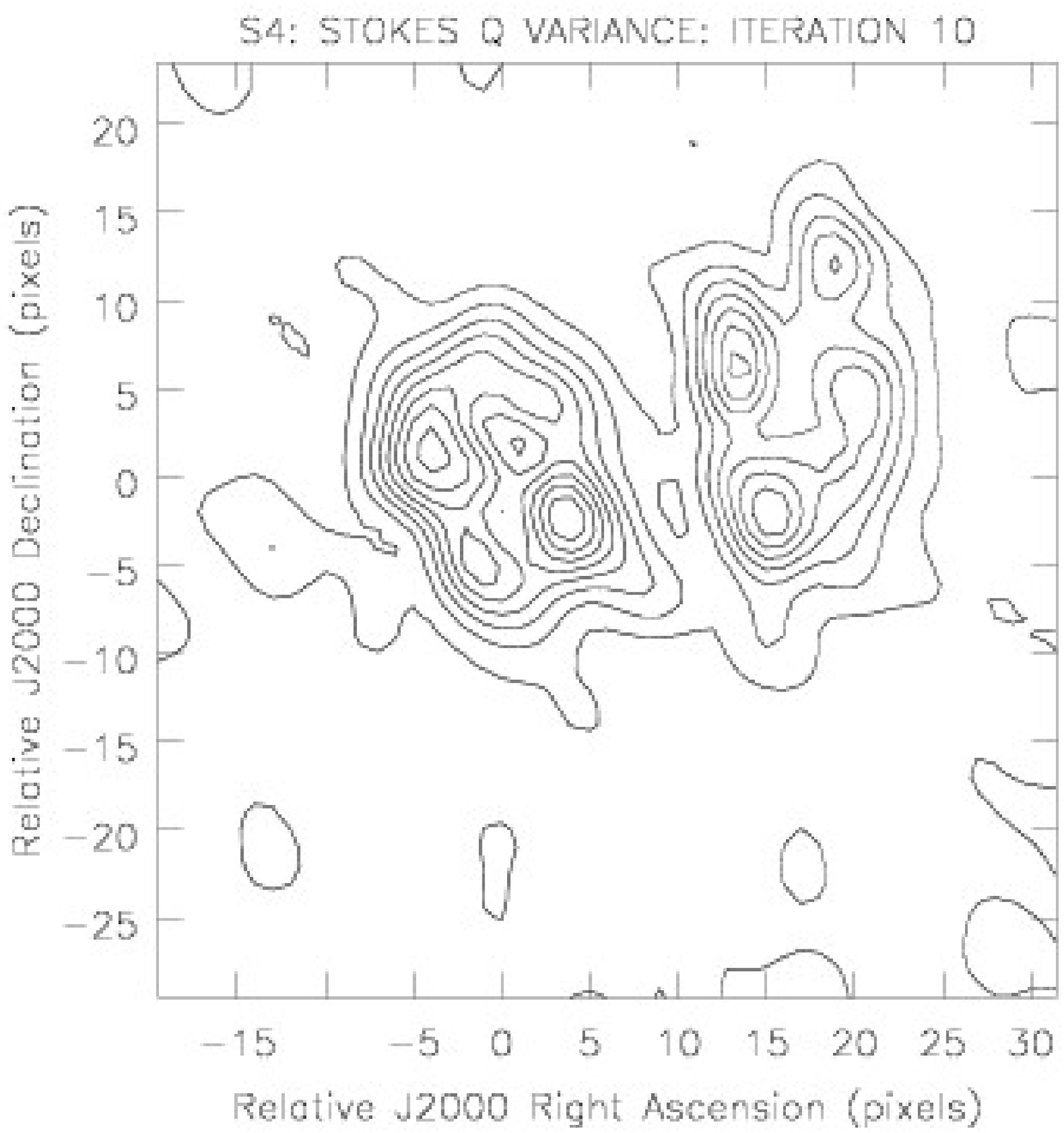} \\ 
\end{array}$ 

\caption{The imaging estimator Stokes $Q$ variance for the final
polarization self-calibration iteration obtained by direct Monte Carlo
simulation (MC; {\it left}), and for the optimum model-based (M3; {\it
center column}) and subsample (S4; {\it right-most column}) bootstrap
methods identified in Table~\ref{tbl-vmse} for run codes A, B, and C
({\it from top to bottom}). The parameters for these bootstrap
resampling codes are listed in Table~\ref{tbl-mcode} and
Table~\ref{tbl-scode} respectively. The run code parameters are given
in Table~\ref{tbl-par} and Table~\ref{tbl-sep}. The contour levels are
plotted at levels \textbraceleft0, 0.1, 0.2,...,1.0\textbraceright \ of the peak variance
values: (A-MC: 1.883,\ A-M3: 1.783,\ A-S4: 1.924,\ B-MC: 2.534,\ B-M3:
2.429,\ B-S4: 3.089,\ C-MC: 3.697,\, C-M3: 3.450,\ C-S4: 5.580)$
\times 10^{-6}$ (Jy/beam)$^2$.}

\label{fig-qvar-abc} 
\end{figure}

\begin{figure}[h] 
\advance\leftskip-1cm
\advance\rightskip-1cm
 $\begin{array}{c@{\hspace{2mm}}c@{\hspace{2mm}}c} 
\includegraphics[width=57mm, height=57mm, trim = 0mm 5mm 10mm 5mm, clip]{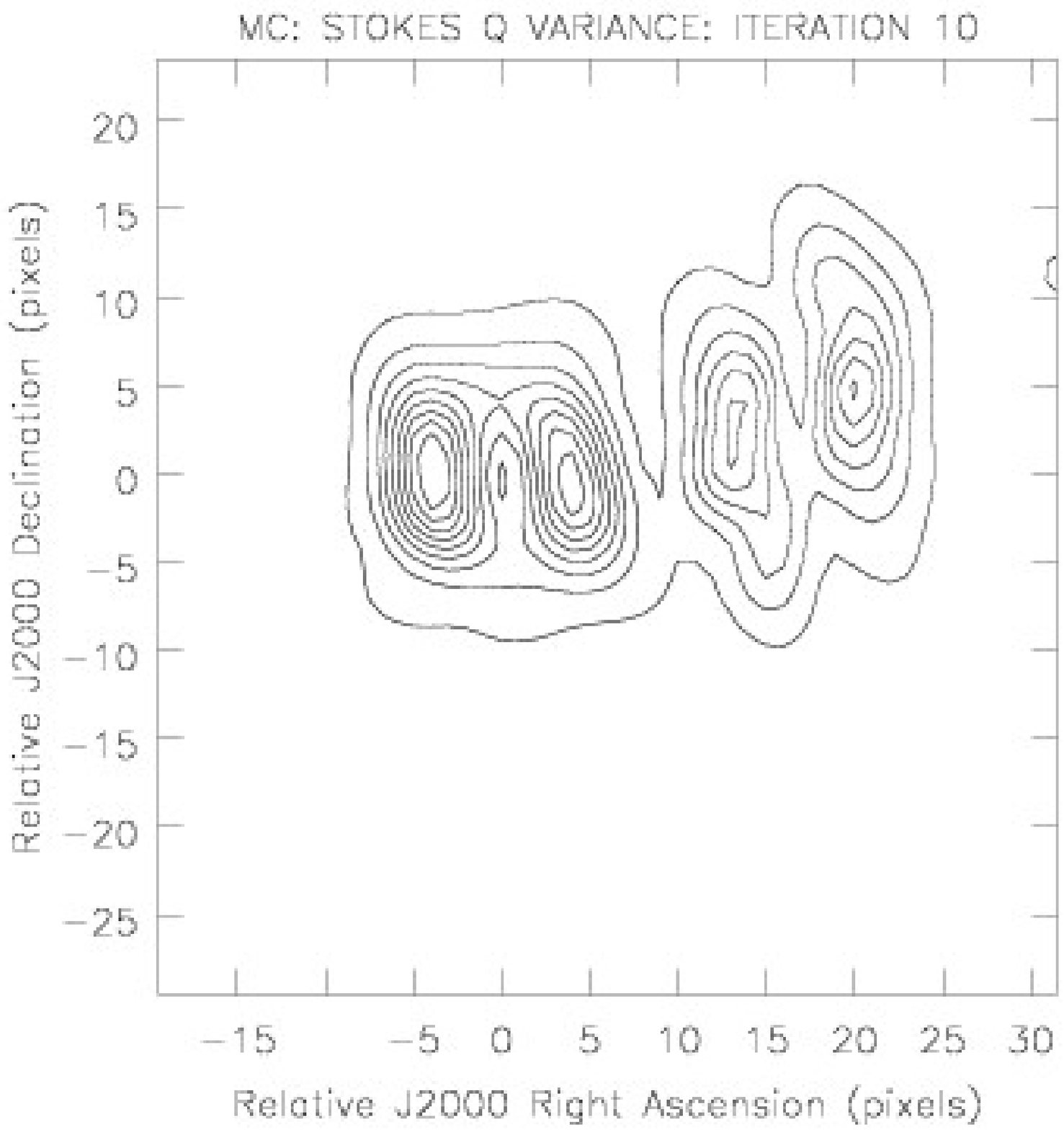} & 
\includegraphics[width=57mm, height=57mm, trim = 0mm 5mm 10mm 5mm, clip]{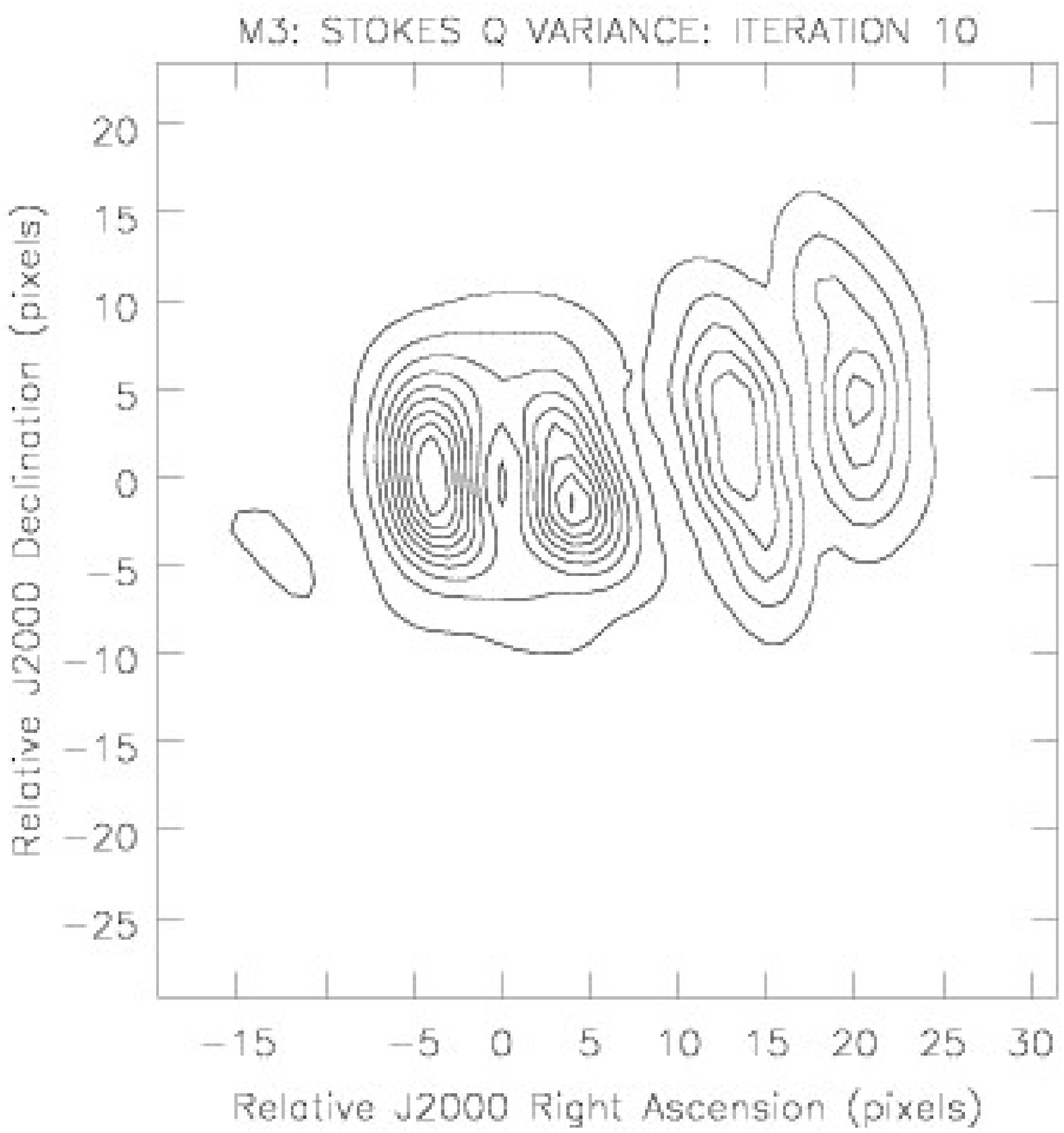} & 
\includegraphics[width=57mm, height=57mm, trim = 0mm 5mm 10mm 5mm, clip]{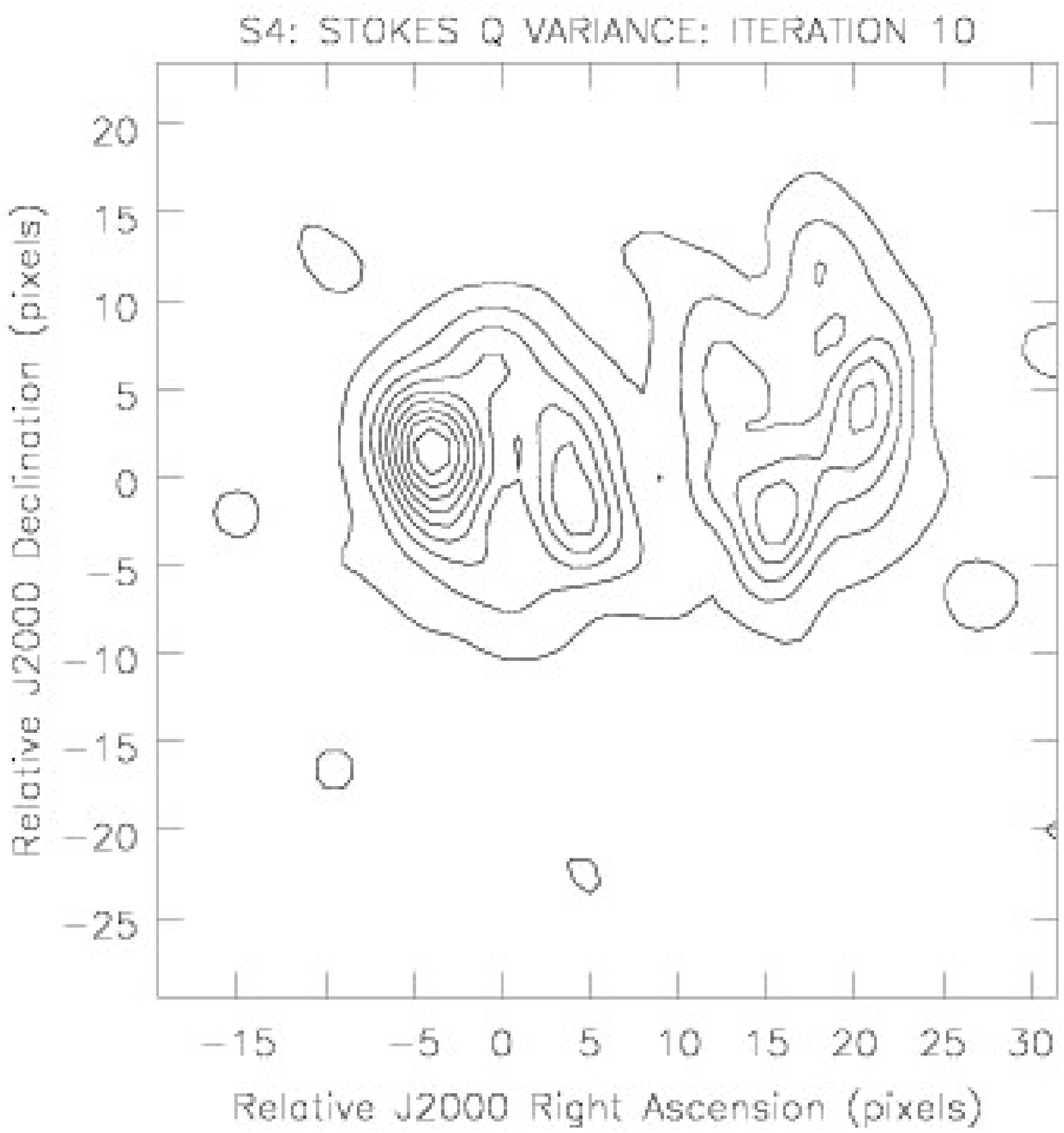} \\ 
\includegraphics[width=57mm, height=57mm, trim = 0mm 5mm 10mm 5mm, clip]{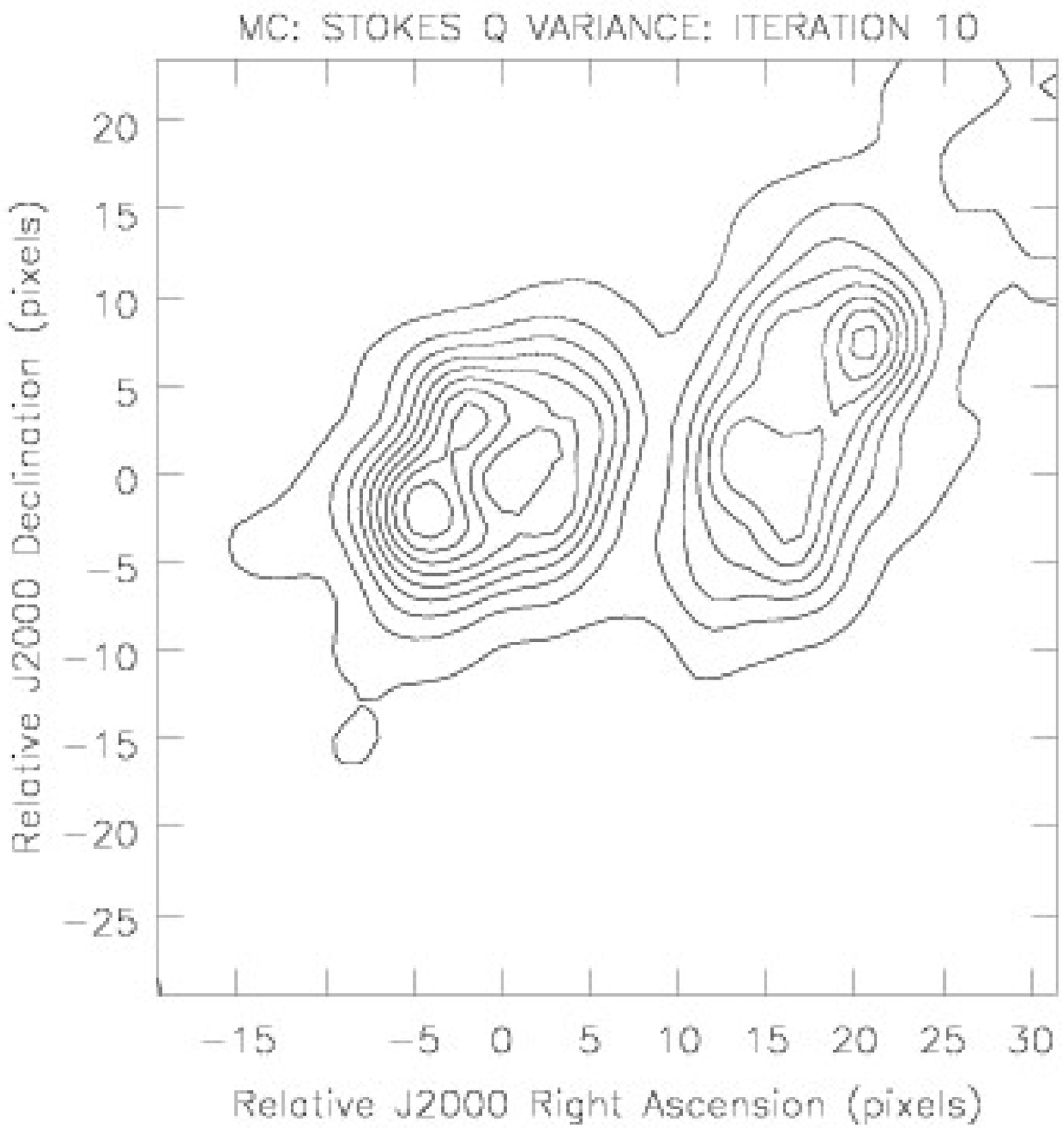} & 
\includegraphics[width=57mm, height=57mm, trim = 0mm 5mm 10mm 5mm, clip]{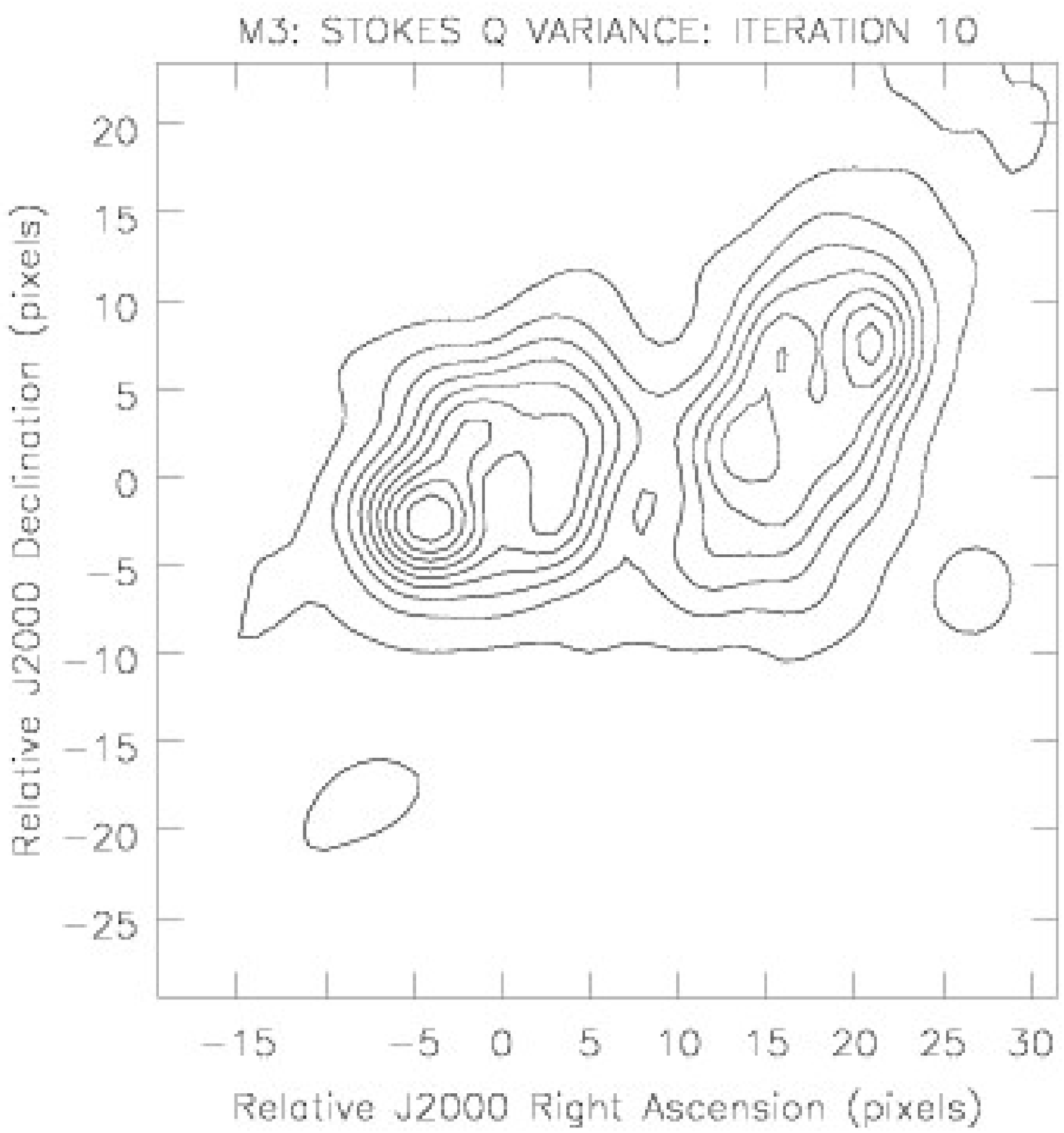} & 
\includegraphics[width=57mm, height=57mm, trim = 0mm 5mm 10mm 5mm, clip]{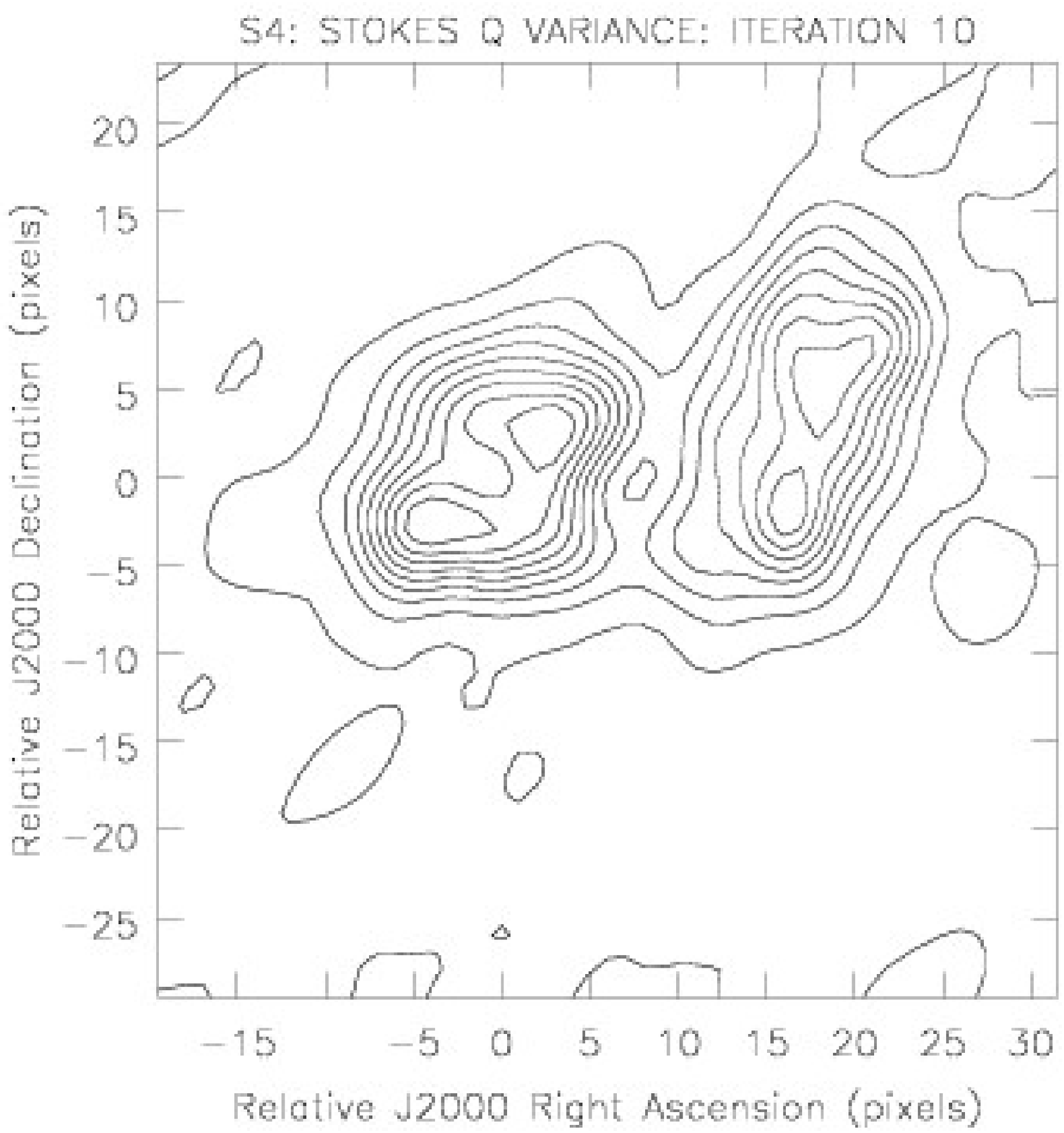} \\ 
\includegraphics[width=57mm, height=57mm, trim = 0mm 5mm 10mm 5mm, clip]{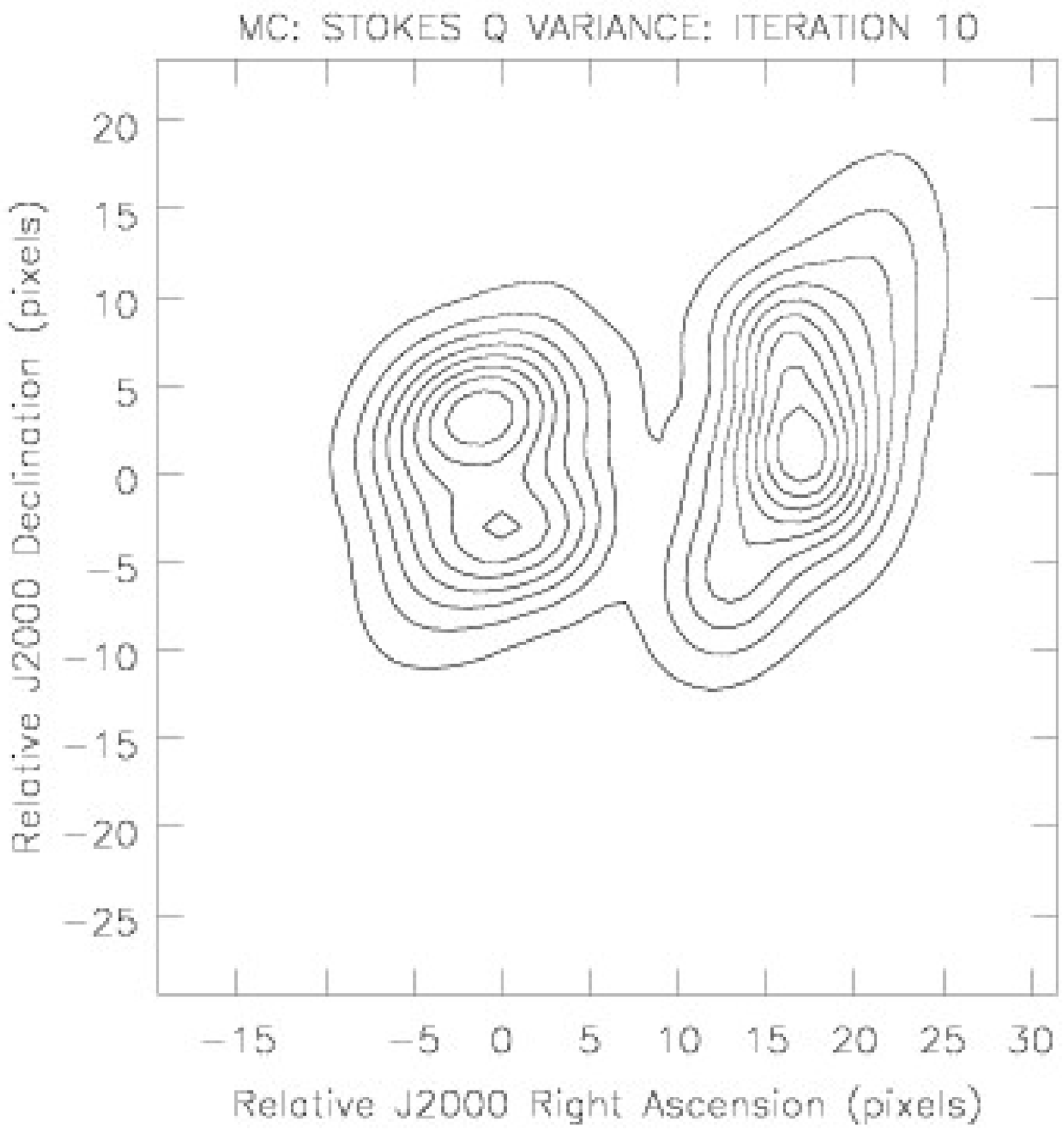} & 
\includegraphics[width=57mm, height=57mm, trim = 0mm 5mm 10mm 5mm, clip]{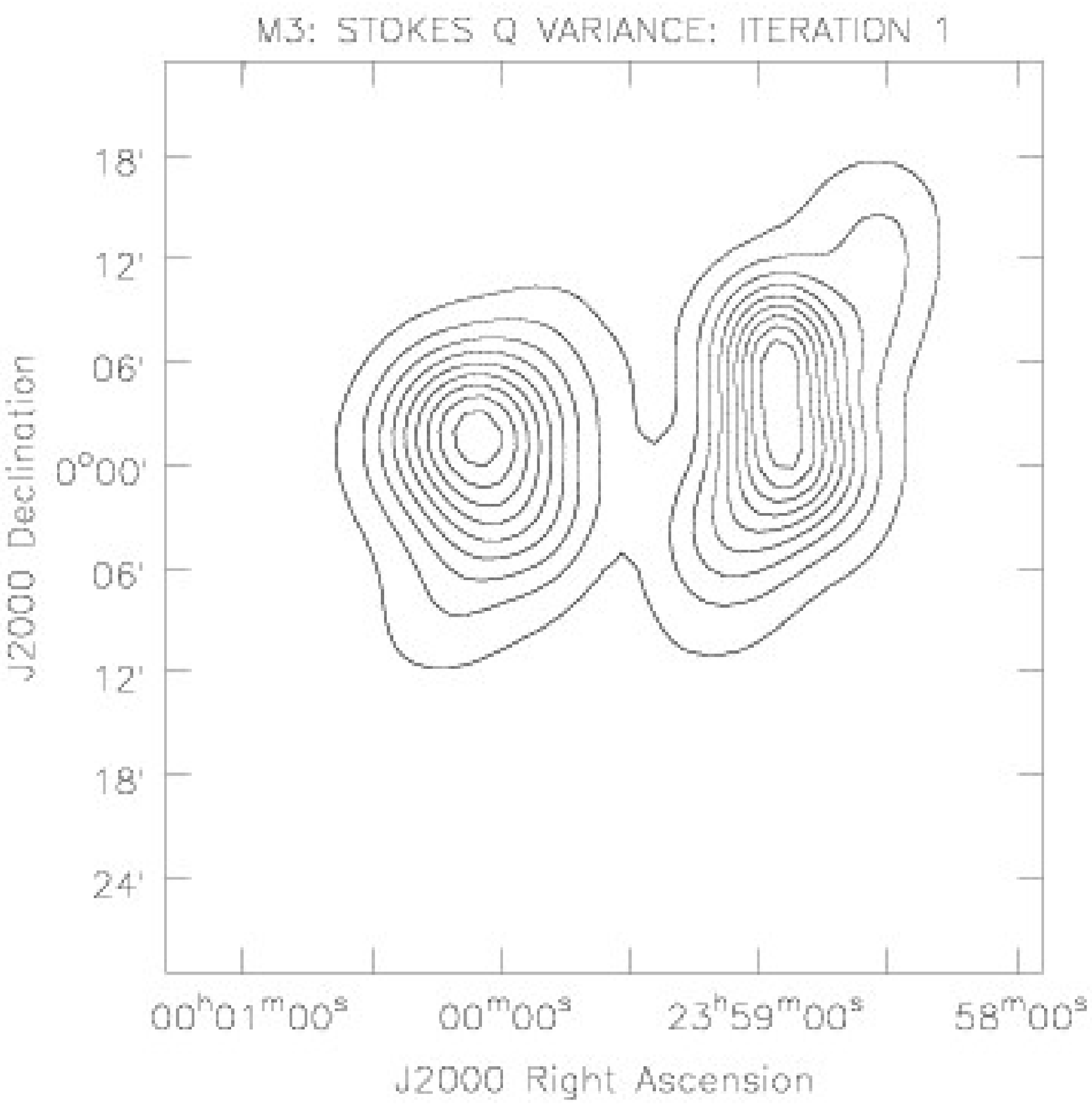} & 
\includegraphics[width=57mm, height=57mm, trim = 0mm 5mm 10mm 5mm, clip]{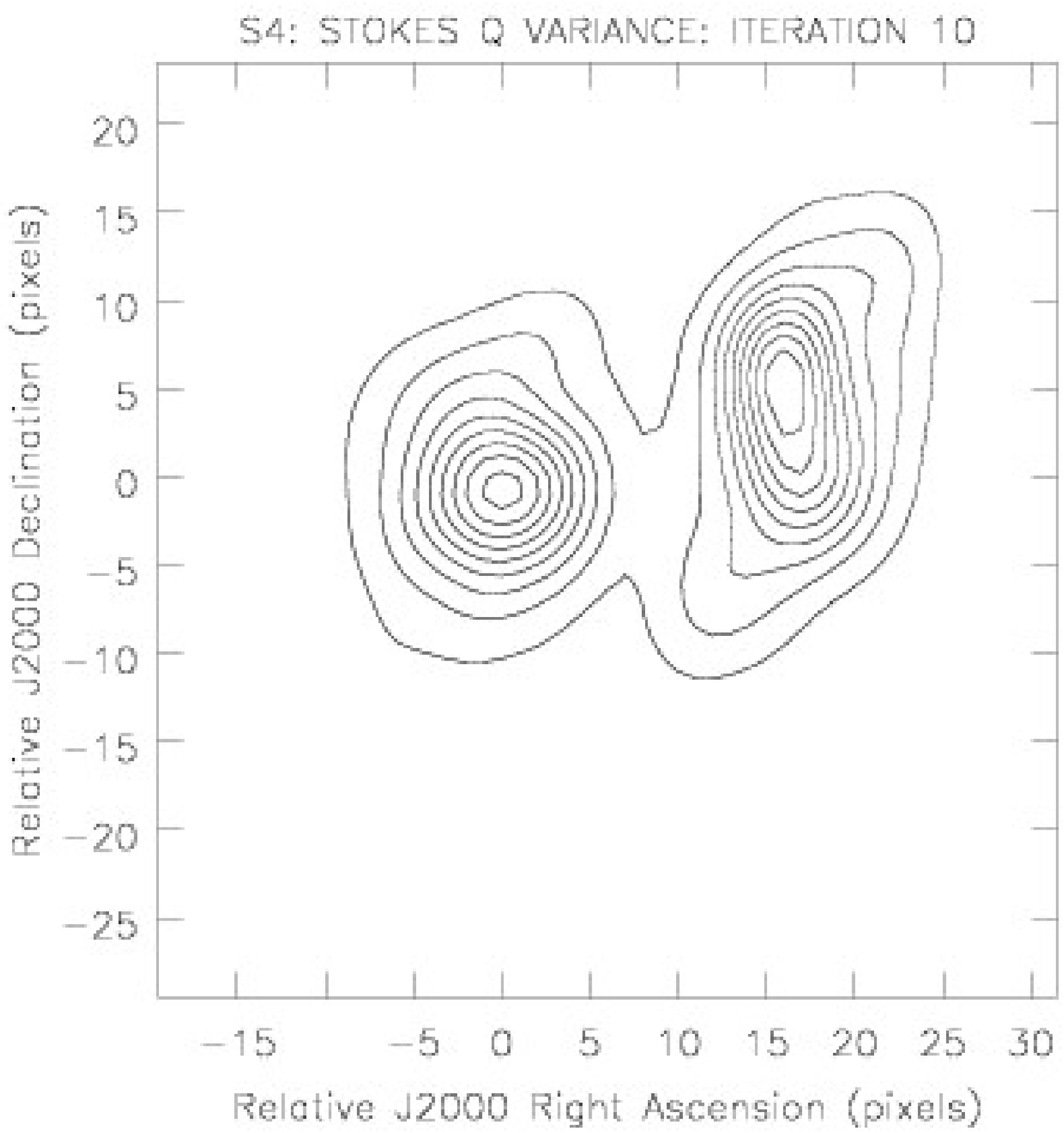} \\ 
\end{array}$ 

\caption{Same as Figure~\ref{fig-qvar-abc}, but for run codes D, E,
and F ({\it from top to bottom}), The contour levels are plotted at
levels \{0, 0.1, 0.2,...,1.0\} of the peak variance values: (D-MC:
12.74,\ D-M3: 13.36,\ D-S4: 18.22,\ E-MC: 6.839,\ E-M3: 7.048,\ E-S4:
13.74,\ F-MC: 44.37,\, F-M3: 51.70,\ E-S4: 114.1)$ \times 10^{-6}$
(Jy/beam)$^2$.}

\label{fig-qvar-def} 
\end{figure}

\begin{figure}[h] 
\advance\leftskip-1cm
\advance\rightskip-1cm
 $\begin{array}{c@{\hspace{2mm}}c@{\hspace{2mm}}c} 
\includegraphics[width=57mm, height=57mm, trim = 0mm 5mm 10mm 5mm, clip]{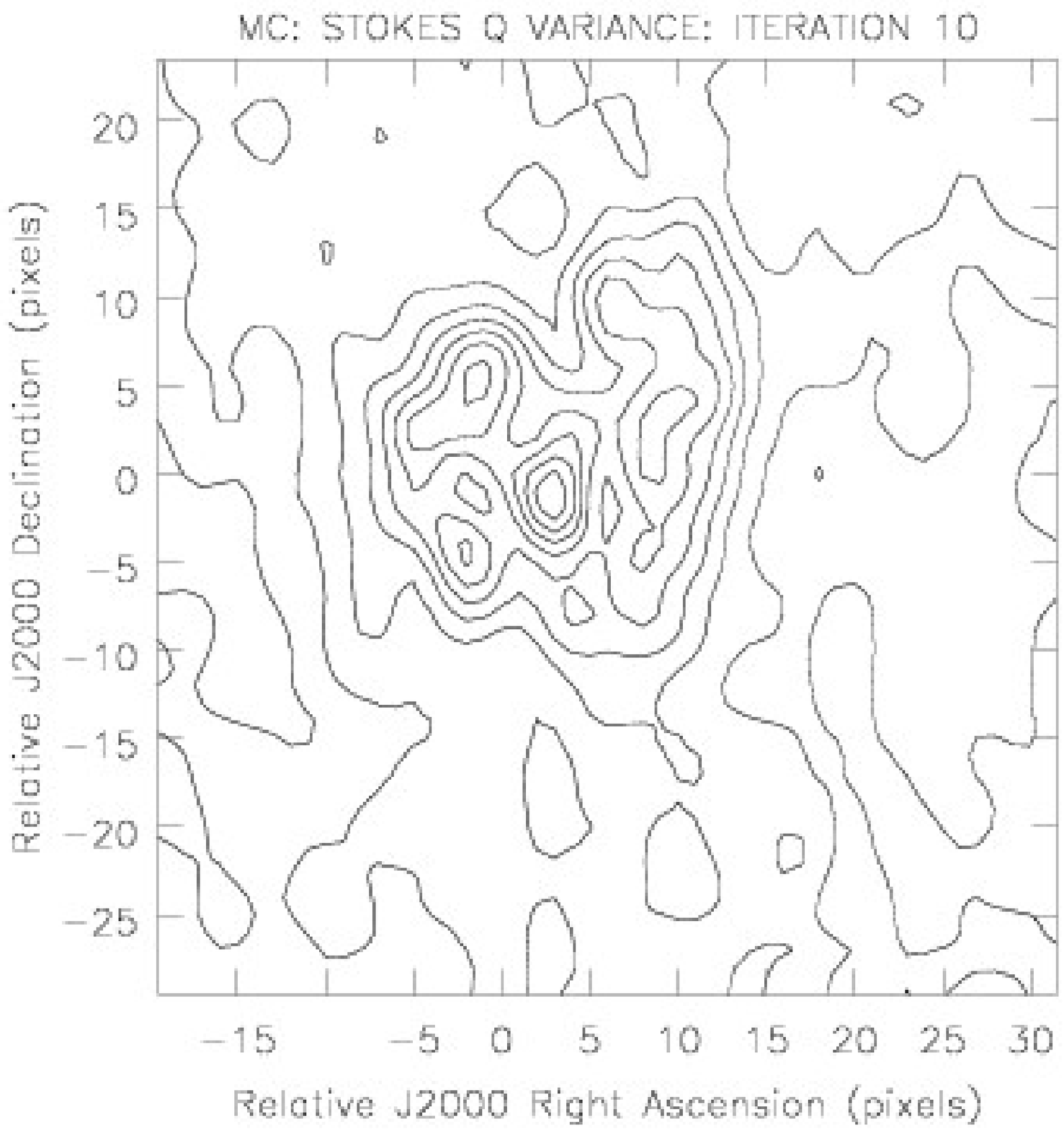} & 
\includegraphics[width=57mm, height=57mm, trim = 0mm 5mm 10mm 5mm, clip]{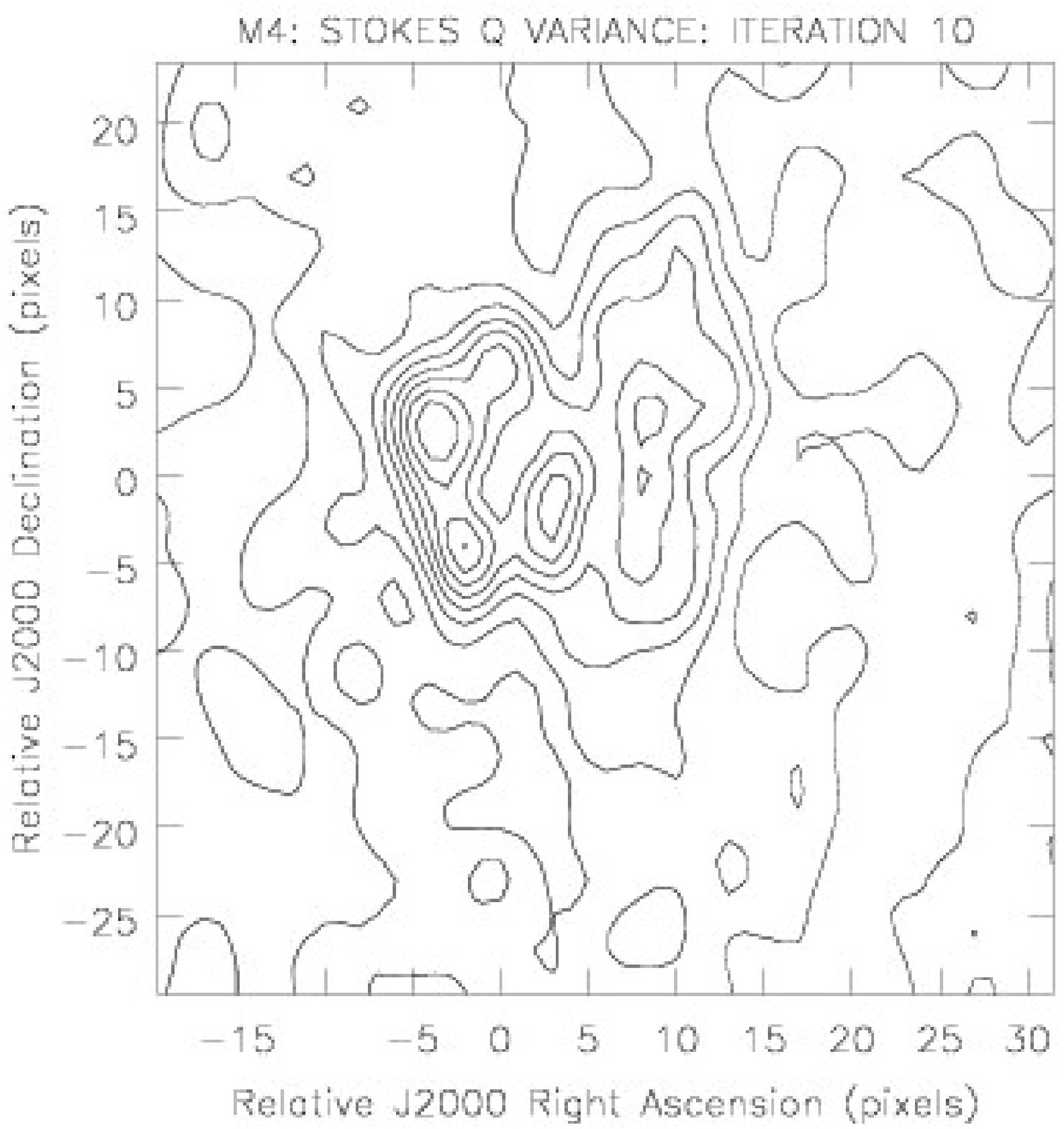} & 
\includegraphics[width=57mm, height=57mm, trim = 0mm 5mm 10mm 5mm, clip]{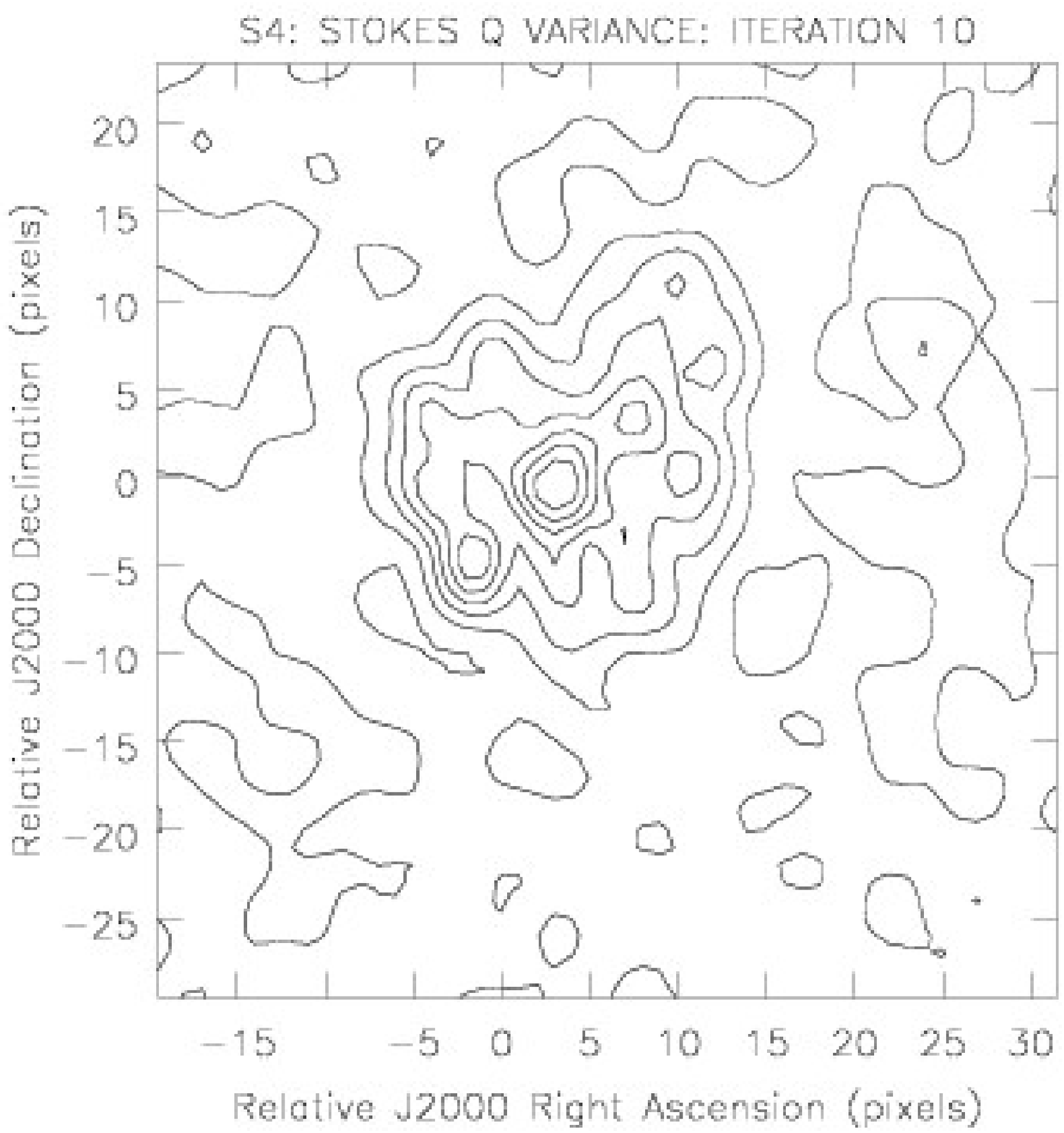} \\ 
\includegraphics[width=57mm, height=57mm, trim = 0mm 5mm 10mm 5mm, clip]{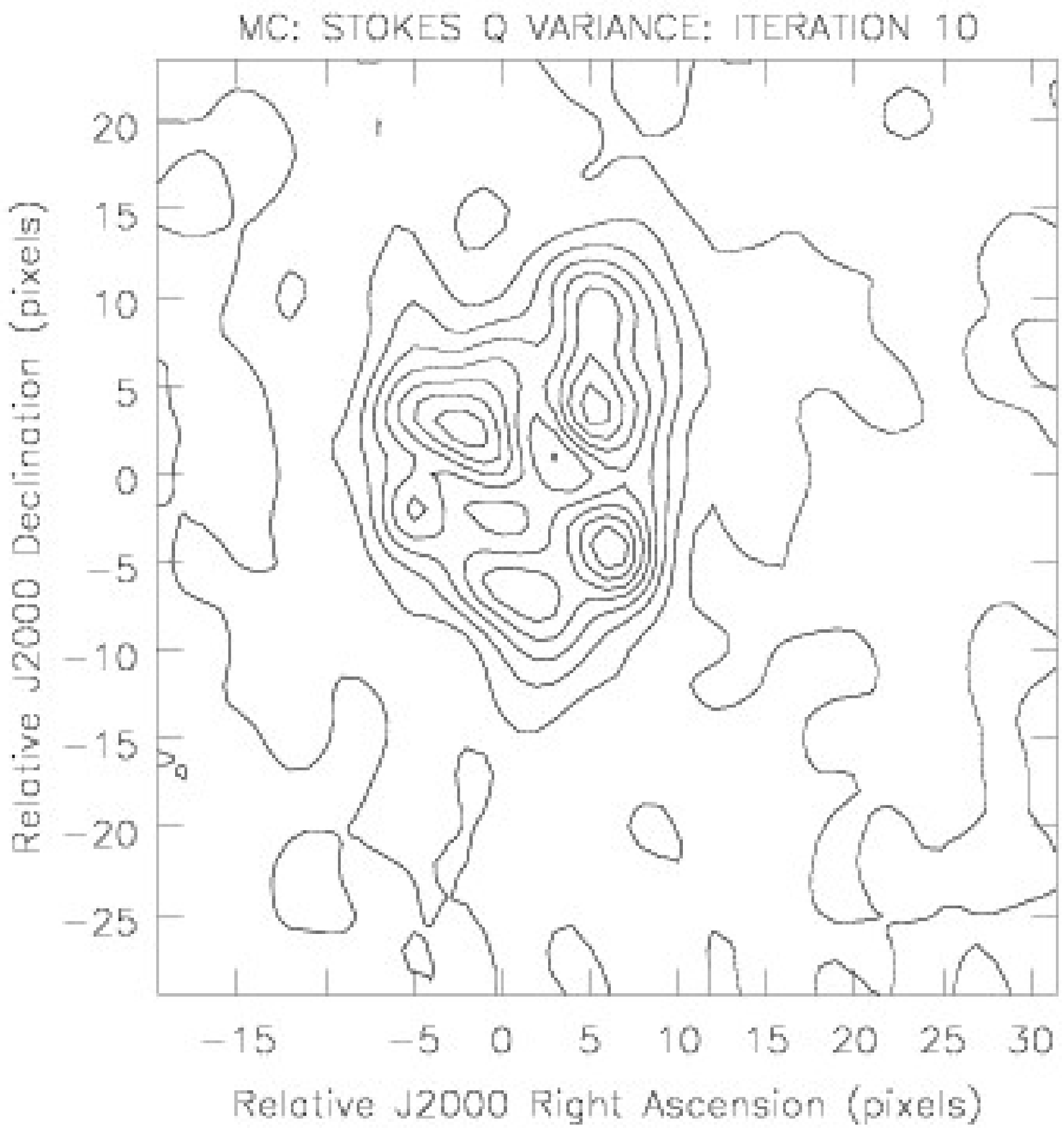} & 
\includegraphics[width=57mm, height=57mm, trim = 0mm 5mm 10mm 5mm, clip]{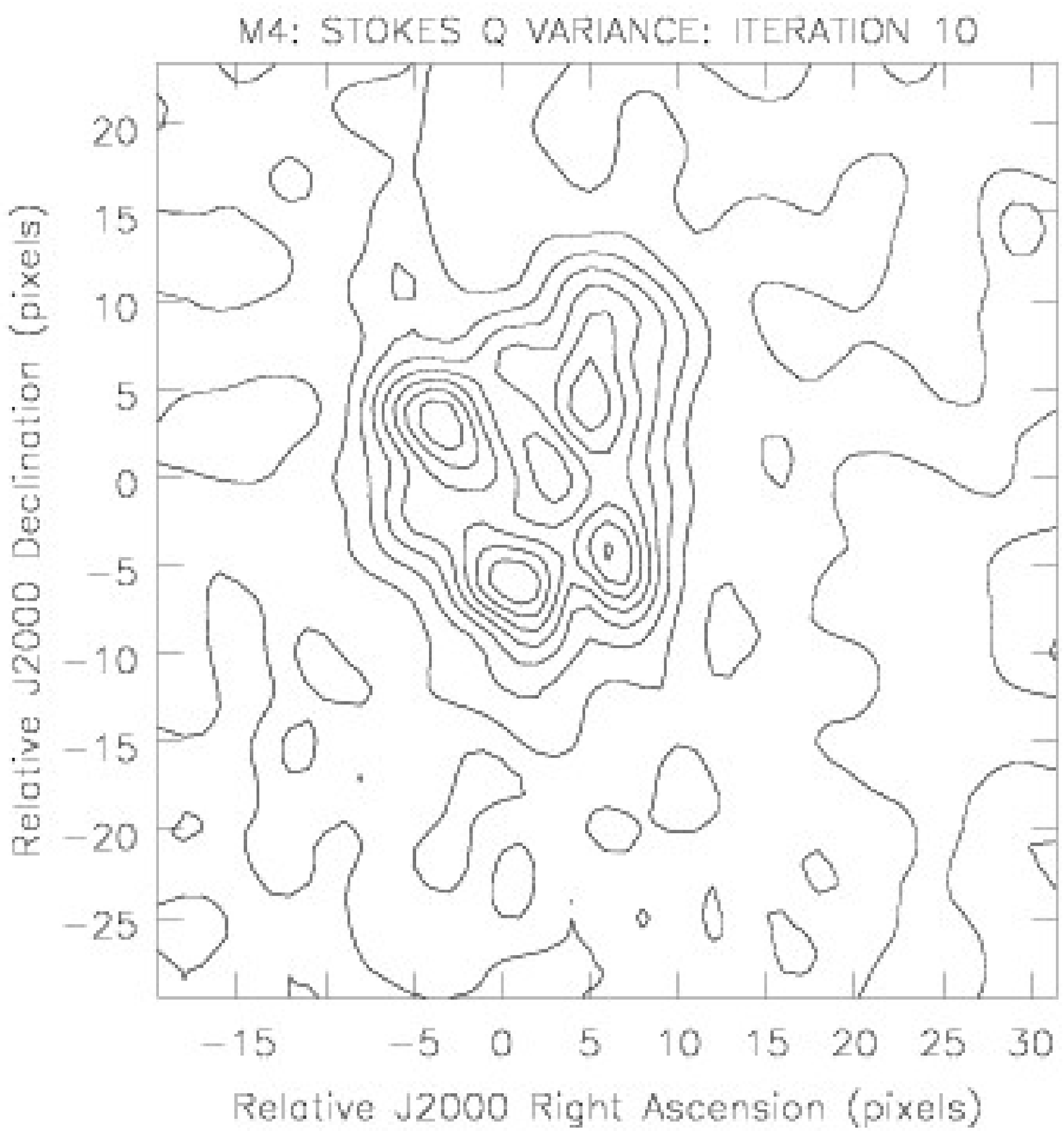} & 
\includegraphics[width=57mm, height=57mm, trim = 0mm 5mm 10mm 5mm, clip]{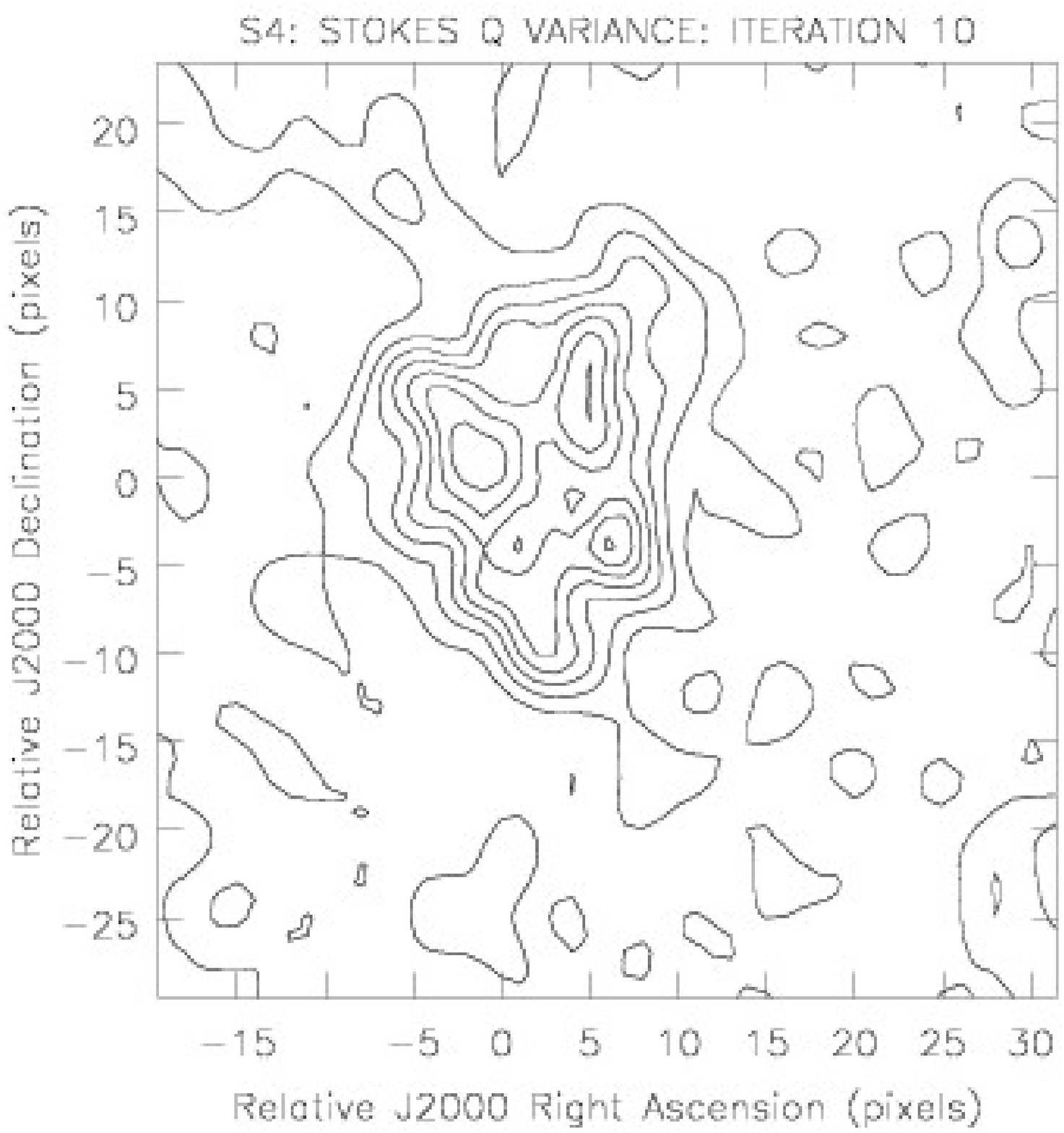} \\ 
\includegraphics[width=57mm, height=57mm, trim = 0mm 5mm 10mm 5mm, clip]{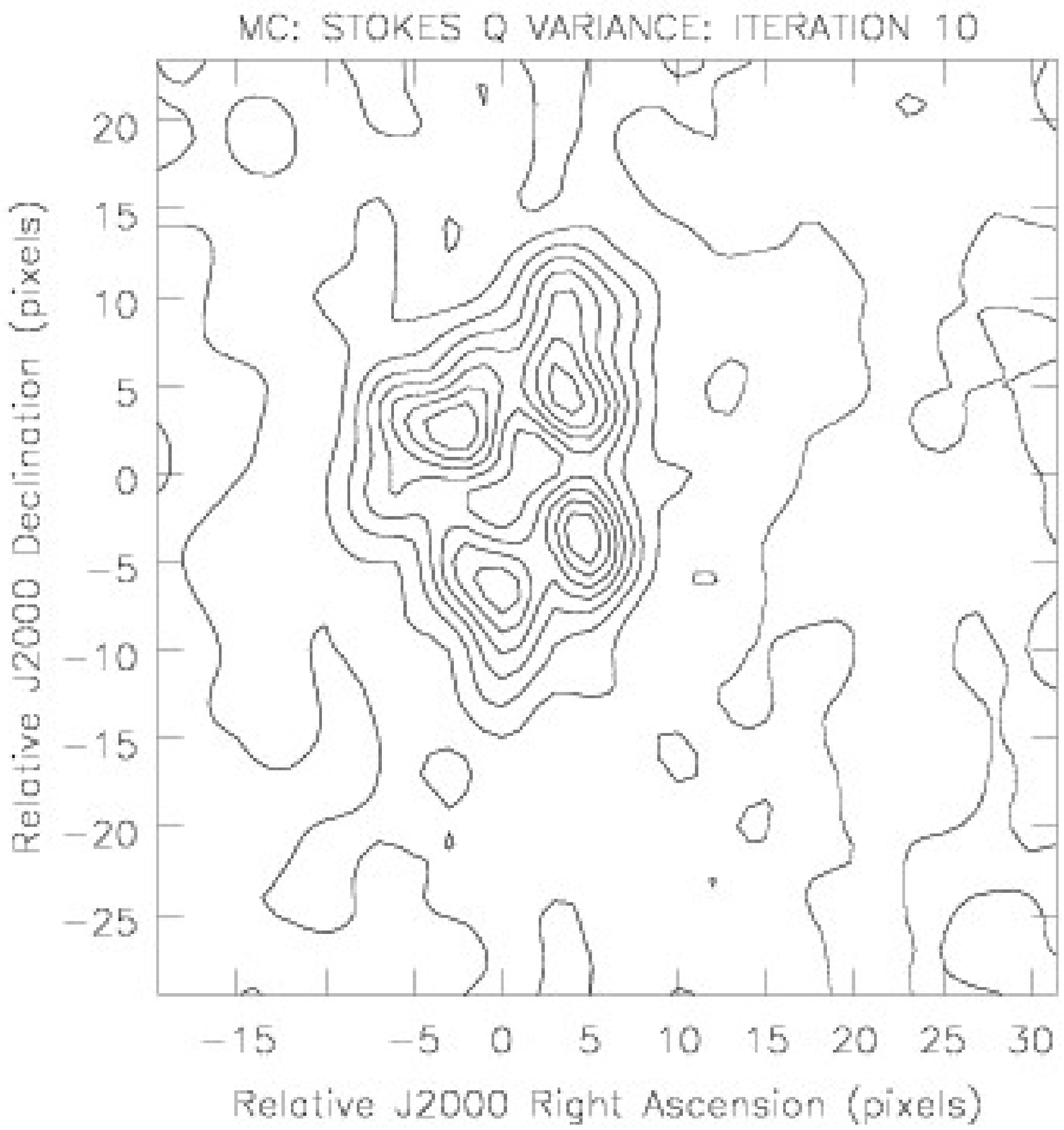} & 
\includegraphics[width=57mm, height=57mm, trim = 0mm 5mm 10mm 5mm, clip]{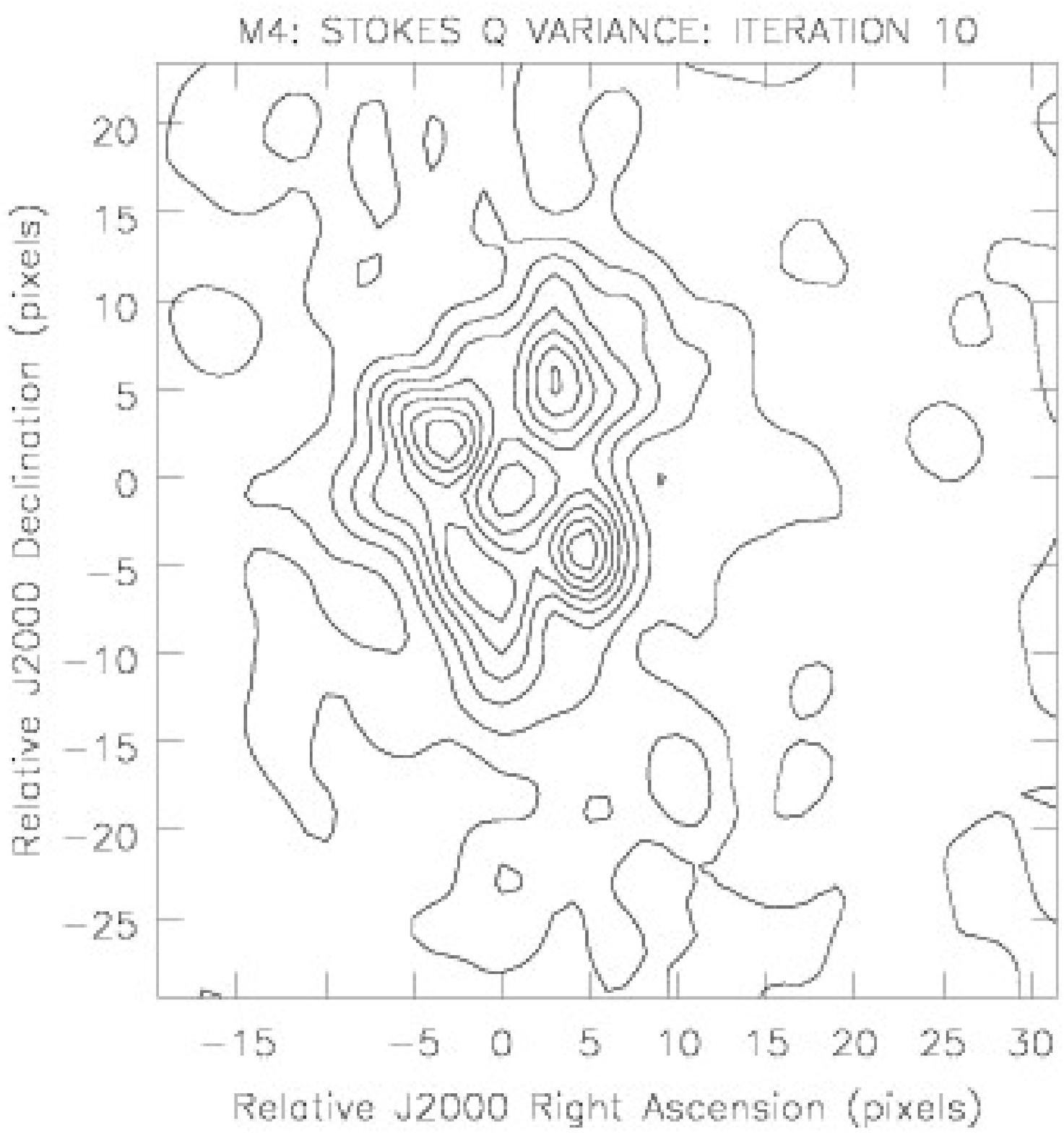} & 
\includegraphics[width=57mm, height=57mm, trim = 0mm 5mm 10mm 5mm, clip]{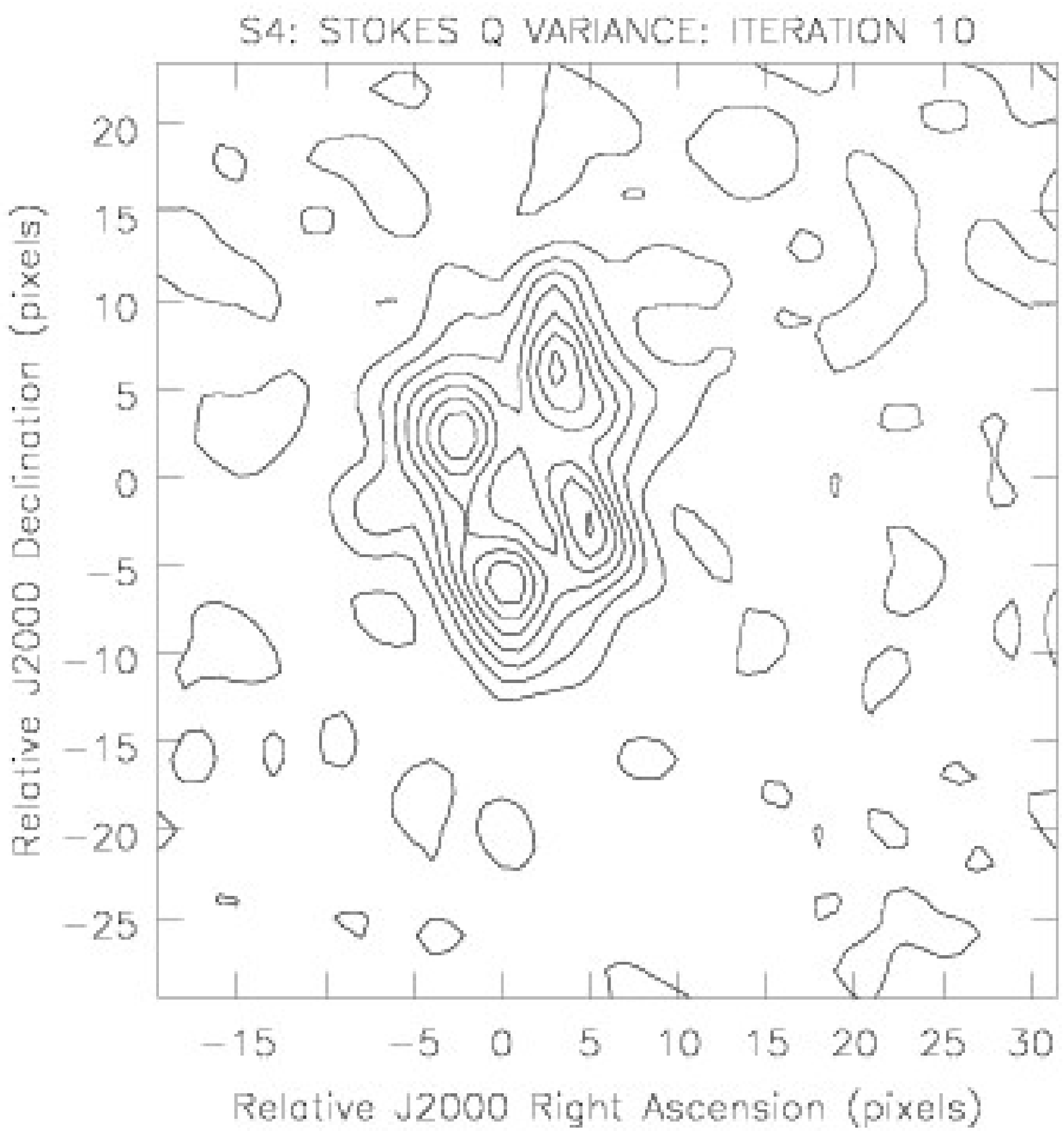} \\ 
\end{array}$ 

\caption{Same as Figure~\ref{fig-qvar-abc}, but for run codes X, Y,
and Z ({\it from top to bottom}), The contour levels are plotted at
levels \{0, 0.1, 0.2,...,1.0\} of the peak variance values: (X-MC:
1.451,\ X-M4: 1.372,\ X-S4: 2.415,\ Y-MC: 1.504,\ Y-M4: 1.438,\ Y-S4:
2.012,\ Z-MC: 1.507,\, Z-M4: 1.650,\ Z-S4: 2.351)$ \times 10^{-6}$
(Jy/beam)$^2$.}

\label{fig-qvar-xyz} 
\end{figure}

In Figures~\ref{fig-qvar-abc} through \ref{fig-qvar-xyz} we plot the
pixel-based Stokes $Q$ variance for the final polarization
self-calibration iteration for each run code \{A-F,\ X-Z\} for both
the direct Monte Carlo reference simulation and the optimal
model-based and subsample bootstrap methods identified in
Table~\ref{tbl-vmse}. In these figures, the Stokes $Q$ variance images
are drawn as contour plots with identical fractional contour levels in
each figure sub-panel.

\section{Discussion}

The results presented in the previous section show that the
applicability of statistical resampling as a tool for imaging fidelity
assessment in radio interferometry is upheld from our initial
evaluation presented in Paper I, when explored over a larger domain
subset, here a greater range of expected image fidelity and source
brightness distribution morphology. The current work continues to
support the conclusion that contemporary statistical resampling
techniques developed for data with long-range statistical dependence
\citep{lah03} are a promising method for estimating the statistical
properties of regularized calibration and imaging estimators used for
the solution of the imaging equation~\ref{eqn-ie} in radio
interferometry. The technique also intrinsically allows an estimate of
image quality per pixel, in contrast to coarser direction-independent
metrics, such as those based on off-source rms $\sigma_{off}$ for
example. As a bootstrap technique, it is also not sensitive to
assumptions about the functional form or parametrization of the parent
PDF of the measured visibility data, and does not require the frequent
idealization of uniform Gaussianity for the total noise
contribution.

As described in the preceding section, we assess each bootstrap method
quantitively against the direct Monte Carlo variance reference image;
this goodness-of-fit statistic $v_{MSE}$ is tabulated in
Table~\ref{tbl-vmse}. An optimal bootstrap method, within a given set,
is the method that agrees most closely with the Monte Carlo result by
this quantitative measure.

Our finding in Paper I that the model-based bootstrap generally strongly
outperforms the subsample bootstrap continues to hold true in the
broader evaluation described here. The statistical performance of the
subsample bootstrap improves with increasing subsample delete fraction
$f_s$, as expected theoretically \citep{dav97}. The subsample
bootstrap performance drops precipitously for low or moderate delete
fraction $f_s < 0.5$ and the optimal subsample bootstrap code is found to
be S4 consistently across all run codes \{A-F,\ X-Z\} in the current
study. The subsample delete fraction for bootstrap code S4 is 0.75,
the highest value of $f_s$ amongst the bootstrap codes
\{S1-S4\}. Despite the poorer statistical performance of the subsample
bootstrap, we note that it has the advantage of ease of implementation
and requires no free user-adjustable parameters if a fixed value $f_s=0.75$ is
adopted.

The optimal model-based bootstrap code is found to be M3 across run
codes \{A-F\} and M4 for run codes \{X-Z\}. The best single bootstrap
over all run codes is M3. The model-based bootstrap used in this study
models the long-term statistical dependence in the data by fitting a
series of independent segmented polynomial fits (of maximum degree
$N_p$) to the visibility time-series on each interferometer baseline
over successive solution intervals of length $\dtb$. This does not
require a priori knowledge of the source brightness distribution,
however. Full details of the implementation of this method can be
found in Paper I. Bootstrap codes M3 and M4 have increasing solution
intervals $\dtb^{M4} > \dtb^{M3}$ (as enumerated in
Table~\ref{tbl-mcode}). The optimality of M4 over M3 for run codes
\{X-Z\} is consistent with the more compact source morphology for
these run codes, as noted in Table~\ref{tbl-sep}, and therefore a
lower mean rate of change of source visibility over time. As shown in
Table~\ref{tbl-vmse}, the relative performance of the model-based
bootstrap shows a weak dependence on $\dtb$, compared to the strong
dependence of the subsample bootstrap performance on the delete
fraction $f_s$. In this sense the model-based bootstrap is also more
robust against sub-optimal bootstrap parameter choices. The current
study confirms that $\dtb$ should be chosen with regard to the
expected mean time rate of change of the visibility function. This can
also in principle be deduced automatically from the actual observed time-variability in
$V_{mn}^{obs}$. The interval $\dtb$ needs to be chosen as the longest
interval over which the piece-wise polynomial model for the
statistical dependence in the visibility time-series holds on each
interferometer baseline.

\begin{figure}
\plotone{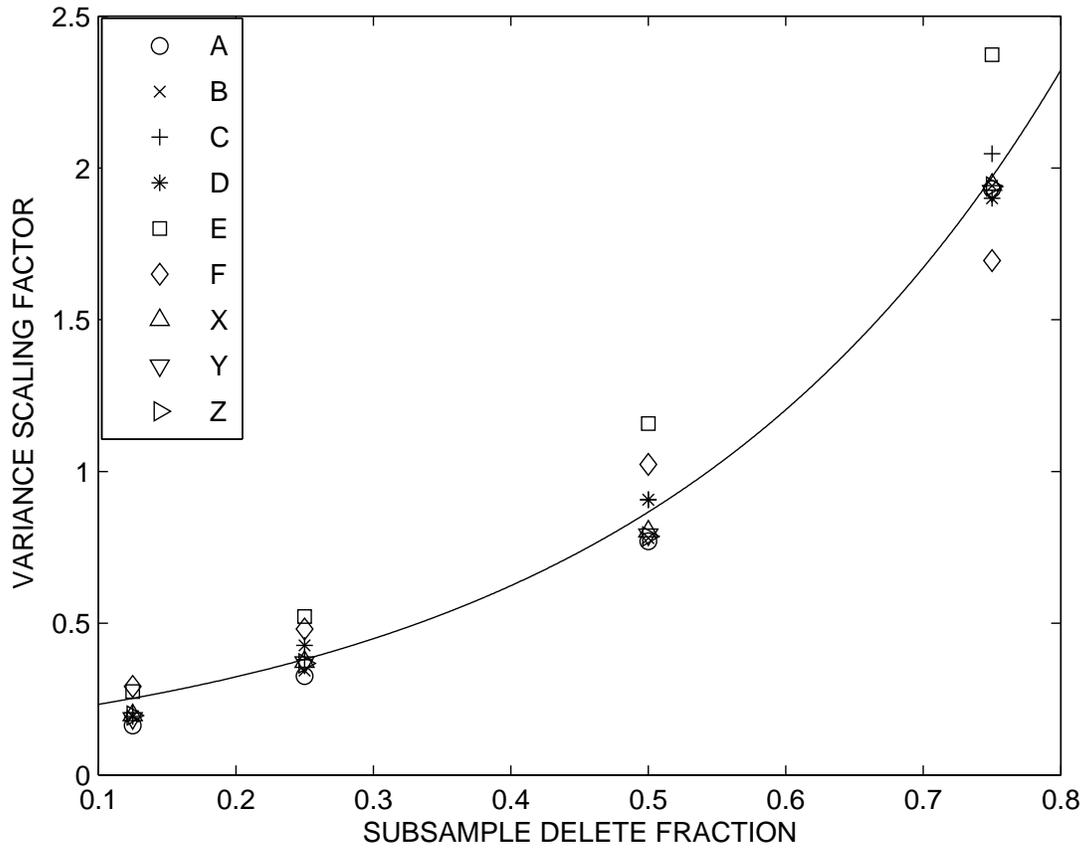}

\caption{The variance scaling factor, $v_f$, plotted against the
subsampling delete fraction, $f_s$\label{fig-vf-fs}, obtained from
subsample bootstrap runs \{S1-S4\} for run codes \{A-F,\ X-Z\}. The
best joint exponential fit is plotted as a solid line, with parameters
as described in the main text.}

\label{fig-vf-scale}

\end{figure}

For the subsample bootstrap, the variance scaling factor $v_f$ is
expected to have a power-law dependence on subsample delete fraction
$f_s$. \citet{dav97} quote an expected relation $v_f \sim
(1-f_s)^{-a}$ for the related delete-d jacknife method. The fitted
scaling factors from the run codes \{A-F,\ X-Z\} in the current study
are plotted against subsample delete fraction $f_s$ in
Figure~\ref{fig-vf-scale}. We find the relation to be broadly
consistent across the different parallactic angle ranges, which have
different numbers of visibilities $N_{vis}$, and the source
morphologies considered in the current work. We find empirically in
the current study that a function of the form $v_f=b e^{-a(1-f_s)}$
provides a better fit to the data (either separately for each run code
or jointly across all run codes) than $v_f=b(1-f_s)^a$ or $v_f=a f_s^2
+ bf_s + c$, although we stress that the current data do not
statistically rule out these models. Further studies with a finer
sampling in $f_s$ are planned to constrain this relation more
closely. Additional work will also be required to determine its broader
applicability; this is an empirical relation in its current form. A joint best
fit across run code \{A-F,\ X-Z\} is drawn as a solid line in
Figure~\ref{fig-vf-scale} in the form:

\begin{equation}
v_f(f_s) = 4.481 e^{-3.287(1-f_s)}
\end{equation}

The subset of runs comprising the direct Monte Carlo simulations over
parallactic angle ranges \{A-F\} provide an opportunity to study the
performance of the polarization self-calibration algorithm introduced
in Paper I. The total MSE for the estimator $\hdm$, averaged over
antenna and polarization receptor basis, is plotted on a logarithmic
scale against self-calibration iteration number in
Figure~\ref{fig-calmse}. As noted, and demonstrated, in Paper I, the
algorithm is expected, from relative information arguments, to
demonstrate rapid convergence for the case of good parallactic angle
coverage. The current study measures the algorithm performance for
decreasing parallactic angle range over run codes
\{A-F\}. Figure~\ref{fig-calmse} shows that there is good convergence
for parallactic angle ranges $\ta \sim 80-140^\circ$ that include source
transit, moderate convergence for $\ta \sim 30-60^\circ$, and poor
convergence for $\ta \sim 0^\circ$. This is consistent with heuristic
rules of thumb for other interferometric polarization calibration
techniques that recommend a minimum parallactic angle range $\ta >
100^\circ$ \citep{lep95}.

\begin{figure}[h] 
\advance\leftskip-1cm
\advance\rightskip-1cm
 $\begin{array}{c@{\hspace{2mm}}c@{\hspace{2mm}}c} 
\includegraphics[width=57mm, height=57mm, trim = 0mm 5mm 10mm 5mm, clip]{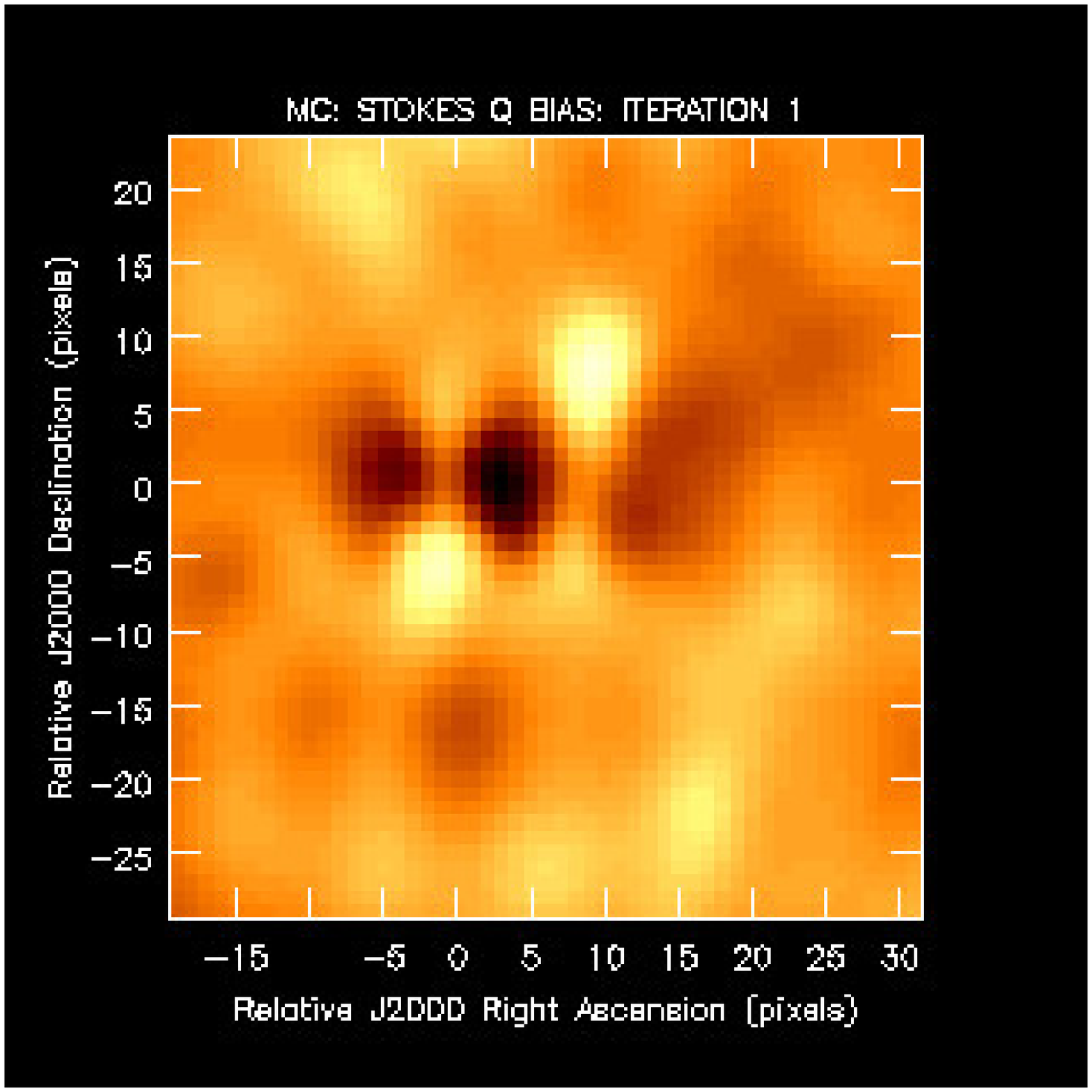} & 
\includegraphics[width=57mm, height=57mm, trim = 0mm 5mm 10mm 5mm, clip]{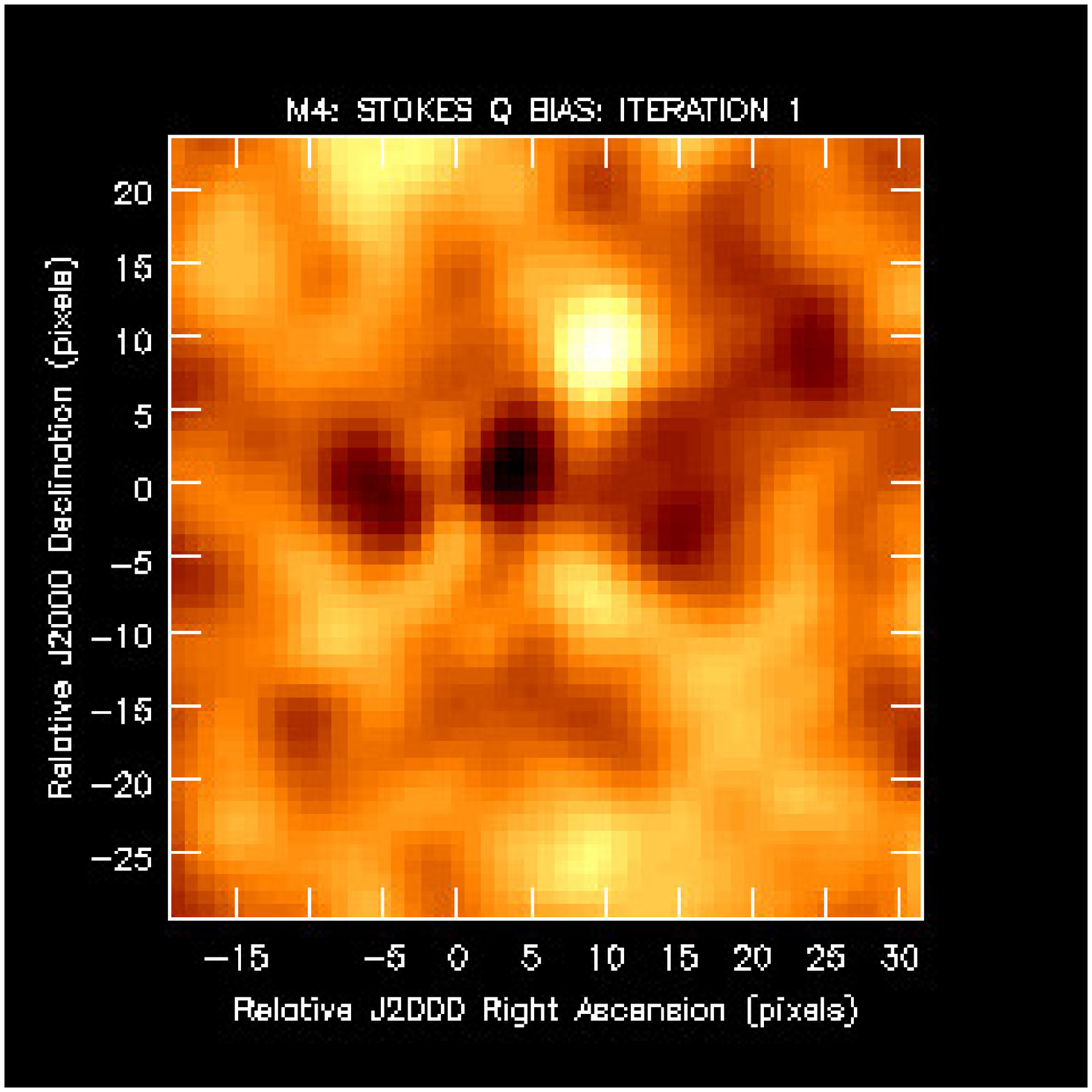} & 
\includegraphics[width=57mm, height=57mm, trim = 0mm 5mm 10mm 5mm, clip]{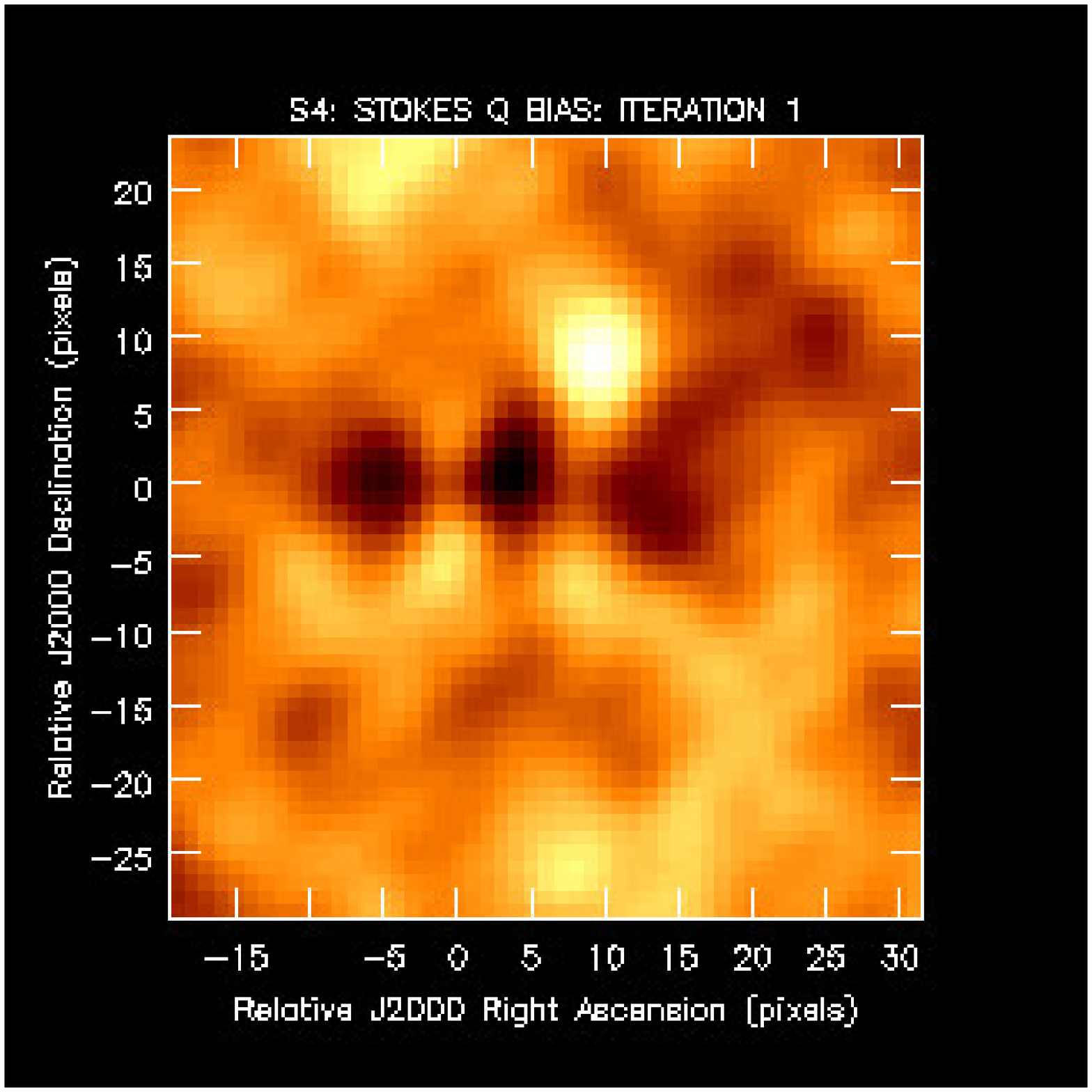} \\ 
\includegraphics[width=57mm, height=57mm, trim = 0mm 5mm 10mm 5mm, clip]{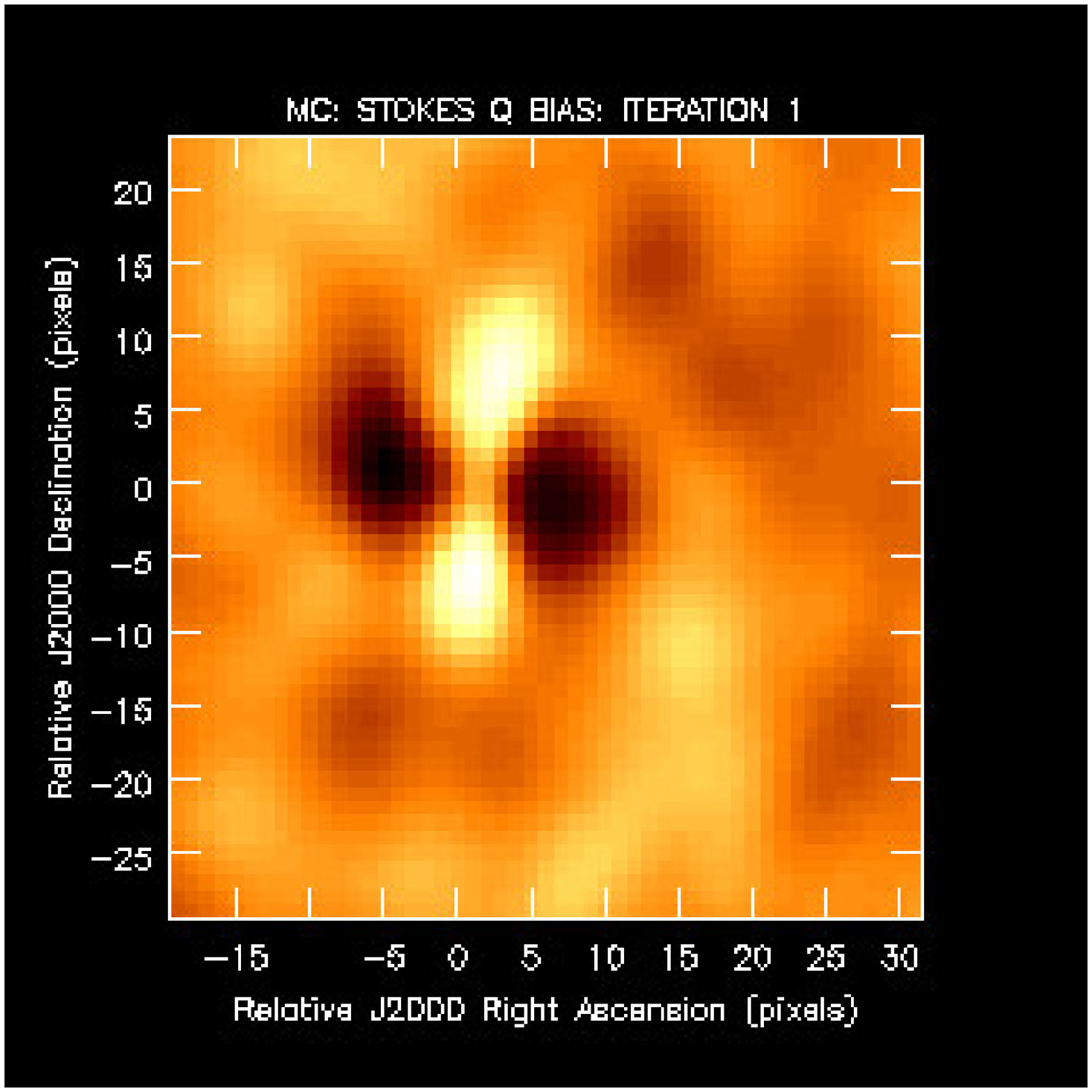} & 
\includegraphics[width=57mm, height=57mm, trim = 0mm 5mm 10mm 5mm, clip]{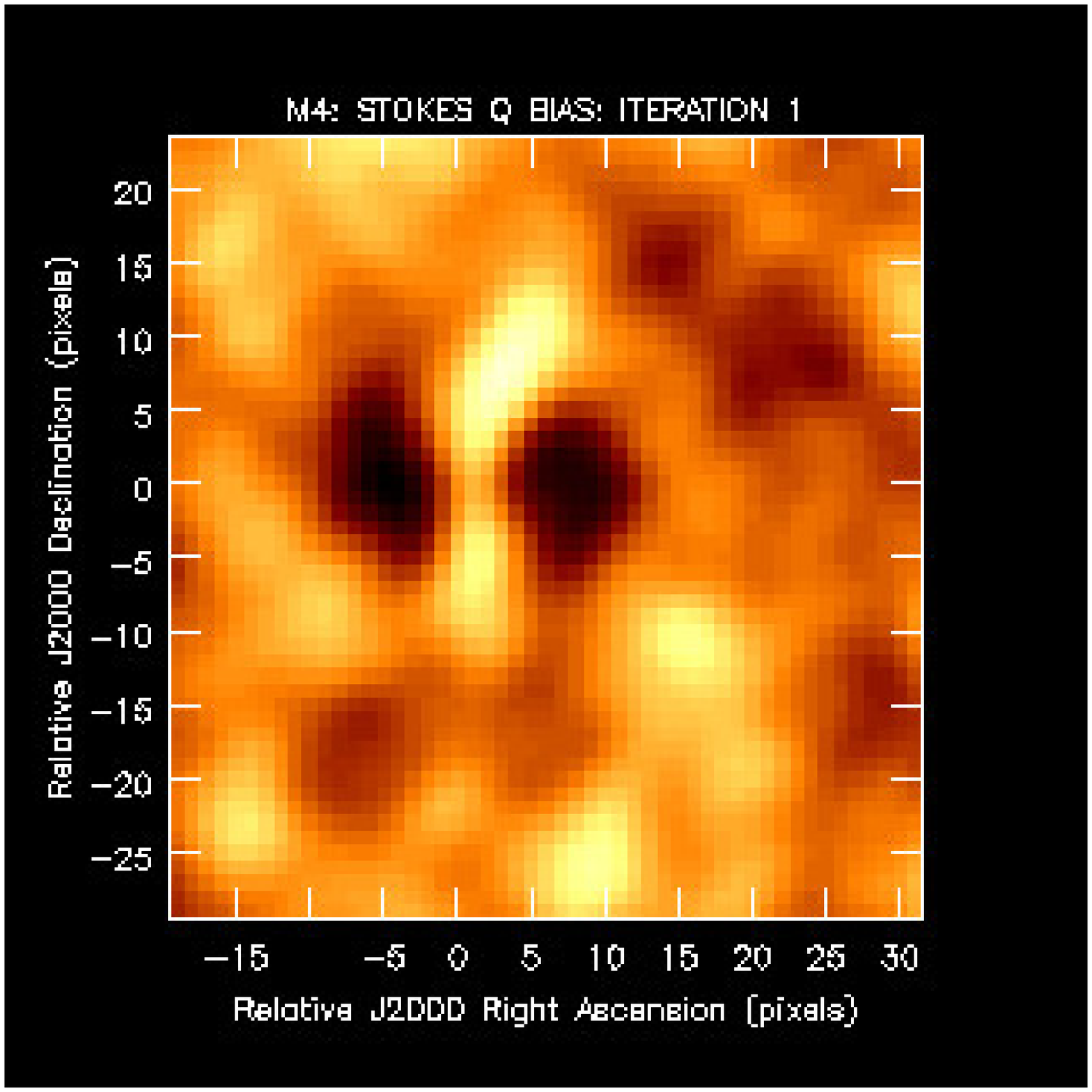} & 
\includegraphics[width=57mm, height=57mm, trim = 0mm 5mm 10mm 5mm, clip]{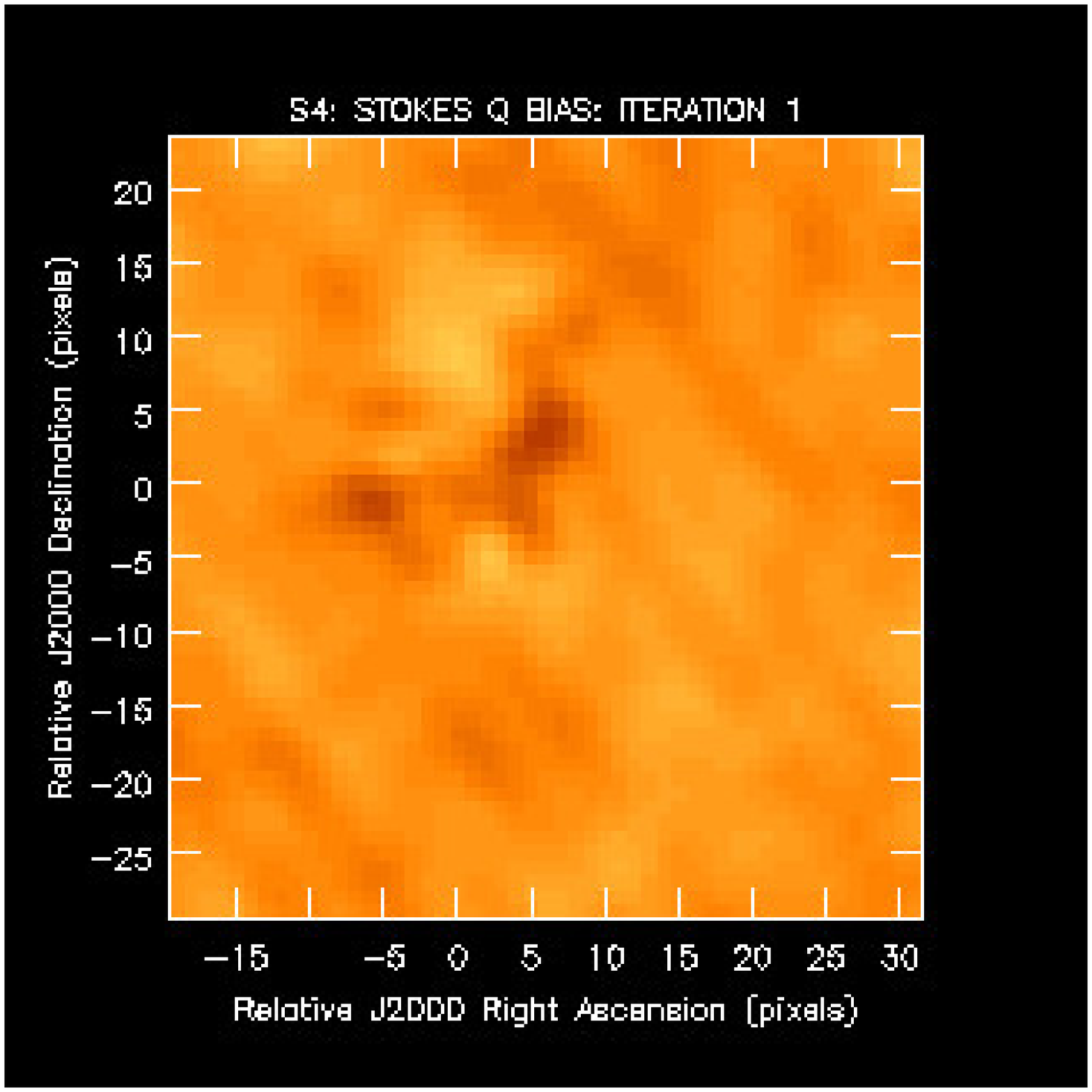} \\ 
\end{array}$ 

\caption{The imaging estimator Stokes $Q$ bias for the first iteration
of polarization self-calibration for run codes X ({\it top}) and Z
({\it bottom}), as measured by direct Monte Carlo simulation (MC; {\it
left}), model-based bootstrap resampling (M4; {\it center}), and
subsample bootstrap resampling (S4; {\it right}) methods. The bootstrap codes
are chosen for their optimality, as indicated in Table~\ref{tbl-vmse}.
The same default color mapping is used for each bootstrap image bias
cube, where increasing color brightness denotes increasing image
bias.}

\label{fig-col-qbias} 
\end{figure}

Although the current study has focused on measuring the variance of
the sampling distribution of the imaging estimator $\hip$, we show
initial results on bias estimation in
Figure~\ref{fig-col-qbias}. These example images are selected for
their expected high levels of bias, as anticipated in the current
study for the case of Stokes $\{Q,U\}$ images at low polarization
self-calibration iteration numbers, where the instrumental
polarization terms will still have high MSE. As for the variance study, we
assess the statistical performance of the bootstrap resampling
estimate of bias against the result obtained by direct Monte Carlo
simulation. Figure~\ref{fig-col-qbias} shows that the optimal variance
estimators, as identified in Table~\ref{tbl-vmse}, show qualified
promise as bias estimators in the high-bias regime. However, we note
that our preliminary experience is that they have lower statistical
performance than variance estimators. \citet{efr93} discuss the
performance of bootstrap bias estimators, and propose techniques that
can significantly improve their statistical performance. We intend to
examine bias estimation in this application in more detail in a future
paper.

We also note that new statistical imaging techniques of the type
considered in this paper allow novel approaches to understanding direction-dependent
radio-interferometric calibration and imaging fidelity, especially
those challenges posed by wide-field imaging at uniform, high dynamic
range, as needed for future leading-edge telescopes such as the
SKA. These techniques offer a new approach, distinct from instrument
simulation which has well-known limitations, to derive
direction-dependent imaging fidelity assessments from single
observations with real prototype arrays.

The computational cost of the bootstrap resampling techniques
evaluated here is approximately $\sim O(10^2)$ times the processing
cost for a single dataset without resampling. This places it well
within contemporary HPC resources at the teraflop scale and those
projected to be available soon in the petascale era \citep{bad08}. The
cost of the algorithm is strongly mitigated by its high degree of
scalability and effective parallel decomposition.

\section{Conclusions}

We draw the following conclusions from the current evaluation of
bootstrap resampling for radio-interferometric imaging fidelity
assessment over a broader domain subset:

\begin{enumerate}

\item{Statistical resampling techniques for dependent data remain a
promising approach to the important problem of pixel-level fidelity
assessment in radio-interferometric calibration and imaging. They show
good statistical performance as assessed against direct Monte Carlo
simulation, and are computationally affordable and scalable on
contemporary and future HPC resources. They have the desirable
statistical properties of independence on the detailed functional form
or parametrization of the parent PDF of the observed visibility data
and of not requiring an analytic solution for the calibration and imaging
inference problem.}

\item{Model-based bootstrap techniques outperform subsampling
bootstrap methods in general in our study, although both are of value
in this application. The subsample bootstrap performs well for delete
fraction $f_s \sim 0.75$, is straight-forward to implement, and has no
adjustable parameters if $f_s$ is adopted as this value. The model-based
bootstrap has superior statistical performance in general. It has free
parameters $(N_p,\ \dtb)$ that can in principle be deduced from the
observed data and against which the method is relatively robust to the
exact values used.}

\item{The subsample bootstrap variance scaling relation $v_f=f(f_s)$
measured in the current study shows relatively limited variability
across the broader domain subset considered here. The current data are
consistent with an exponential function: $v_f=be^{-a(1-f_s)}$.}

\item{The polarization self-calibration algorithm introduced in Paper
I is found to converge at a faster than exponential rate for good
parallactic coverage $\ta$. Its performance for lower parallactic
angle ranges is found to deteriorate for $\ta < 100^\circ$ and to break
down in the limit as $\ta \to 0$ as expected.}

\item{An initial examination of statistical resampling for bias
estimation shows qualified promise in the high-bias regime in this discipline,
but further evaluation is required.}

\end{enumerate}

\acknowledgments

This material is based upon work supported by the National Science
Foundation under Grant Nos. AST-0506745 and AST-0431486. Any opinions,
findings, and conclusions or recommendations expressed in this
material are those of the authors and do not necessarily reflect the
views of the National Science Foundation. This work was also partially
supported by the National Center for Supercomputing Applications under
AST-050003 and used the system tungsten.ncsa.uiuc.edu.

\clearpage

\begin{deluxetable}{lrrrr}
\tabletypesize{\scriptsize}
\tablecaption{Parallactic angle ranges and run codes\label{tbl-par}}
\tablewidth{0pt}
\tablehead{
\colhead{} & \colhead{$|\ta_1|$} & \colhead{$|\ta_2|$} &  
\colhead{$|\bta|$} & \colhead{$N_{vis}$} \\
\colhead{Run code} & \colhead{(deg)} & \colhead{(deg)} &
\colhead{(deg)} & \colhead{}
}
\startdata
A   &   155  &  123  & 139   & 222750 \\
B   &   155  &   67  & 111   & 182250 \\
C   &   155  &    4  &  79.5 & 141750 \\
D   &   121  &    0  &  60.5 & 101250 \\
E   &    59  &    0  &  29.5 &  60750 \\
F   &     3  &    0  &   1.5 &  20250 \\
\enddata
\tablecomments{$\ta_1$ and $\ta_2$ are the parallactic angle ranges at
the central antenna (Pie Town) before and after source transit. The
mean value is denoted as $\bta$. The number of visibilities in
each simulated dataset is listed under column $N_{vis}$.  }
\end{deluxetable}

\begin{deluxetable}{lrr}
\tabletypesize{\scriptsize}
\tablecaption{Simulation source model component separations\label{tbl-sep}}
\tablewidth{0pt}
\tablehead{
\colhead{} & \colhead{$\Delta \alpha$} & \colhead{$\Delta \delta$} \\
\colhead{Run code} & \colhead{(mas)} & \colhead{(mas)} 
}
\startdata
A & -0.5      & +0.1 \\ \hline
X & -0.25     & +0.05 \\
Y & -0.125    & +0.025 \\
Z & -0.0625   & +0.0125 \\
\enddata
\end{deluxetable}

\begin{deluxetable}{lcc}
\tabletypesize{\scriptsize}
\tablecaption{Model-based bootstrap run parameters\label{tbl-mcode}}
\tablewidth{0pt}
\tablehead{
\colhead{Code} & \colhead{max($N_p$)} & \colhead{$\dtb$ (sec)} 
}
\startdata
M1 & 1 & 60 \\
M2 & 1 & 150 \\
M3 & 1 & 300 \\
M4 & 2 & 900 \\
\enddata
\tablecomments{max($N_p$) is the maximum degree of the model polynomial used in each bootstrap model interval $\dtb$.}
\end{deluxetable}

\begin{deluxetable}{lc}
\tabletypesize{\scriptsize} 
\tablecaption{Subsample bootstrap run parameters\label{tbl-scode}} 
\tablewidth{0pt} 
\tablehead{
\colhead{Code} & \colhead{$f_s$} } 
\startdata 
S1 & 0.125 \\
S2 & 0.25 \\ 
S3 & 0.5 \\
S4 & 0.75 \\
\enddata
\tablecomments{$f_s$ is the fraction of data deleted in each subsample realization}
\end{deluxetable}

\begin{deluxetable}{lrrrrrrrr}
\tabletypesize{\scriptsize}
\tablecaption{Bootstrap performance: variance scale factor $v_f$\label{tbl-vf}}
\tablewidth{0pt}
\tablehead{
\colhead{Run code} & \colhead{M1} & \colhead{M2} & \colhead{M3} &
\colhead{M4} & \colhead{S1} & \colhead{S2} & \colhead{S3} & \colhead{S4} 
}
\startdata
A &   1.00 &   1.00 &   0.99 &   1.02 &   0.16 &   0.33 &   0.77 &   1.93\\
B &   0.95 &   0.99 &   1.01 &   1.03 &   0.19 &   0.34 &   0.78 &   1.94\\
C &   0.97 &   1.02 &   1.00 &   1.03 &   0.20 &   0.38 &   0.91 &   2.05\\
D &   0.98 &   0.98 &   0.98 &   0.99 &   0.19 &   0.43 &   0.91 &   1.90\\
E &   0.96 &   0.98 &   0.95 &   1.01 &   0.28 &   0.52 &   1.16 &   2.37\\
F &   0.91 &   0.94 &   1.03 &   0.98 &   0.29 &   0.48 &   1.02 &   1.70\\
\hline
X &   0.97 &   0.98 &   1.00 &   1.00 &   0.20 &   0.37 &   0.80 &   1.94\\
Y &   0.96 &   1.00 &   1.01 &   1.02 &   0.19 &   0.37 &   0.79 &   1.92\\
Z &   0.97 &   1.00 &   1.02 &   1.02 &   0.20 &   0.37 &   0.79 &   1.94\\
\enddata
\tablecomments{The bootstrap performance measure $v_f$ is defined in
the main text.}
\end{deluxetable}

\begin{deluxetable}{lrrrrrrrrc}
\tabletypesize{\scriptsize}
\tablecaption{Bootstrap performance: variance mean-squared error $v_{MSE}$\label{tbl-vmse}}
\tablewidth{0pt}
\tablehead{
\colhead{} & \colhead{M1} & \colhead{M2} & \colhead{M3} &
\colhead{M4} & \colhead{S1} & \colhead{S2} & \colhead{S3} & \colhead{S4} &
\colhead{min$(v_{MSE})$} \\
\colhead{Run code} & \colhead{} & \colhead{} & \colhead{} &
\colhead{} & \colhead{} & \colhead{} & \colhead{} & \colhead{} &
\colhead{$\times 10^{-15}$} 
}
\startdata
A &  1.42 &  1.05 &  1.00 &  1.07 &  3.18 &  2.53 &  1.73 &  1.52 & 3.40 \\
B &  1.36 &  1.12 &  1.00 &  1.14 &  5.81 &  3.87 &  2.14 &  1.64 & 4.73 \\
C &  1.40 &  1.17 &  1.00 &  1.15 &  5.10 &  3.65 &  2.32 &  1.73 & 6.67 \\
D &  1.53 &  1.18 &  1.00 &  1.13 &  5.51 &  5.49 &  3.48 &  2.73 & 16.4 \\
E &  1.39 &  1.27 &  1.00 &  1.41 &  5.43 &  5.09 &  4.05 &  2.03 & 20.0 \\
F &  2.30 &  1.47 &  1.00 &  2.28 &  14.8 &  12.3 &  10.7 &  7.94 & 985 \\
\hline
X &  1.28 &  1.10 &  1.02 &  1.00 &  4.90 &  3.70 &  1.72 &  1.22 & 3.39 \\
Y &  1.45 &  1.35 &  1.17 &  1.00 &  5.40 &  4.49 &  2.16 &  1.51 & 2.91 \\
Z &  1.42 &  1.23 &  1.14 &  1.00 &  5.93 &  4.27 &  2.05 &  1.38 & 2.97 \\

\enddata

\tablecomments{The right-most column is the minimum $v_{MSE}$ measured
across bootstrap type for each run code. The values for bootstrap
codes \{M1-M4,\ S1-S4\} in each row are scaled by this minimum value,
to allow the relative performance of the different bootstrap methods
for a given run code to be more easily compared.}

\end{deluxetable}



\clearpage

\clearpage

\clearpage

\clearpage

\end{document}